\documentclass[english,12pt]{article}
\usepackage{graphicx}
\usepackage{epsfig}
\usepackage{times}
\usepackage{cite}
\usepackage{color}
\usepackage[T1]{fontenc}
\usepackage[latin9]{inputenc}
\usepackage{amsmath}
\usepackage{graphicx}
\usepackage{amssymb}
\usepackage{babel}

\usepackage{ulem} 
\usepackage{color}

\def\pcomma{$^,$}
\usepackage[margin=1in]{geometry}
\begin{document}
\hfill{\bf Date:} {\today}
\begin{center}
\section*{PHYSICS OPPORTUNITIES WITH MESON BEAMS} 
\end{center}
\begin{center}
William~J.~Briscoe$^a$\pcomma\footnote[1]{Electronic address: briscoe@gwu.edu}, 
Michael D\"oring$^a$\pcomma\footnote[2]{Electronic address: doring@gwu.edu}, 
Helmut~Haberzettl$^a$\pcomma\footnote[3]{Electronic address: helmut@gwu.edu}, 
D.\ Mark Manley$^b$\pcomma\footnote[4]{Electronic address: manley@kent.edu},\\ 
Megumi Naruki$^c$\pcomma\footnote[5]{Electronic address: m.naruki@scphys.kyoto-u.ac.jp}, 
Igor~I.~Strakovsky$^a$\pcomma\footnote[6]{Electronic address: igor@gwu.edu}, 
Eric S.\ Swanson$^d$\pcomma\footnote[7]{Electronic address: swansone@pitt.edu}
\end{center}
\hspace{0.15in}

$^a$ The George Washington University, Washington, DC 20052, USA\\
$^b$ Kent State University, Kent, OH 44242, USA \\
$^c$ Kyoto University, Kyoto 606-8502, Japan \\
$^d$ University of Pittsburgh, Pittsburgh, PA 15260, USA \\
\noindent

\begin{center}{\large\bf Abstract}\end{center}


Over the past two decades, meson photo- and electroproduction
data of unprecedented quality and quantity have been measured at
electromagnetic facilities worldwide. 
By contrast, the meson-beam data for the same hadronic final states
are mostly outdated and largely of poor quality, or even non-existent,
and thus
provide inadequate input to help interpret, analyze, and exploit the full
potential of the new electromagnetic data. To reap the full benefit of the
high-precision electromagnetic data, new high-statistics data from
measurements
with meson beams, with good angle and energy coverage for a wide range of
reactions, are critically needed to advance our knowledge in
baryon and
meson spectroscopy and other related areas of hadron physics. To address
this
situation, a state-of-the-art meson-beam facility needs to be constructed.
The present paper summarizes unresolved issues in hadron physics and
outlines the vast opportunities and advances that only become possible with
such a facility.

PACS numbers: 
 13.75.Gx,     
 13.75.Jz,     
 14.20.Gk,     
 14.20.Jn,     
 14.40.Be,     
 14.40.Df,     
 14.40.Rt      
\section{Introduction}
\label{sec:intro}


Great strides have been made over the last two decades to increase our knowledge of baryon spectroscopy with the help of meson photo- and electroproduction data of unprecedented quality and quantity coming out of major electromagnetic (EM) facilities such as JLab, MAMI, ELSA, SPring-8, BEPC, and others. These advances in our understanding have benefited greatly from modern energy-dependent coupled-channel analysis methods that incorporate simultaneous treatments of the primary hadronic final states of the electromagnetic processes under consideration. It should be clear, however, that the effect of the final-state interactions coming out of such analyses is critically dependent on the quality of the underlying hadronic data. Regrettably, the meson-beam data for these final states are mostly outdated and largely of poor quality, or even non-existent, and thus limit us in fully exploiting the full potential of the new electromagnetic data. To reap their full benefit, new high-statistics data from measurements with mesons beams are critically needed to complement the existing wealth of electromagnetic data and make both electromagnetic and hadronic data sets of commensurate quality and quantity. 


The center-of-mass energy range up to 2.5 GeV is rich in opportunities for physics with pion and kaon beams to study baryon and meson spectroscopy questions complementary to the electromagnetic programs underway at electromagnetic facilities. This White Paper highlights some of these opportunities and describes how facilities with high-energy and high-intensity meson beams can contribute to a full understanding of the high-quality data now coming from electromagnetic facilities. We emphasize that what we advocate here is not a competing effort, but an experimental program that provides the hadronic complement of the ongoing electromagnetic program, to furnish the common ground for better and more reliable phenomenological and theoretical analyses based on high-quality data.

On April 7, 2012, a workshop on \textit{Physics with Secondary Hadron Beams 
in the 21st Century} was held at Ashburn, VA~\cite{gwu12}.  The workshop 
aimed to bring together experts in spectroscopy and neutron physics to 
discuss how advances in these two areas will benefit the 
 proposed electron-ion collider at JLab. 
An electron-ion collider (EIC) will likely be one of the future large accelerator facilities for high-energy and nuclear physics \cite{EIC}. There are currently five proposals under active development worldwide, including eRHIC at BNL and MEIC at JLab.  The creation of a state-of-the-art hadron physics complex to study QCD at the deepest level with an EIC
provides the unique infrastructure for a meson-beam facility to complete our picture of the hadron spectrum of QCD at the same time.   A number of the topics mentioned in this White Paper are addressed in the summary of the recent DNP Town Meeting on QCD and Hadron Physics \cite{temple2015}, which notes (on page 28) that meson beams are being considered.


\section{Opportunities with Pion Beams}

Most of our current knowledge about the bound states of three light quarks 
has come from partial-wave analyses (PWAs) of $\pi N \to \pi N$ 
scattering~\cite{hoehler79,cutkosky79a,cutkosky79b,arndt85,arndt91,arndt94,
arndt95,arndt06}. Measurements of $\pi N$ elastic scattering are mandatory for determining absolute
$\pi N$ branching ratios.  Without such information, it is likewise impossible to determine
absolute branching ratios for other decay channels.
A summary of resonance properties [pole positions (masses 
and widths), branching ratios, helicity couplings to $\gamma N$, etc.] is 
provided in the \textit{Review of Particle Physics} (RPP)~\cite{pdg}.  

The information on resonance properties obtained from analyses of experimental 
data provides fundamental information about QCD in the nonperturbative region.  
A variety of quark models~\cite{isgur77,isgur78,isgur79a,isgur79b,bowler81,
kalman82,amaral83,liu83,carlson83,bhaduri84,murthy84,sartor85,mattis85,
karliner86,capstick86,weber88,neeman88,ferraris95,dziembowski96,zuckert97} 
have been used to interpret these results. 
Dyson-Schwinger approaches provide a picture of baryons in
terms of quarks and gluons, incorporating dynamical chiral symmetry \cite{Roberts:2011cf,Wilson:2011aa,Chen:2012qr}.
Results from lattice gauge 
theory calculations are constantly improving~\cite{Lang:2012db, Edwards:2011jj,Engel:2013ig} and are therefore becoming more relevant to experiment.

A comparison of the experimental results and models led to the well-known 
conundrum known as the ``missing resonances'' problem~\cite{koniuk80}. Put 
simply, the models (and lattice-gauge calculations) predict far more states 
than are observed experimentally.  (These missing resonances, however, do not appear at all in the quark-diquark model.  See, for example, Refs.\cite{goldstein80,goldstein88}.)
The reason for this, it is conjectured, is their weak coupling to the $\pi N$
channel, which supplies the bulk of our information about baryonic resonance
states.  A desire to test this hypothesis by looking for resonances in reactions that do not involve $\pi N$ in either the initial or final state was a major reason behind the construction of Hall B and the CLAS facility at JLab. 
This is part of a global spectroscopy effort also pursued at MAMI (Mainz, Germany), ELSA (Bonn, Germany), SPring-8 (Japan), and other facilities. 


This joint global effort has boosted the field of baryon spectroscopy
through data of unprecedented accuracy. 
 The discoveries made in the photoproduction program have triggered much
theoretical interest probing hypotheses of resonance nature including, e.g.,
multiquark states, hadronic molecules,
crypto-exotics, hybrid baryons, chiral symmetry restoration in the
resonance spectrum, and string models of resonances (AdS/QCD). In $\eta$ photoproduction on the neutron, a much-debated narrow structure at center-of-mass energy $W{\sim}1.65$~GeV was discovered. New photoproduction data are constantly being measured, in particular double-polarization observables that will pave the way to ``complete experiments''.

The photo- and electroproduction programs have advanced our
understanding of the resonance region. Because the data from these programs have reached unprecedented
levels of precision, it is timely to review the analysis techniques used
to extract partial waves, multipoles, and, ultimately, resonances. Modern techniques include pion-induced data  in
global coupled-channel approaches to search for even the faintest
resonance signals.
Coupled-channel energy-dependent analyses have proven to be the most efficient tools to search for resonances. On one hand, this is because $N^\ast$ and $\Delta^\ast$ resonances are broad and overlapping. On the other hand, resonances manifest themselves as poles in the complex plane of the scattering energy, where positions are the same irrespective of the analyzed reaction. Also, energy-dependent methods are required to provide an amplitude that can be analytically continued into the complex plane to extract the mathematically well-defined pole positions and residues that correspond to resonance masses, widths, and branching ratios~\cite{pdg}.
Residues also contain a phase that gives information about the decay properties of a resonance~\cite{hoehler79}, even for the EM helicity couplings~\cite{Workman:2013rca, Ronchen:2014cna, Kamano:2013iva}.

It is for these reasons that all major current analysis efforts use energy-dependent coupled-channel schemes. Some of these efforts are briefly discussed in Sec.~\ref{baryon_analyses}. As alluded to in the Introduction, to describe the hadronic final-state interactions of the photon-induced reactions, coupled-channel analysis schemes critically depend on data from pion-induced reactions. The latter data sets, unfortunately, are frequently of inferior quality and thus do not provide constraints commensurate with the quality of the recent photo- and electroproduction data. Undoubtedly, the interpretation of electromagnetic experiments would benefit greatly from having improved pion-induced data at our disposal. This need provides the major motivation for this White Paper.



The world data on $\pi N\to \eta N,\,K\Lambda, \,K\Sigma$ were collected in Ref.~\cite{Ronchen:2012eg} and date back to more than 20 or 30 years ago. In many cases, systematic uncertainties were not reported (separately from statistical uncertainties), and in many cases it is known that systematic uncertainties were underestimated~\cite{Nefkens}.  These problems of pion-induced reaction data have led to the emergence of many different analyses that claim a different resonance content. While many analyses agree on the 4-star resonances that are visible in elastic $\pi N$ scattering~\cite{arndt06}, there is no conclusive agreement on resonances that couple only weakly to the $\pi N$ channel, starting with the $N(1710)1/2^+$ and ending with resonances found by the Bonn-Gatchina group in $K\Lambda$ photoproduction~\cite{anisovich12a,anisovich12b} but not confirmed so far. In Sec.~\ref{sec:reactions}, the status of pion- and photon-induced reaction data is reviewed. 

Due to the problematic data situation, it is also difficult to 
determine rigorously the statistical significance of resonance signals.
The final goal in baryon spectroscopy is not only the determination of the baryonic resonance content, but also its statistical significance, including statistically meaningful uncertainties for pole positions and residues. 
New measurements of pion-induced reactions are necessary to reach this goal.
If the quality of hadronic data 
matched that of photoproduction data, much better statistically sound
statements on the significance of resonance signals could be made.
In addition, analyses of 
photoproduction data in modern multichannel methods~\cite{shklyar05,
shklyar07,julia-diaz07,sarantsev09,anisovich11,huang12,anisovich12a,
anisovich12b,shrestha12} require comparable high-quality pion-induced data for the same final states to determine absolute branching ratios and partial widths.

Missing states that couple weakly to $\pi N$ are presently searched for in $\gamma N\to\eta N,\,K\Lambda,\,K\Sigma$ and related reactions. For the same final states, multichannel photoproduction analyses need eight independent observables at fixed c.m.\ energy $W$ and scattering angle (or fixed $t$) while the pion-induced hadronic amplitude needs but four (where a fourth observable is necessary to remove a sign ambiguity~\cite{Wolfenstein1954,Lesquen1972,Supek1993,Klempt2010,jackson2014}). A search for missing states in pion-induced production of $\eta N$, $KY$, and other final states thus provides a promising and complementary source of information for baryon spectroscopy, without presenting any major experimental hurdles.


Baryon spectroscopy is not a self-contained field; it needs to seek the connection to first-principle QCD calculations such as lattice simulations, discussed in Sec.~\ref{sec:lattice}. So far, lattice QCD calculations in the baryon sector are restricted to the hadronic masses. (Electromagnetic properties of excited states will follow in the future as anticipated, e.g., in Ref.~\cite{Agadjanov:2014kha}.) It is, however, the hadronic partial waves that will allow for comparison to QCD in the future. As discussed, the determination of the hadronic amplitude requires improved data from pion- and kaon-induced reactions.

More precise pion-induced reaction data close to the $\pi N$ threshold are also called for, as discussed in Sec.~\ref{sec:ChPT}. These are crucial for chiral perturbation theory (ChPT) and are especially important for the determination of the low-energy constants. The latter not only provide a consistent picture for hadron dynamics but also assist in making nuclear {\it ab~initio} calculations. In particular, these constants serve as input for the nuclear lattice calculations of the Hoyle state that explain the generation of heavy elements 
\cite{Epelbaum:2011md, Epelbaum:2013wla}.

In summary, better data from hadron-induced reactions will
significantly contribute to answer the same fundamental questions that originally
motivated the photoproduction program: the missing resonance problem,
amplitudes for comparison with {\it ab~initio} calculations, and low-energy
precision physics. A program with hadron beams provides complementary
information with large impact in the extraction of the amplitudes from
observables. With a more precise knowledge of the amplitude it is expected
that much-debated concepts, such as the
aforementioned multiquark hypotheses, hadronic molecules, hybrid states,
chiral symmetry restoration, chiral solitons, or string models,
can be confirmed or ruled out.

\subsection{Baryon Spectroscopy Analyses}
\label{baryon_analyses}
Several analysis groups work actively on disentangling the baryon spectrum.
In dynamical coupled channel approaches like the ANL-Osaka (formerly EBAC), the J\"ulich-Athens-Washington, and other approaches~\cite{Kamano:2013iva,Ronchen:2014cna,Ronchen:2012eg,huang12,Tiator:2010rp,kamano2013}, one solves three-dimensional Lippmann-Schwinger-type scattering equations with off-shell dependence originating from their driving terms. In some cases \cite{Ronchen:2014cna,Ronchen:2012eg,huang12}, covariance is maintained via a covariant Blankenbecler--Sugar-type reduction (for covariant four-dimensional approaches, see Refs.~\cite{Lahiff:1999ur, Borasoy:2005ie}). 

Various approaches exist to reduce the scattering integral equations to
matrix equations in coupled channels by utilizing on-shell
approximations. Real, dispersive expression may be formulated to account
for intermediate propagating states. Such contributions are relevant for the reliable analytic continuation required to search for resonance poles and residues. 

Analyses of this type are pursued in the GWU/INS (SAID) approach in the
Chew-Mandel\-stam formulation~\cite{arndt06,Workman:2012jf}, by the
Bonn--Gatchina group in the N/D
formulation~\cite{anisovich12a,anisovich12b}, by the Kent
State group~\cite{shrestha12}, and by the Zagreb group~\cite{Batinic:2010zz} in the
Carnegie-Mellon-Berkeley (CMB) formulation.  The Giessen group uses a $K$-matrix
formalism~\cite{Shklyar:2004ba,shklyar13} while the MAID approach employs a
unitary isobar formalism in which the final-state interaction is taken from the
SAID approach~\cite{Drechsel:2007if}. The GW SAID $N^\ast$ program consists of $\pi N \to \pi N$, $\gamma N \to \pi N$, and $\gamma^\ast N \to \pi N$ \cite{SAID2}.
Dispersive approaches and unitary isobar
analyses on meson electroproduction have been performed at JLab~\cite{Aznauryan:2012ba} where also two-pion electroproduction is analyzed~\cite{Aznauryan:2005tp}.

The Bonn-Gatchina approach, formulated with covariant
amplitudes~\cite{Anisovich:2004zz}, performs combined analyses of all known data
on single and double-meson photon- and pion-induced reactions; four new states  were reported recently \cite{anisovich12a}.
In the Bonn-Gatchina approach, multi-body final
states  are analyzed in an event-by-event maximum-likelihood method that  
takes fully into account all correlations in the multidimensional phase space.

The Giessen group recently included an analysis of $\pi\pi N$ data in the form
of invariant-mass projections~\cite{Shklyar:2014kra}, similar to the previous
work of the EBAC~\cite{Kamano:2008gr} group while the Kent State group uses 
isobar-model amplitudes from an event-based maximum-likelihood analysis \cite{manley84}.  In addition to their energy-dependent solutions, the Kent State group recently published single-energy solutions to $\pi N$ scattering~\cite{shrestha12a} and $\overline{K}N$ scattering \cite{zhang13}. Single-energy solutions (SES) provide a more model-independent representation of the amplitude. While less suited to search for the broad resonances, SES provide the possibility to search for additional structures in the amplitude not captured by energy-dependent methods.  Reference\cite{shrestha12a} gives an overview of the world database for $\pi^- p \to \eta n$ and $\pi^- p \to K^0 \Lambda$ and Ref.~\cite{zhang13} gives an overview of the world database for 
$\overline{K}N \to \overline{K}N$, $\overline{K}N \to \pi\Lambda$, and  $\overline{K}N \to \pi\Sigma$.

In the GWU/INS (SAID) approach, the interaction is parametrized without the need
of explicit resonance propagators~\cite{Workman:2012jf}. Resonance poles are
generated only if required by data, which makes this framework particularly
model independent for baryon spectroscopy. Also, currently only the SAID group provides partial waves directly from $\pi N$ elastic scattering data. These waves are widely used as input by other groups in the analyses of deep inelastic scattering (DIS), neutrino production, and other reactions.  The SAID output is applicable for both low-energy studies (spectroscopy) and high-energy research as well.
For instance, the theoretical interpretation of DIS experimental results
for target fragmentation, as well as for many other strong
interaction processes, may be
considered either in quark-gluon terms of QCD 
\cite{gribov87,gribov88,anisovich93} or in hadronic terms.
(It might be very interesting to investigate their interconnections.) The hadronic description require a detailed knowledge of the nucleon-nucleon and
meson-nucleon interactions at lower energies; e.g., one can consider
extraction of the pion structure function at small Bjorken $x$ for the process of leading
neutrino production in DIS (see Ref.~\cite{kopeliovich96,kopeliovich07}).
Existing data from the H1 and ZEUS experiments at HERA are in good agreement
with the predictions using the SAID $\pi N$ amplitudes \cite{arndt06} as
input.

\subsection{Status of Data and Analyses for Specific Reactions}
\label{sec:reactions}

Measurements of final states involving a single pseudoscalar meson and a 
spin-1/2 baryon are particularly important.  The reactions involving $\pi N$ 
channels include:

\begin{tabular}{ll}
$\gamma p \to \pi^0 p$,~~~~~~~~~~ & $\pi^- p \to \pi^0 n$, \\
$\gamma p \to \pi^+ n$, & $\pi^- p \to \pi^- p$, \\
$\gamma n \to \pi^- p$, & $\pi^+ p \to \pi^+ p$, \\
$\gamma n \to \pi^0 n$. &  \\
\end{tabular}

\begin{figure}[htpb]
\begin{center} 
	\includegraphics[angle=90, width=0.30\textwidth ]{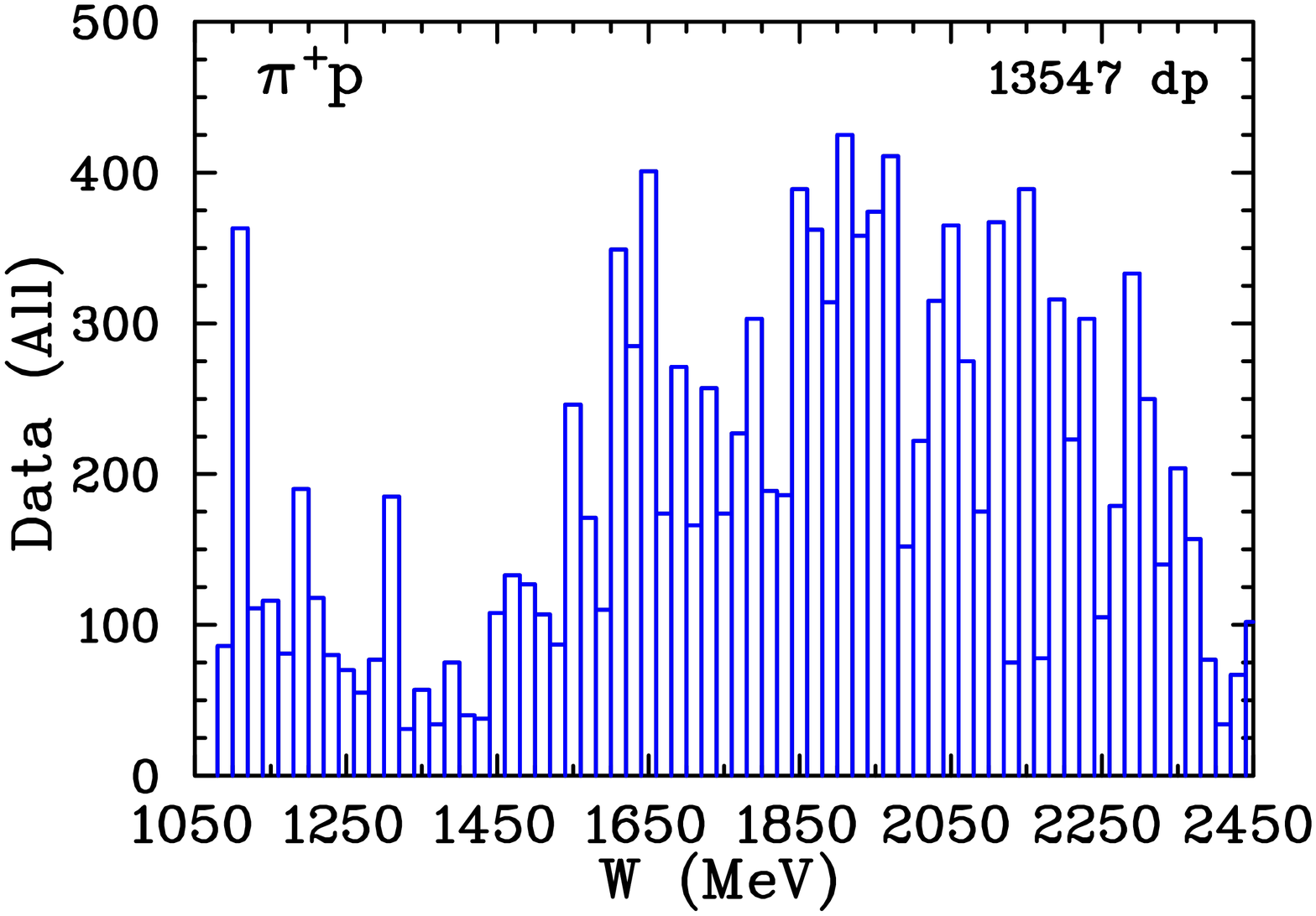}
	\includegraphics[angle=90, width=0.30\textwidth ]{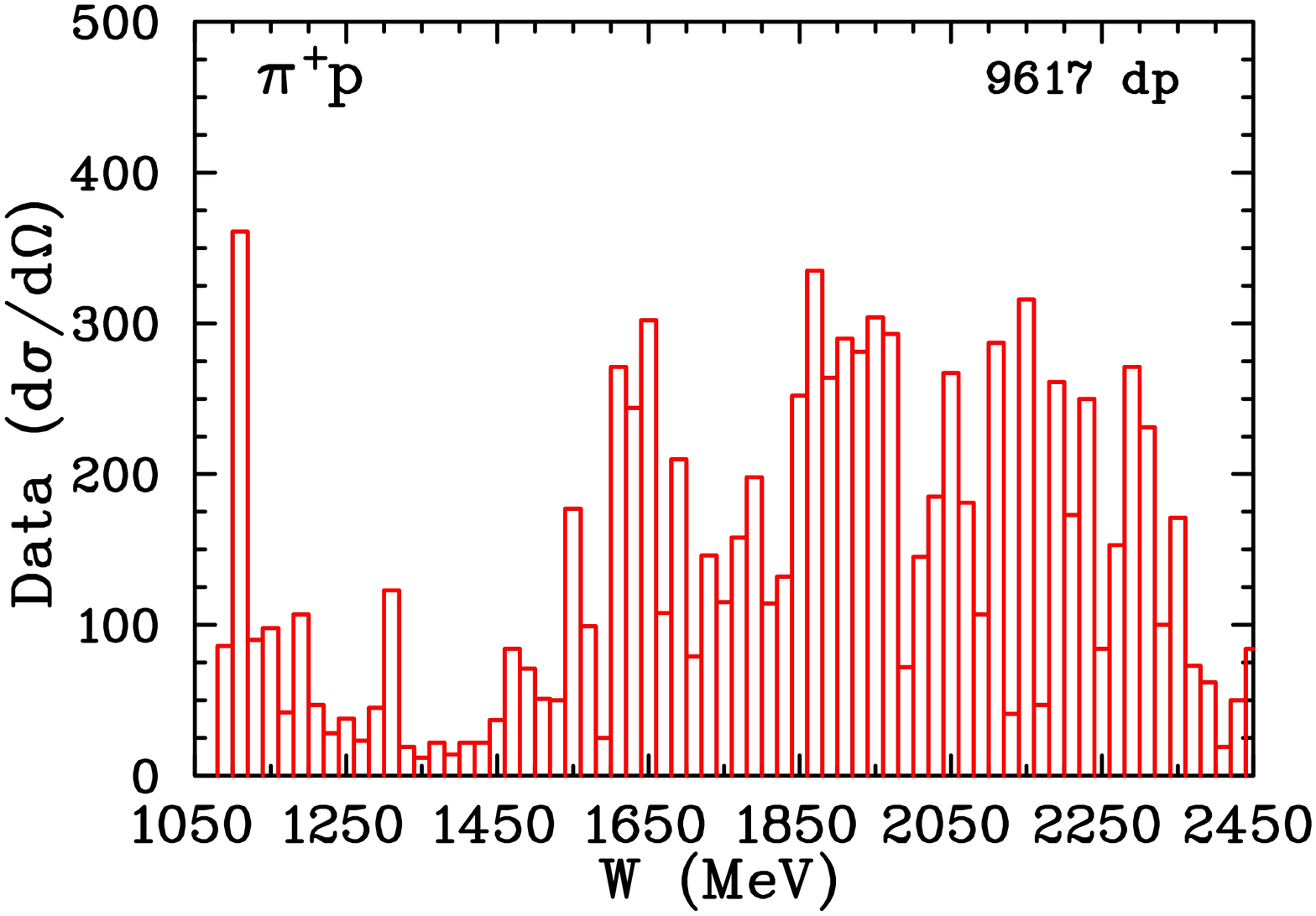}
	\includegraphics[angle=90, width=0.30\textwidth ]{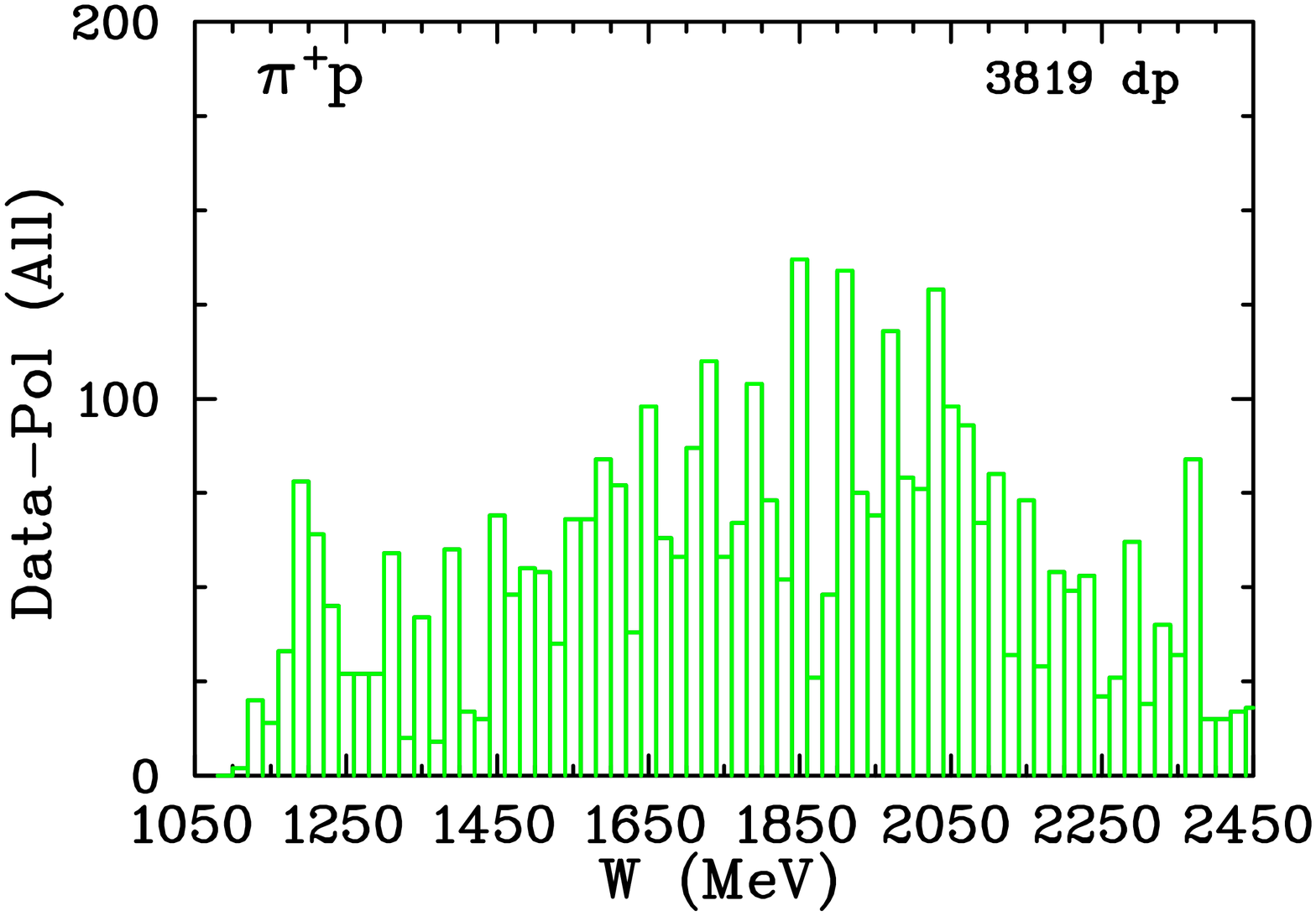}
	\includegraphics[angle=90, width=0.30\textwidth ]{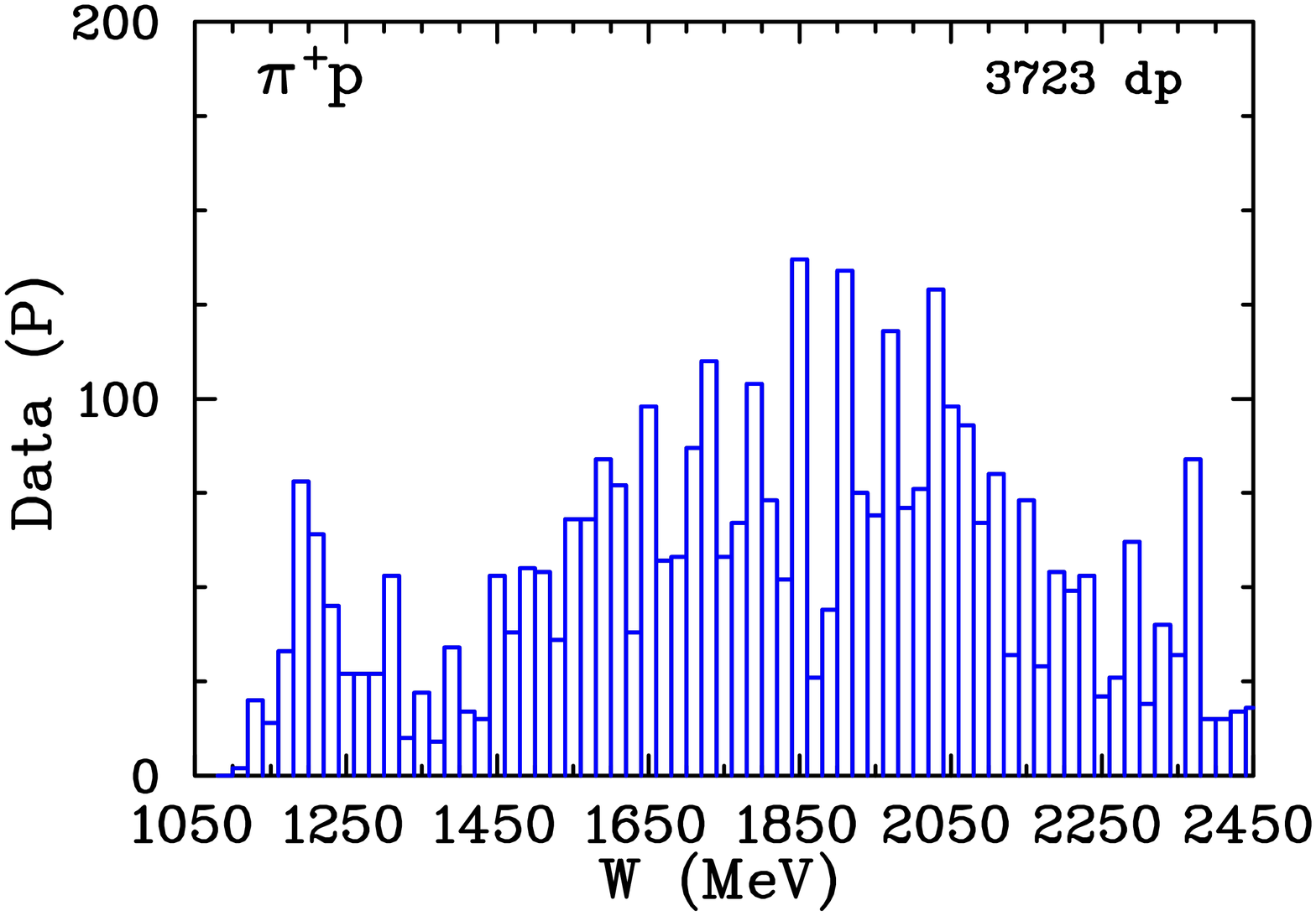}
	\includegraphics[angle=90, width=0.30\textwidth ]{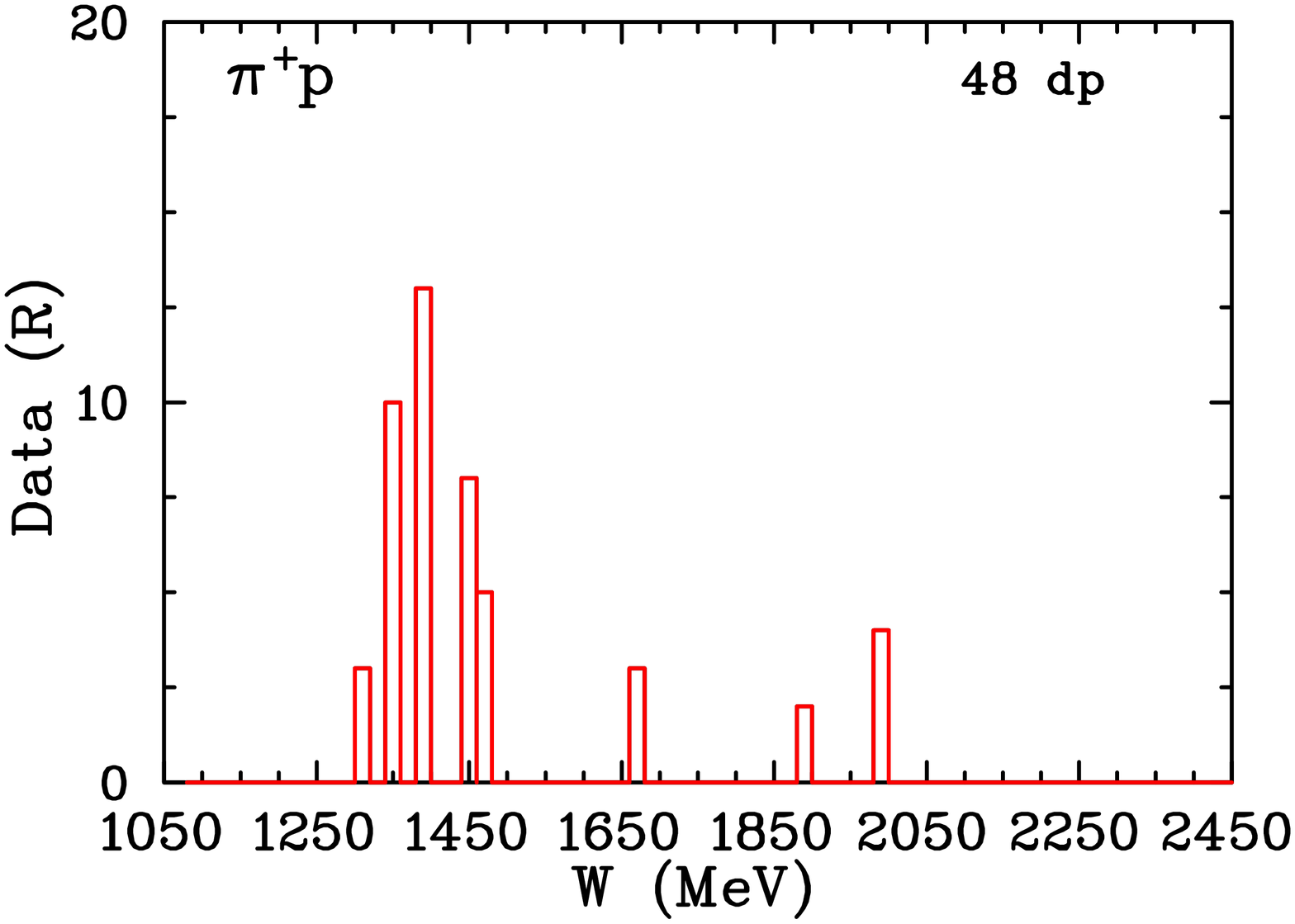}
	\includegraphics[angle=90, width=0.30\textwidth ]{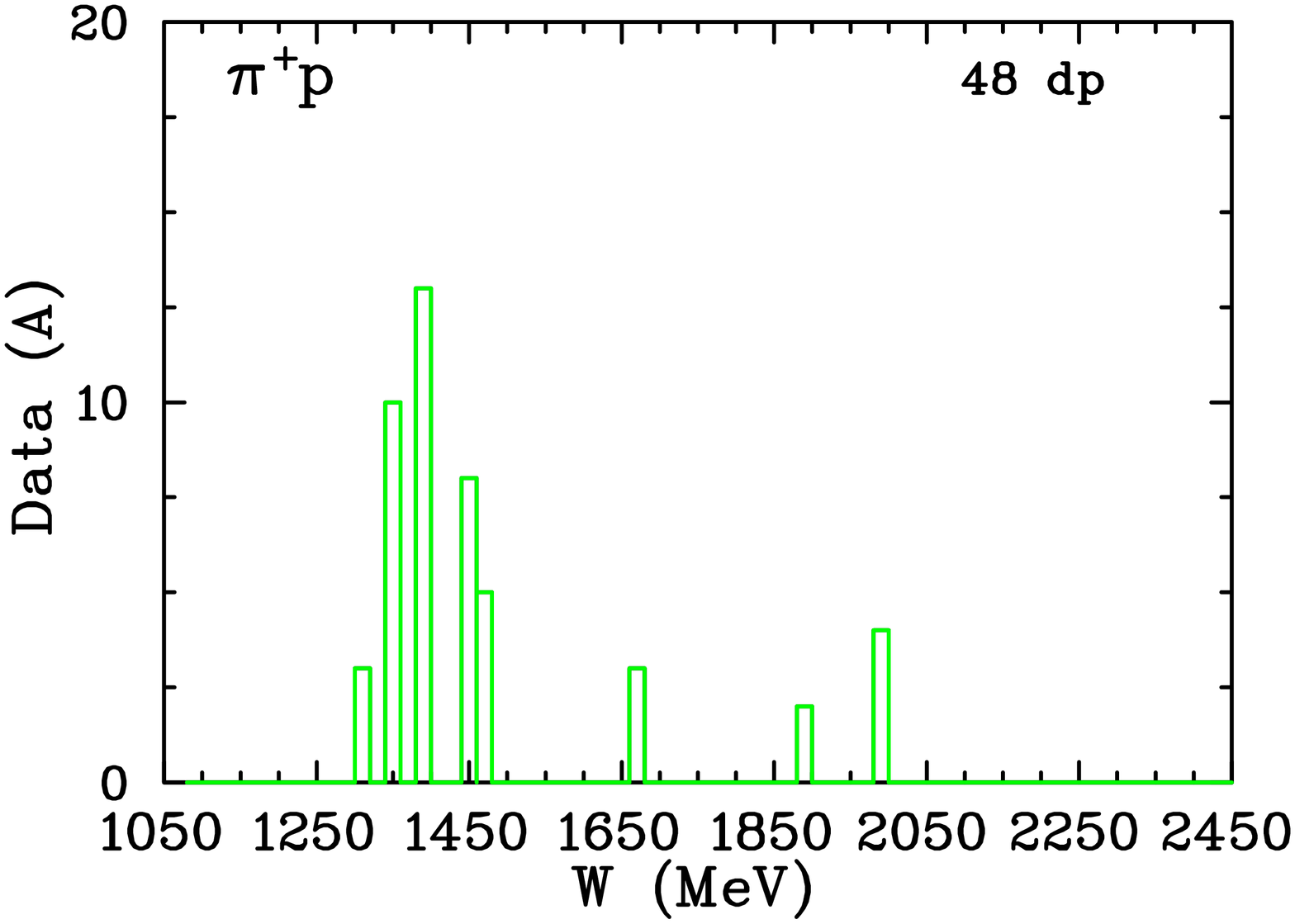}
\end{center}
\centerline{\parbox{0.80\textwidth}{
\caption[] {\protect\small (Color on-line) Data available for $\pi^+ p \to \pi^+ p$ 
	as a function of center-of-mass energy $W$ \cite{SAID}.  The number of 
	data points (dp) is given in the upper righthand side of 
	each subplot.  Row 1: The first subplot (blue) shows the 
	total amount of data available for all observables, the 
	second plot (red) shows the amount of differential 
	cross-section ($d\sigma/d\Omega$) data available, the third 
	plot (green) shows the amount of polarization data available.  
	Row 2: The first subplot (blue) shows the total amount of $P$ 
	observables data available, the second plot (red) shows the 
	amount of $R$ spin observable data available, the third plot 
	(green) shows the amount of $A$ spin observable  data available.}
	\label{fig:pi_plus.eps} } }
\end{figure}

\begin{figure}[htpb]
\begin{center} 
	\includegraphics[angle=90, width=0.30\textwidth ]{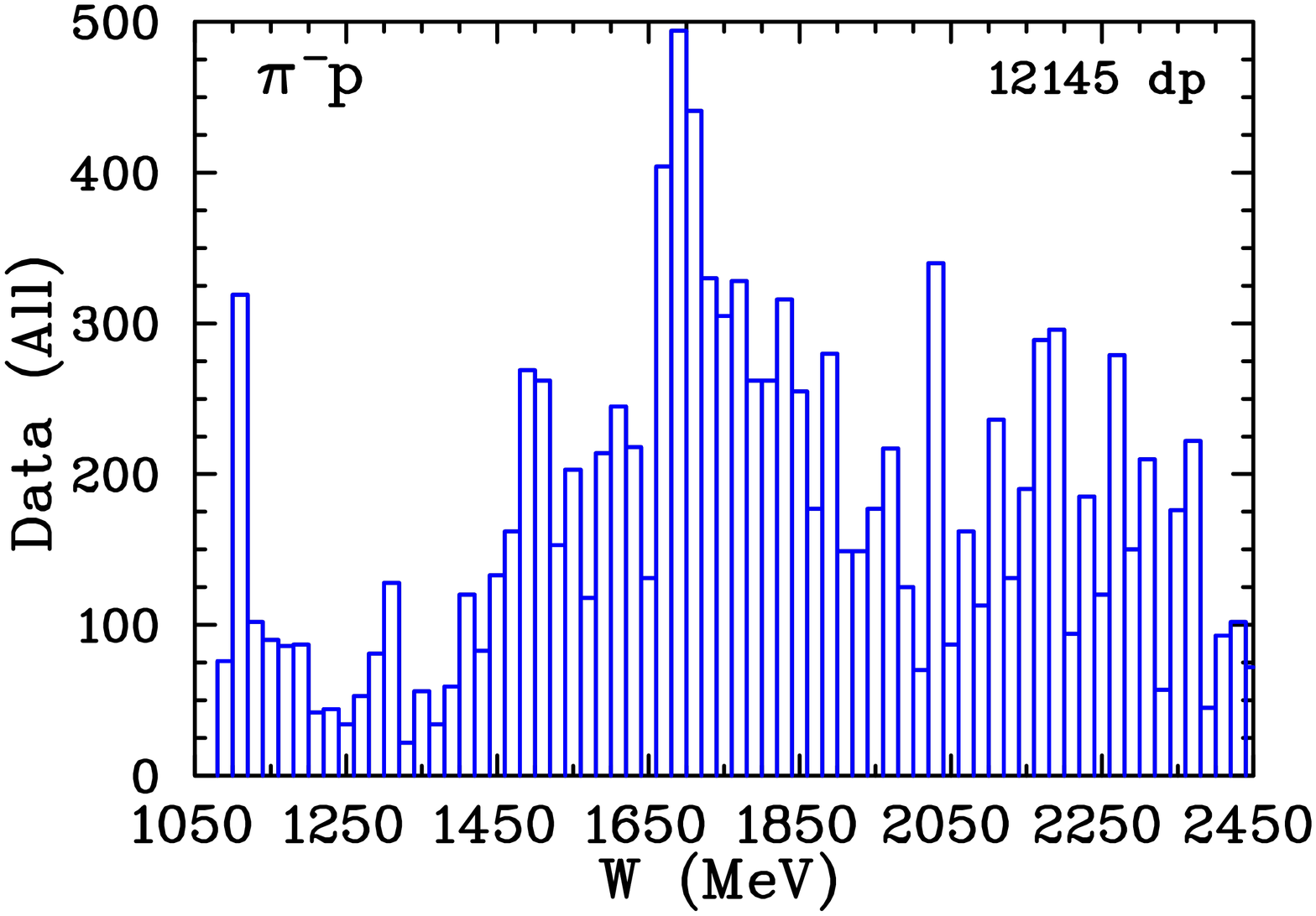}
	\includegraphics[angle=90, width=0.30\textwidth ]{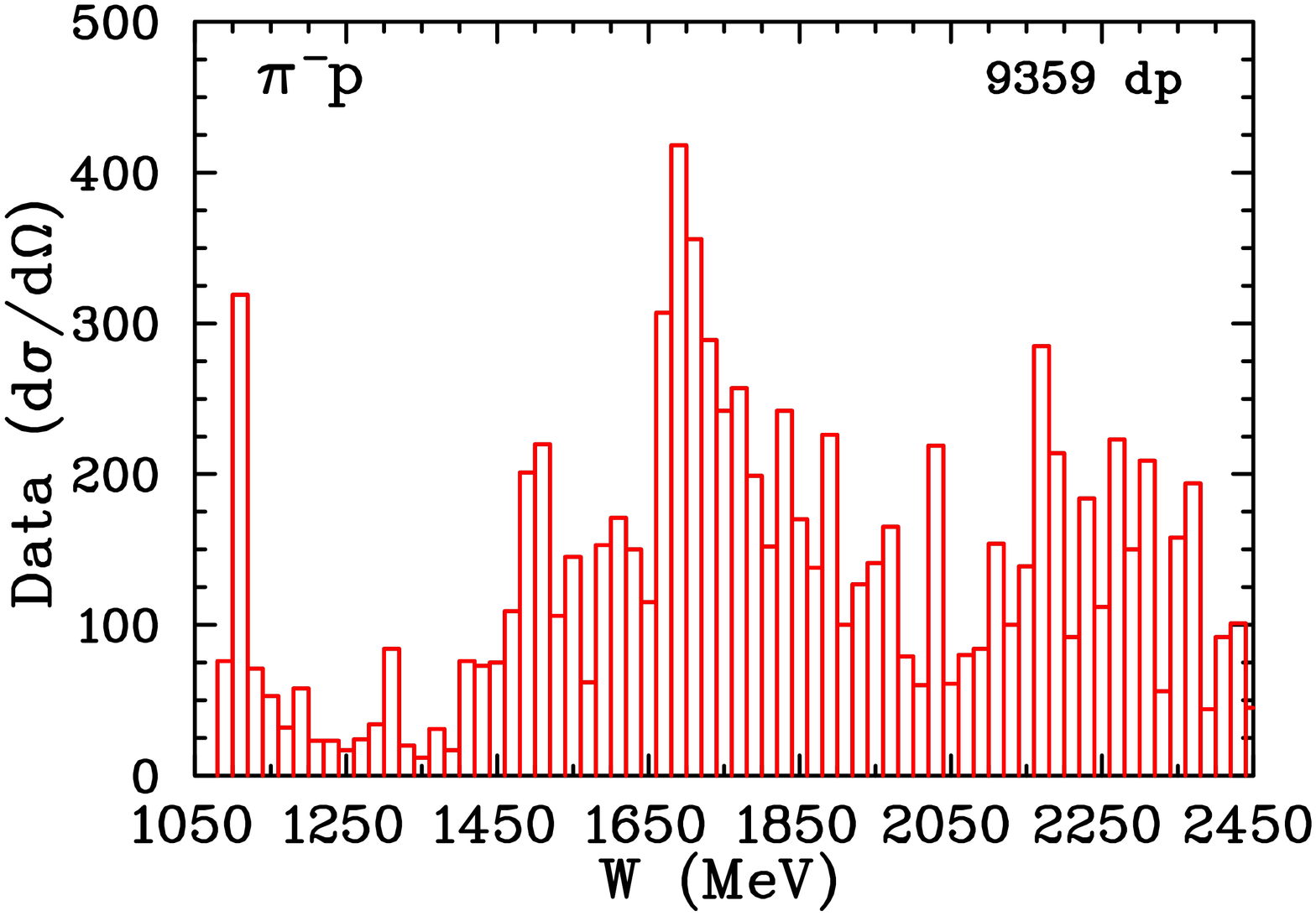}
	\includegraphics[angle=90, width=0.30\textwidth ]{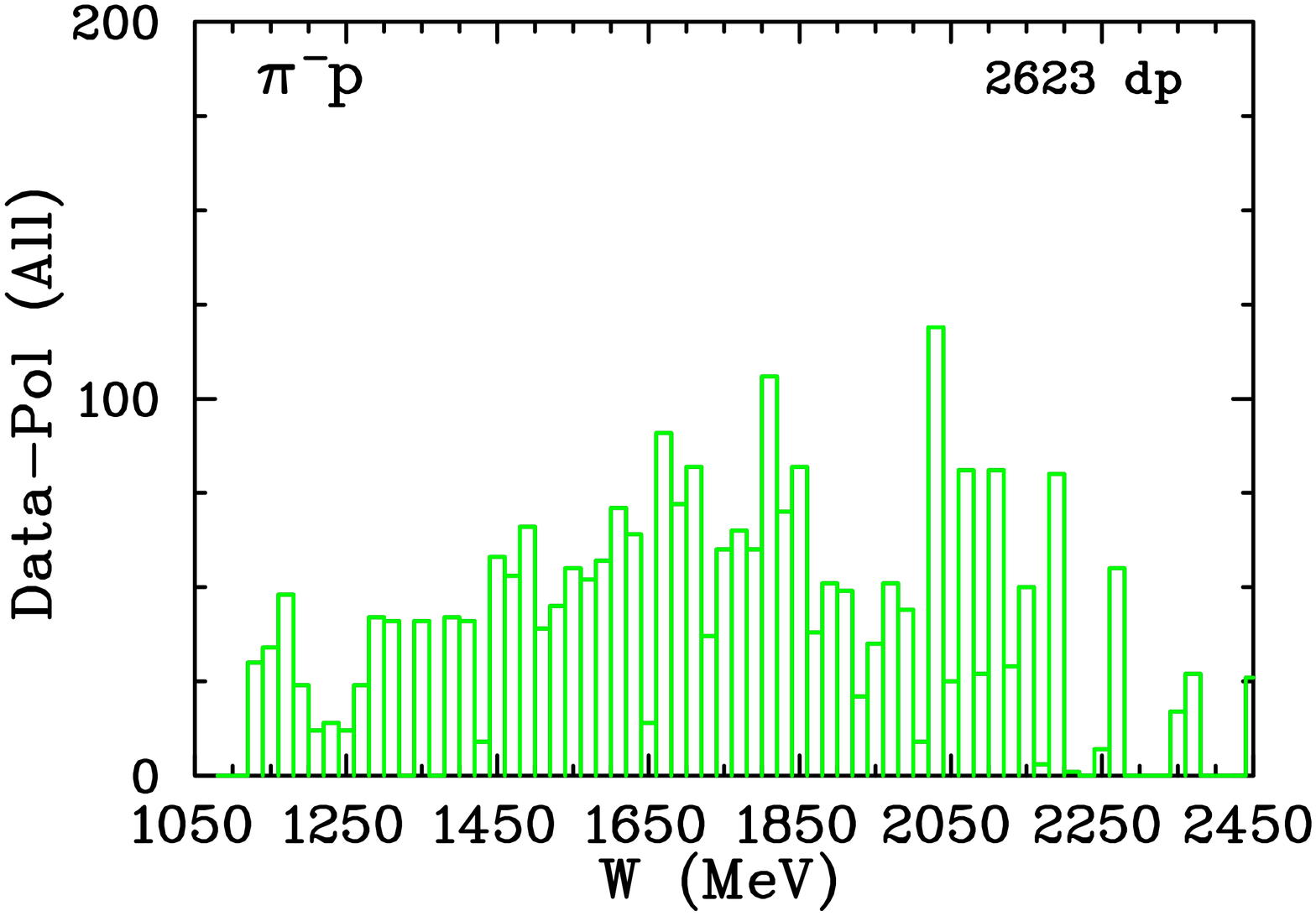}
	\includegraphics[angle=90, width=0.30\textwidth ]{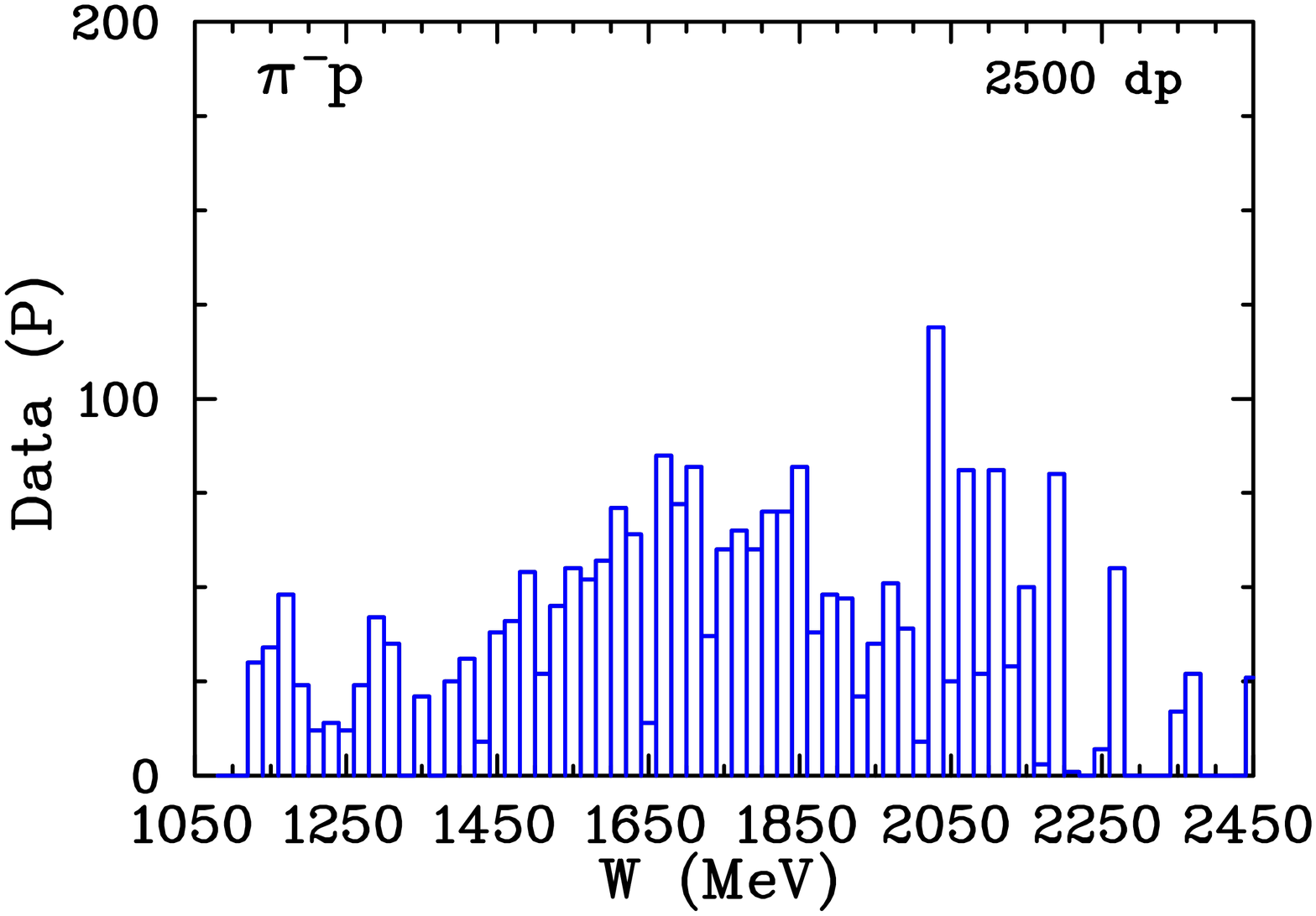}
	\includegraphics[angle=90, width=0.30\textwidth ]{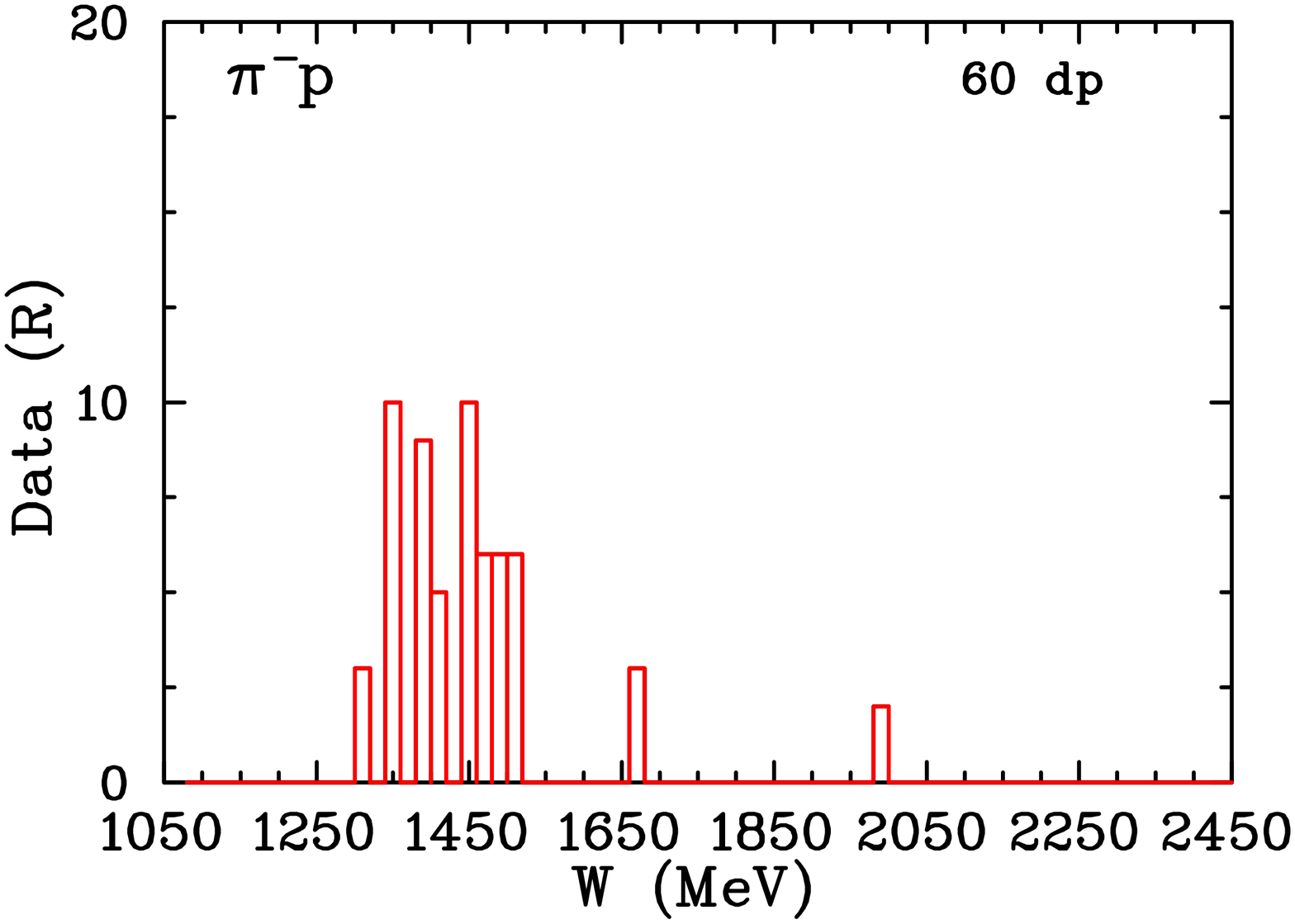}
	\includegraphics[angle=90, width=0.30\textwidth ]{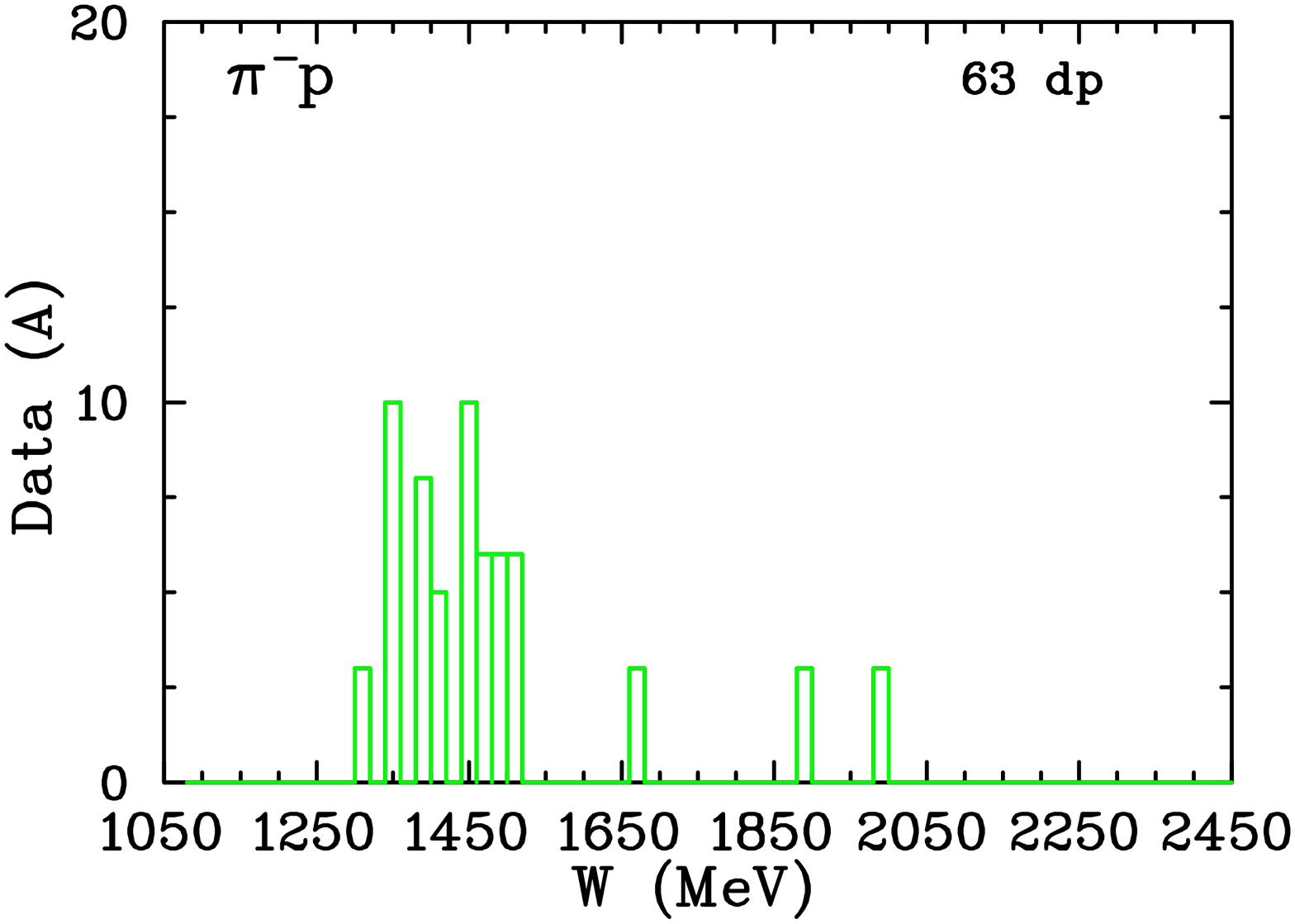}
\end{center}
\centerline{\parbox{0.80\textwidth}{
\caption[] {\protect\small (Color on-line) Data available for $\pi^- p \to 
	\pi^- p$ as a function of center-of-mass energy $W$ \cite{SAID}.  The 
	number of data points (dp) is given in the upper righthand 
	side of each subplot.  Row 1: The first subplot (blue) 
	shows the total amount of data available for all observables, 
	the second plot (red) shows the amount of differential 
	cross-section ($d\sigma/d\Omega$) data available, the third 
	plot (green) shows the amount of polarization data available.  
	Row 2: The first subplot (blue) shows the total amount of $P$ 
	observables data available, the second plot (red) shows the 
	amount of $R$ spin observable data available, the third plot 
	(green) shows the amount of $A$ spin observable data available.}
	\label{fig:pi_minus.eps} } }
\end{figure}

\begin{figure}[htpb]
\begin{center} 
	\includegraphics[angle=90, width=0.30\textwidth ]{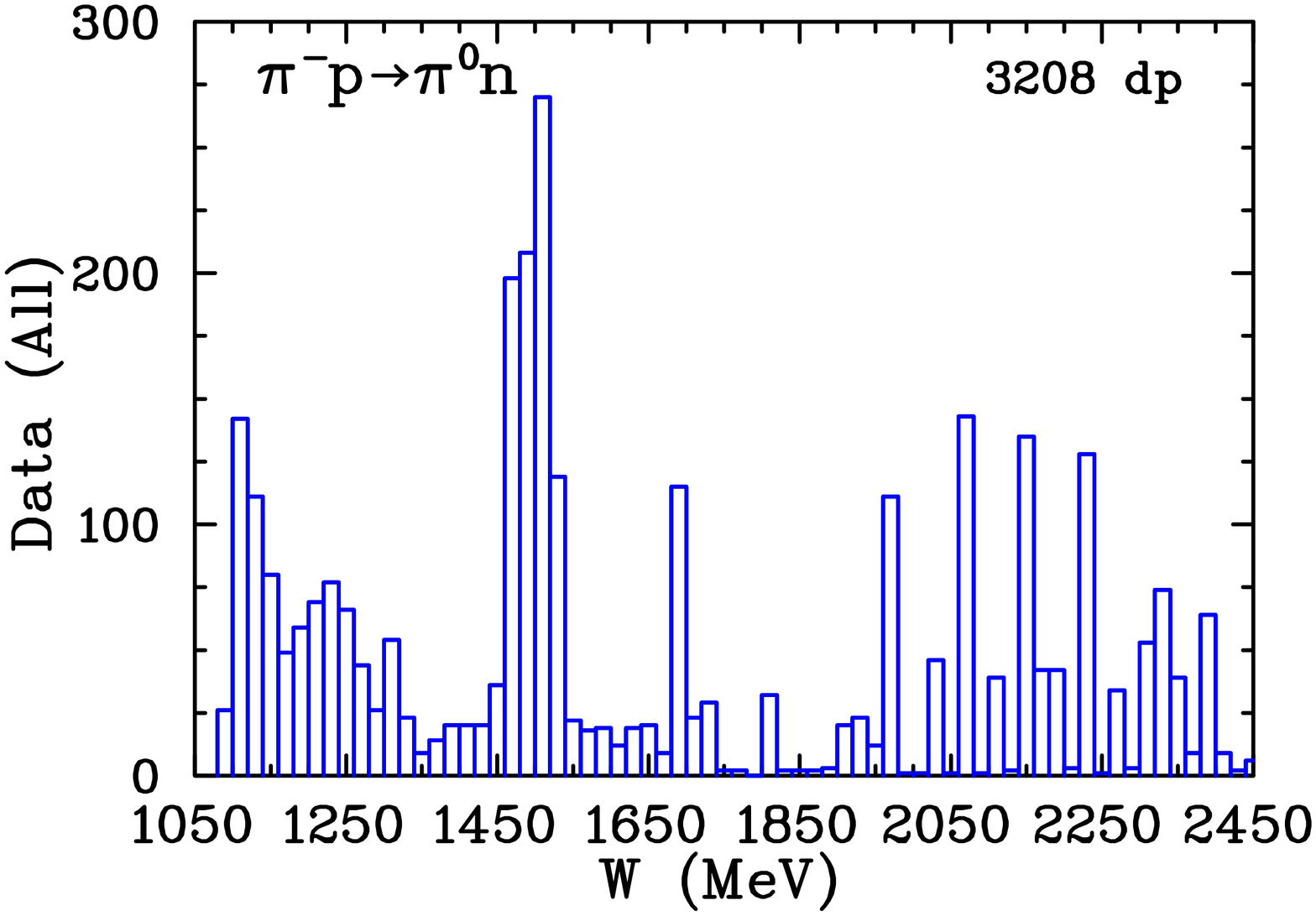}
	\includegraphics[angle=90, width=0.30\textwidth ]{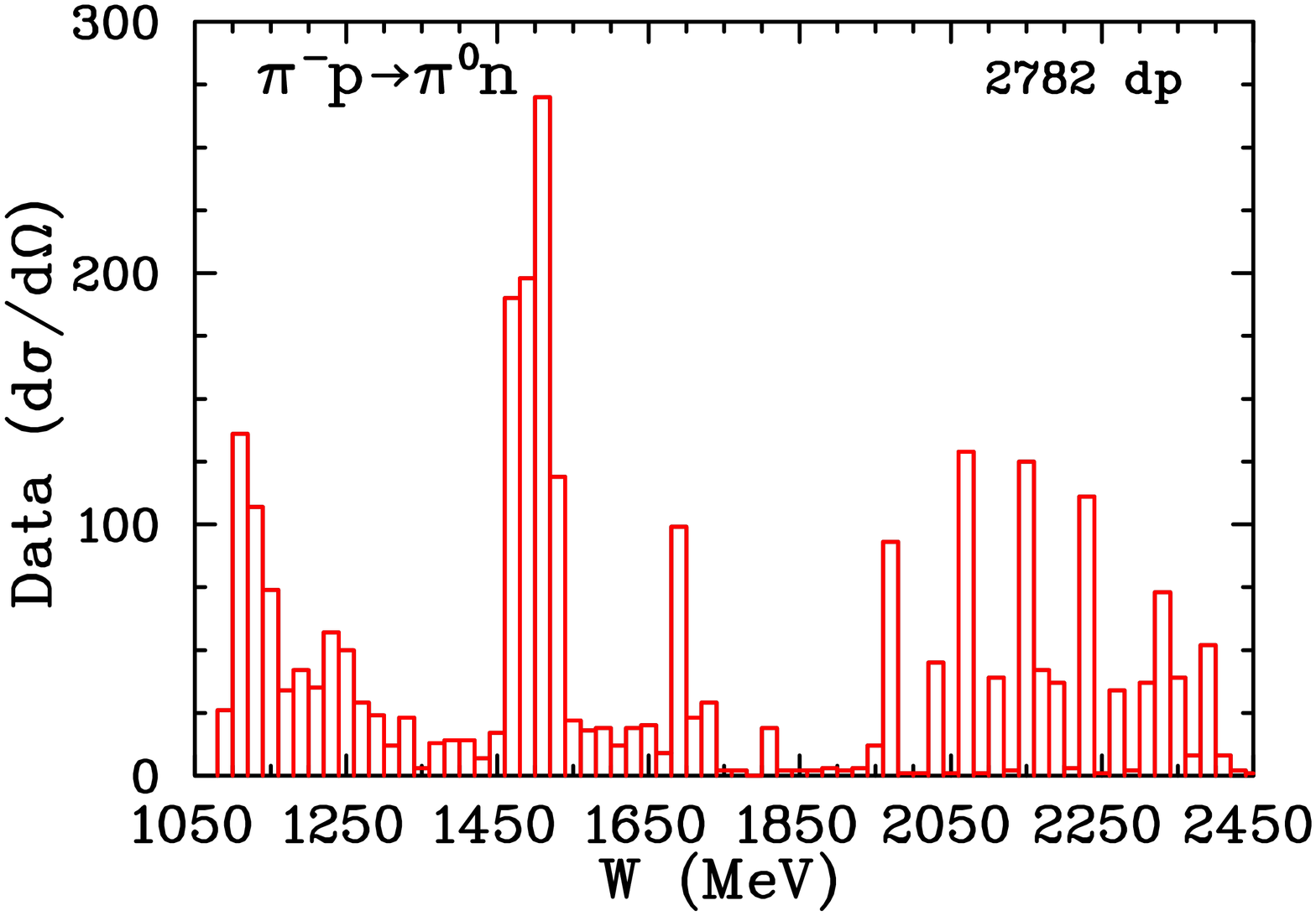}
	\includegraphics[angle=90, width=0.30\textwidth ]{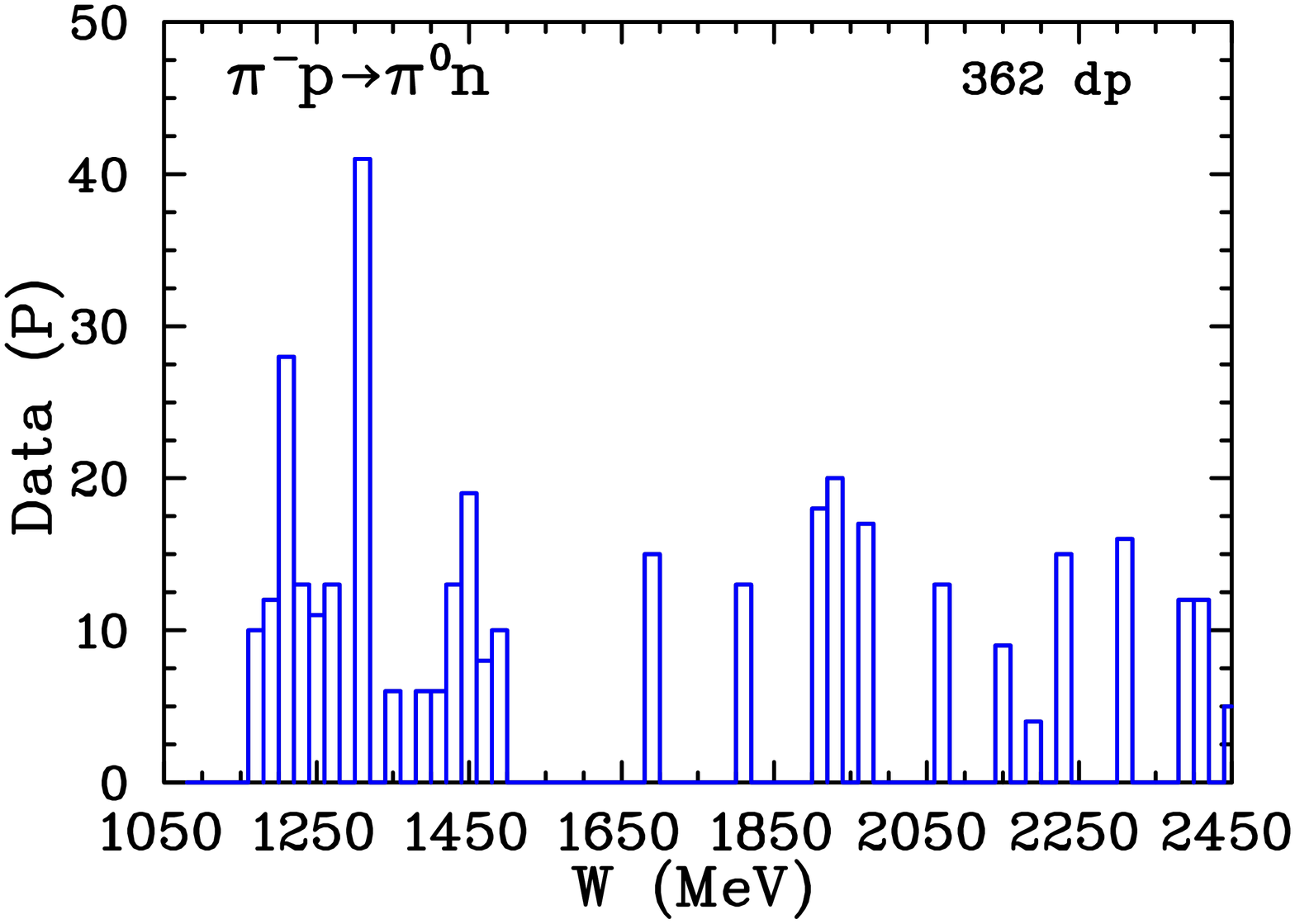}
\end{center}
\centerline{\parbox{0.80\textwidth}{
\caption[] {\protect\small (Color on-line) Data available for $\pi^-p\to\pi^0n$ 
	as a function of center-of-mass energy $W$ \cite{SAID}.  The number of 
	data points (dp) is given in the upper righthand side of each 
	subplot.  The first subplot (blue) shows the total amount of 
	data available for all observables, the second plot (red) 
	shows the amount of differential cross-section 
	($d\sigma/d\Omega$) data available, the third plot (blue) 
	shows the amount of polarization data available.}
	\label{fig:cex.eps} } }
\end{figure}

\begin{figure}[htpb]
\begin{center} 
	\includegraphics[angle=90, width=0.30\textwidth ]{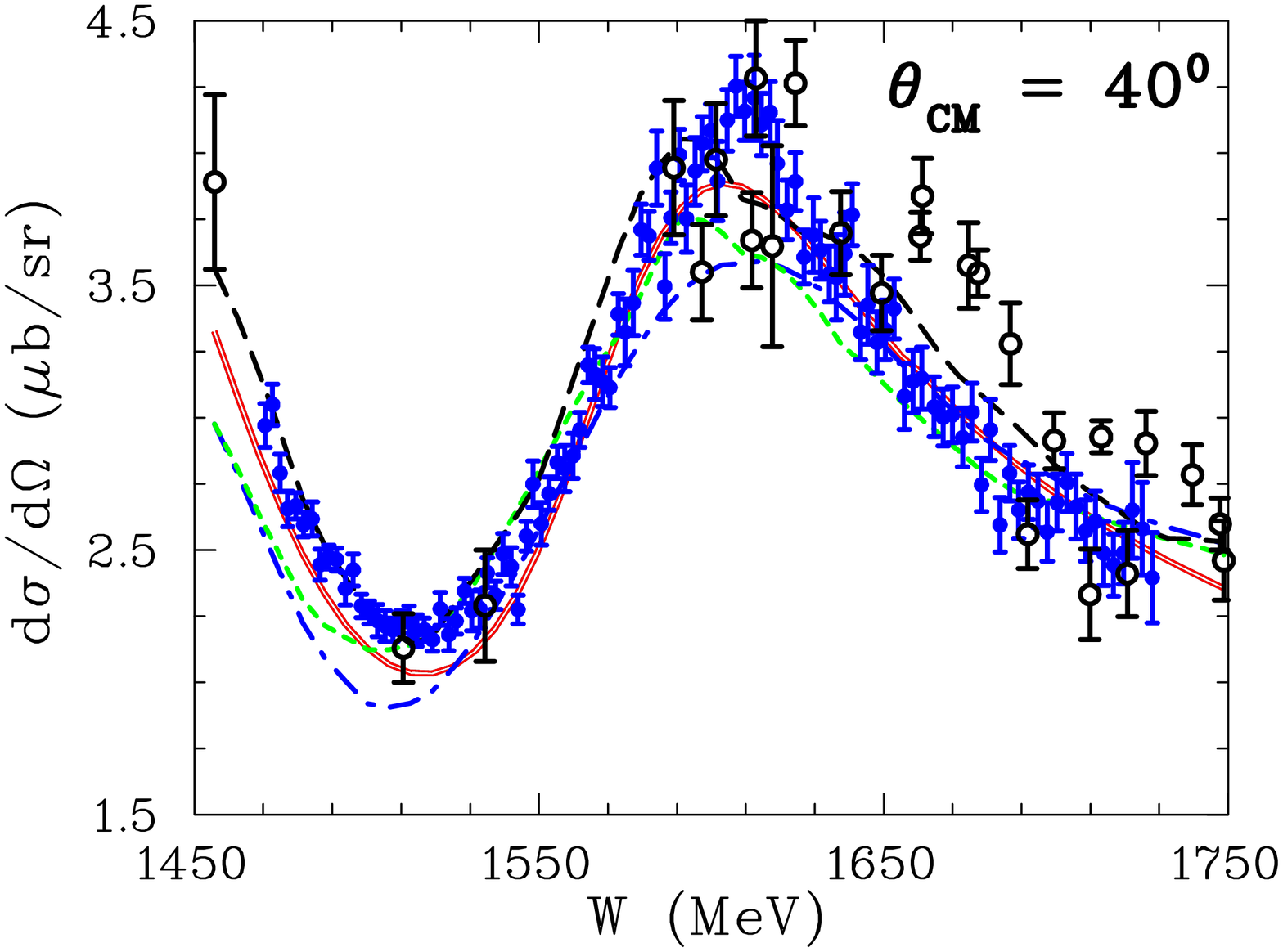}
	\includegraphics[angle=90, width=0.30\textwidth ]{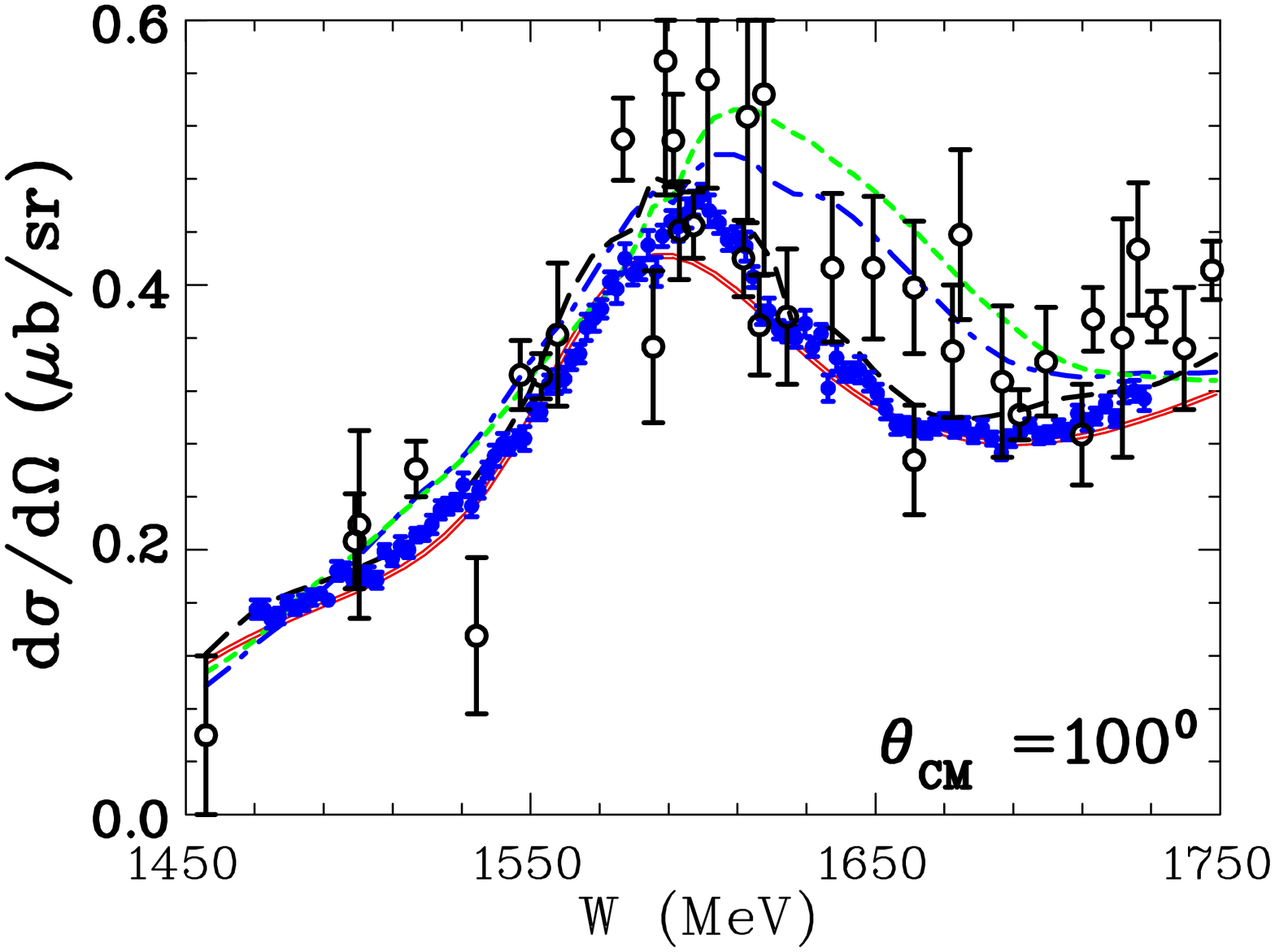}
	\includegraphics[angle=90, width=0.30\textwidth ]{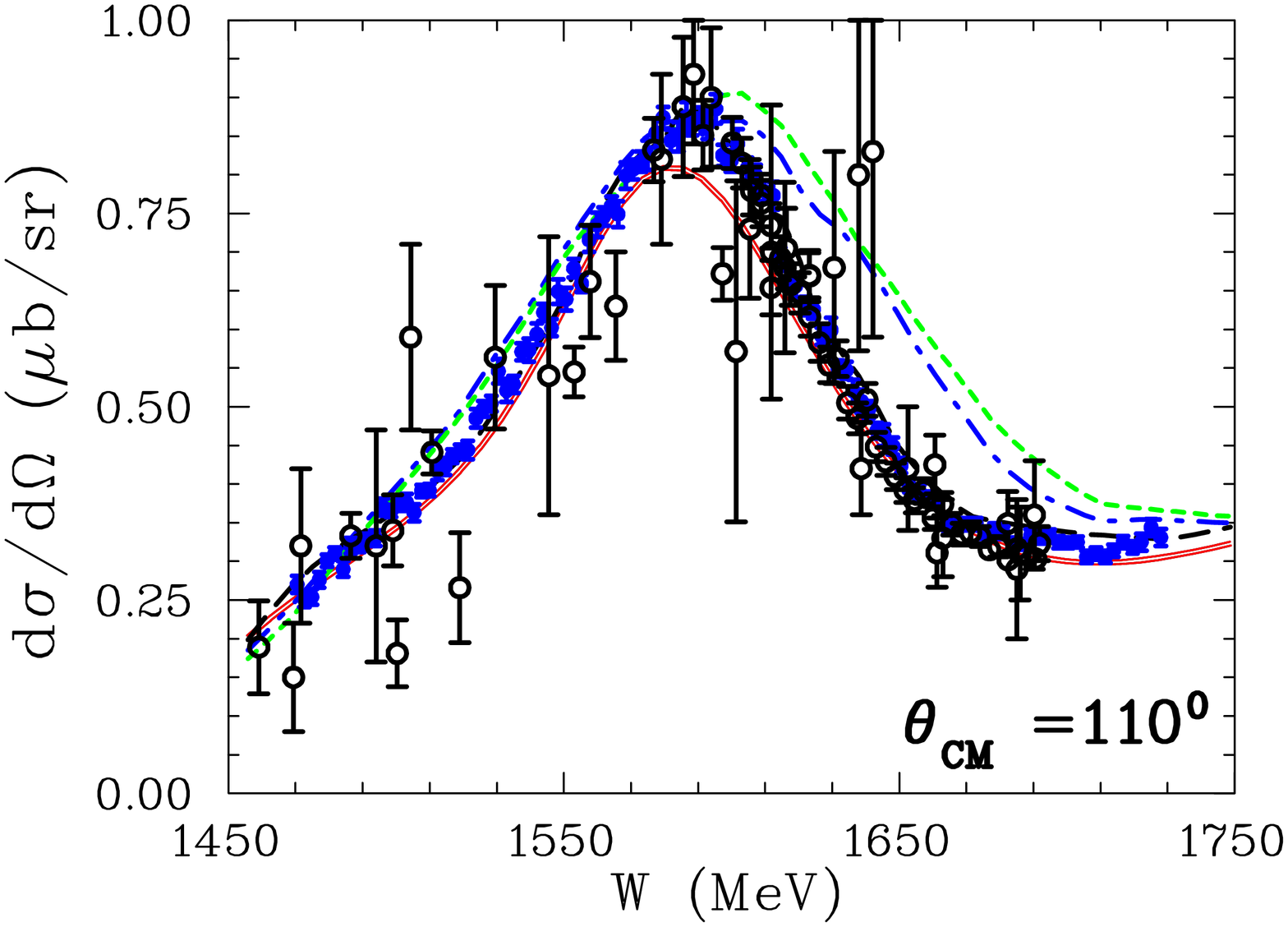}
	\includegraphics[angle=90, width=0.30\textwidth ]{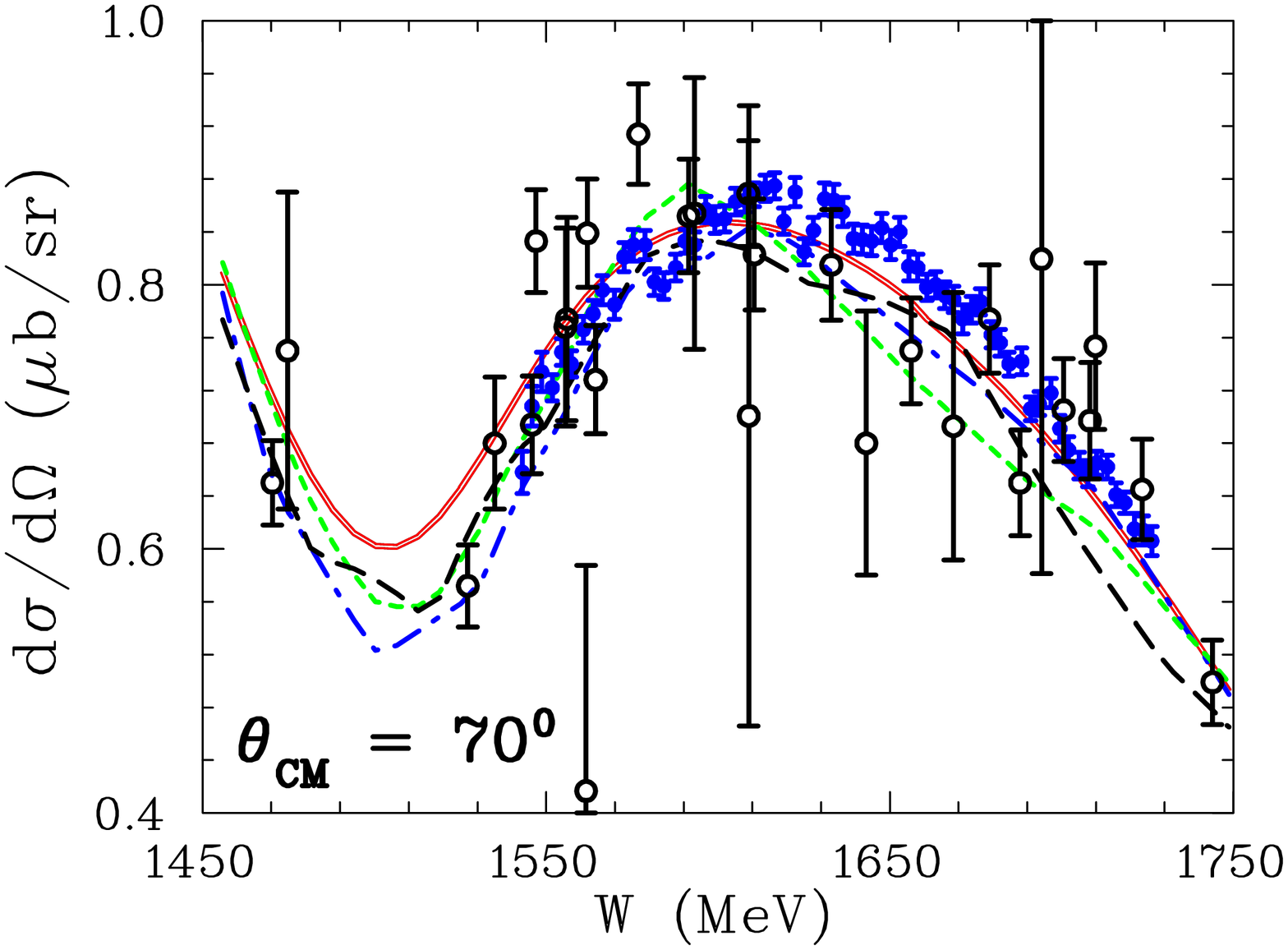}
	\includegraphics[angle=90, width=0.30\textwidth ]{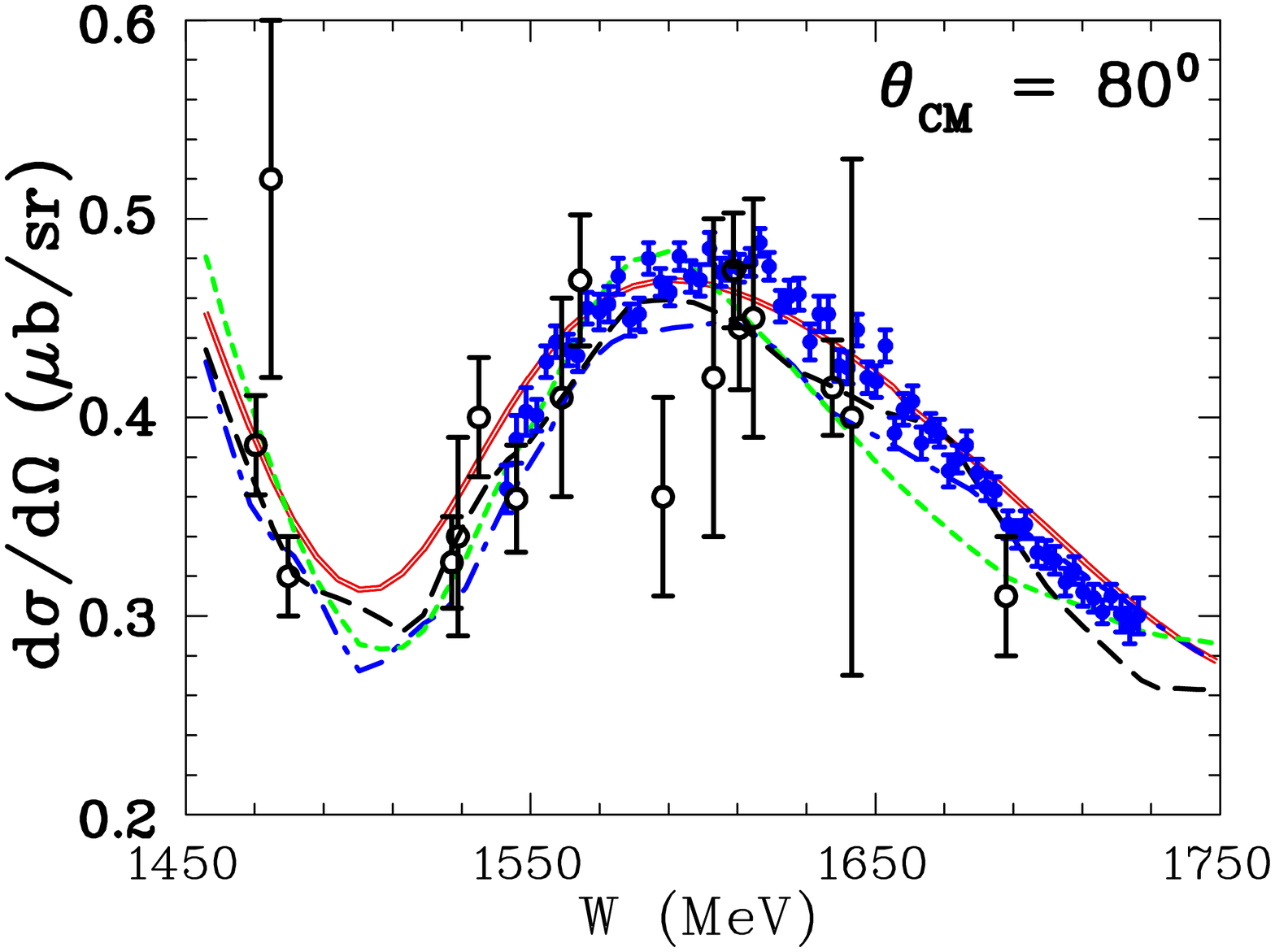}
	\includegraphics[angle=90, width=0.30\textwidth ]{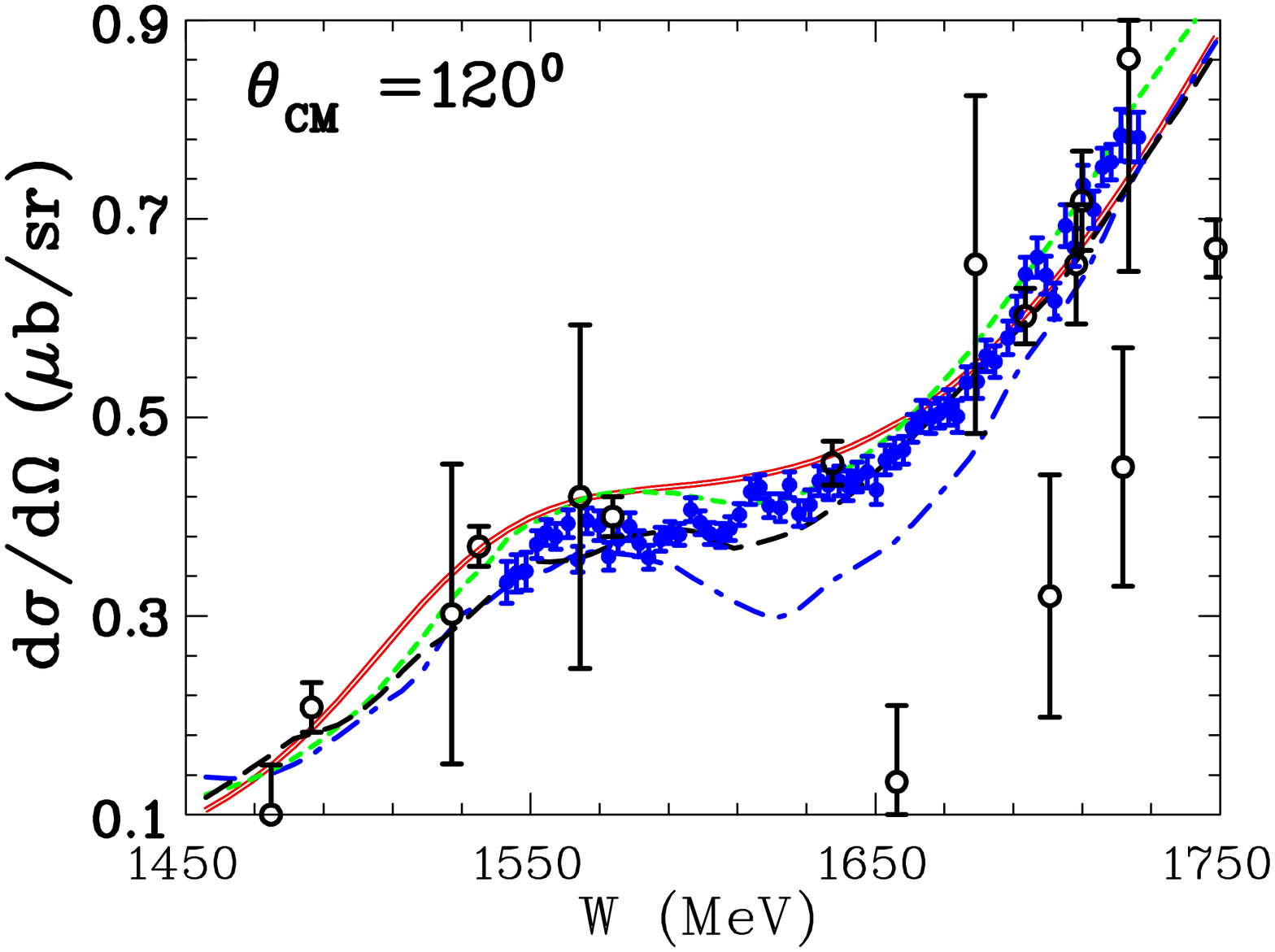}
\end{center}
\centerline{\parbox{0.80\textwidth}{
\caption[] {\protect\small (Color on-line) Differential cross sections for selected angles in the center-of-mass frame
for $\pi^- p$ (top panel) and $\pi^+ p$ (bottom panel) elastic scattering.  New EPECUR data (statistical uncertainties only) are plotted as  filled
	blue circles~\protect\cite{N1686} with previous 
	measurements~\protect\cite{SAID} (statistical uncertainties only)
	presented as open black circles. The data from earlier
	experiments are within bins of $\Delta\theta_{CM} = \pm$1$^\circ$.
	An existing GWU INS DAC fit, WI08~\protect\cite{arndt06} 
	is plotted with a red double solid curve while the older KH80~\protect\cite{hoehler79}, KA84~\protect\cite{KA84}, 
        and CMB~\protect\cite{cutkosky79b} fits are plotted as blue dash-dotted, green short dashed, and black dashed curves, respectively.} \label{fig:pinprecise} } }
\end{figure}

The databases for these reactions are larger than for any of the other reactions discussed here.  
Figures~\ref{fig:pi_plus.eps}, \ref{fig:pi_minus.eps}, and \ref{fig:cex.eps} 
summarize the available data below center-of-mass energy $W = 2.5$~GeV for $\pi^+p\to\pi^+p$, 
$\pi^-p\to\pi^-p$, and $\pi^-p\to\pi^0n$, respectively.  The $\pi N$ elastic scattering data \cite{arndt06} allowed the establishment of the 4-star resonances~\cite{pdg}. Still, many of the data were taken long ago and suffer from systematic uncertainties. In addition, available data for $\pi N$ elastic scattering are incomplete.  As Figs.~\ref{fig:pi_plus.eps} and \ref{fig:pi_minus.eps} show, very few measurements of the $A$ and $R$ polarization observables are available for $\pi^- p \to \pi^- p$ and $\pi^+ p \to \pi^+ p$ and then only for a few energies and angles.  Similarly, as Fig.~\ref{fig:cex.eps} indicates, there are no $A$ and $R$ data available at all for 
$\pi^- p \to \pi^0 n$ (and very few $P$ data).  Measurements of these observables are needed to construct truly unbiased partial-wave amplitudes. 
The dramatic improvement in statistics made possible in modern experimental 
physics (EPECUR) is demonstrated (for medium energies) in 
Fig.~\ref{fig:pinprecise}. The black data represent the current 
pion-nucleon database~\cite{SAID}, often with conflicting measurements. 
The new EPECUR data~\cite{N1686} are shown in blue. A similar improvement 
of the data for low energies is called for.
The importance of elastic $\pi N$ scattering for theory is discussed in Sec.~\ref{sec:ChPT}.

\begin{figure}[htpb]
\begin{center} 
	\includegraphics[angle=90, width=0.30\textwidth ]{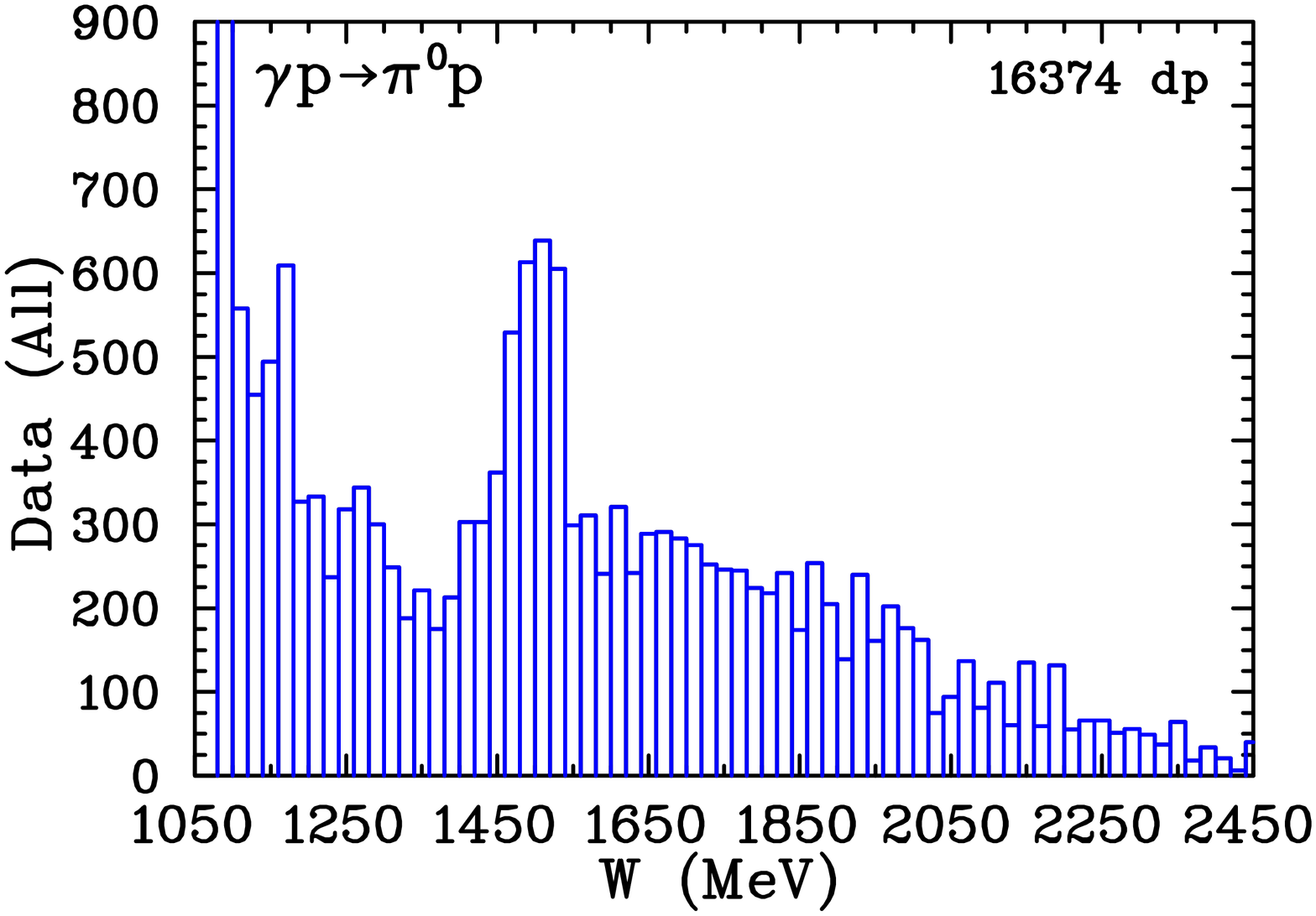}
	\includegraphics[angle=90, width=0.30\textwidth ]{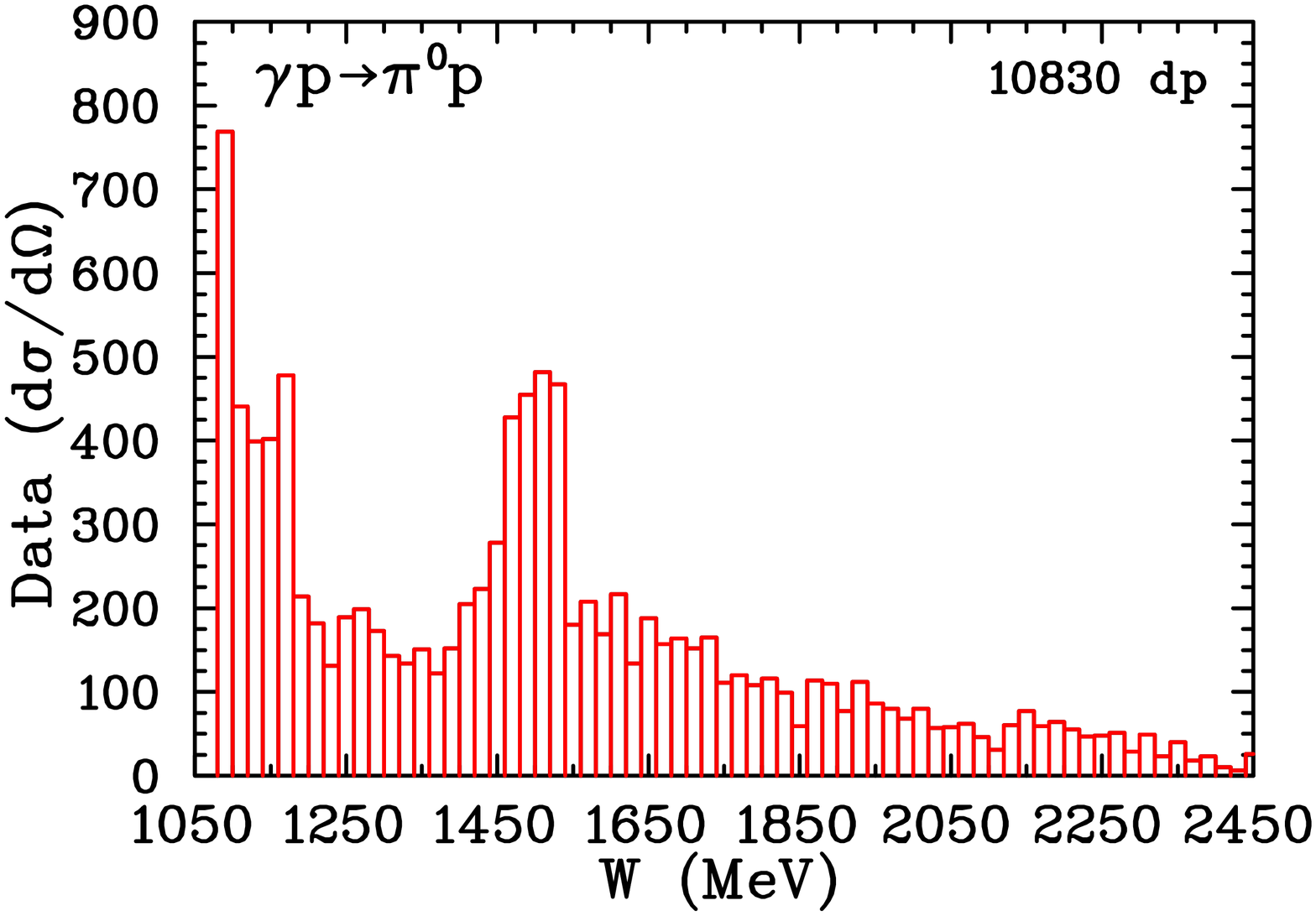}
	\includegraphics[angle=90, width=0.30\textwidth ]{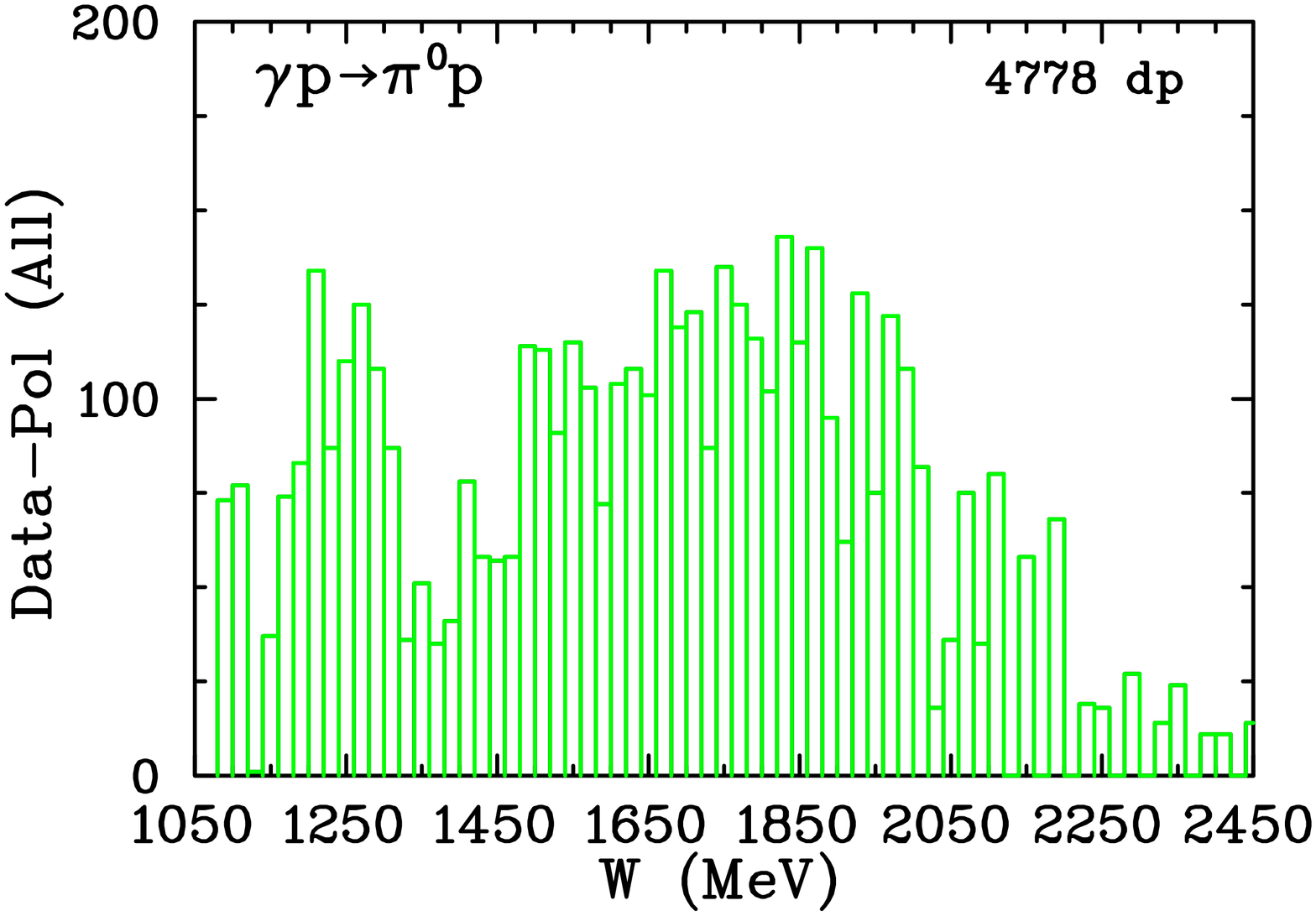}
	\includegraphics[angle=90, width=0.30\textwidth ]{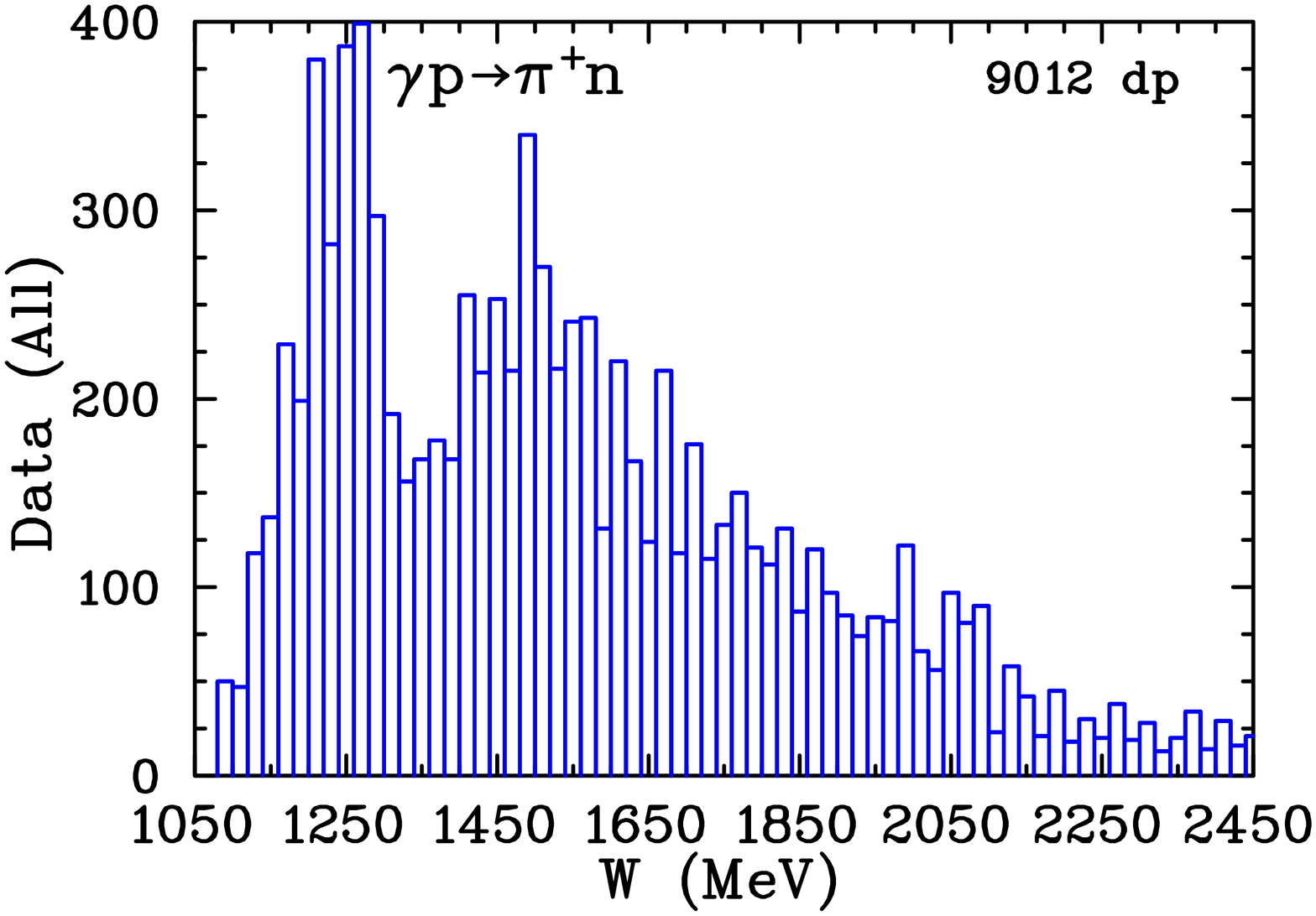}
	\includegraphics[angle=90, width=0.30\textwidth ]{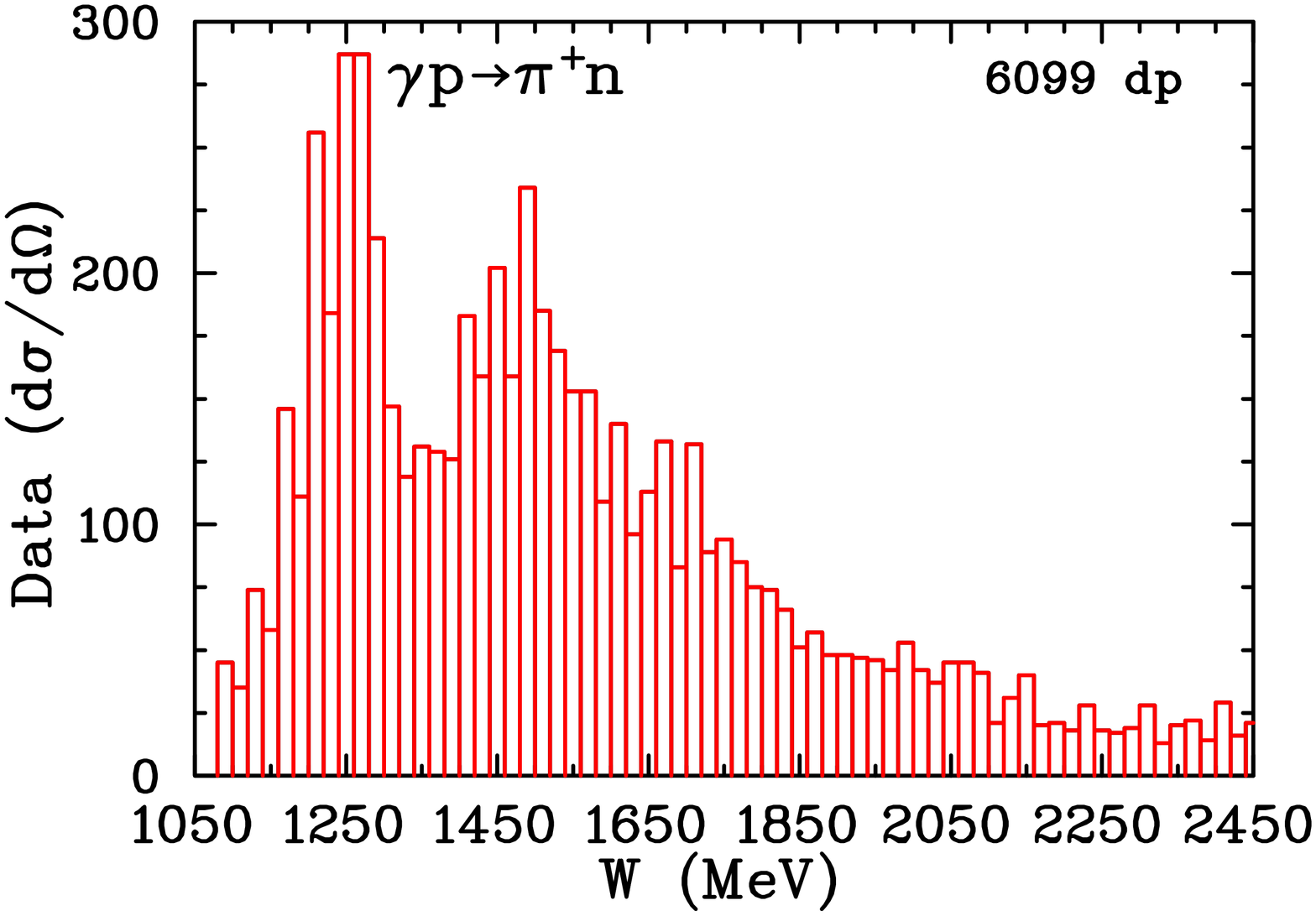}
	\includegraphics[angle=90, width=0.30\textwidth ]{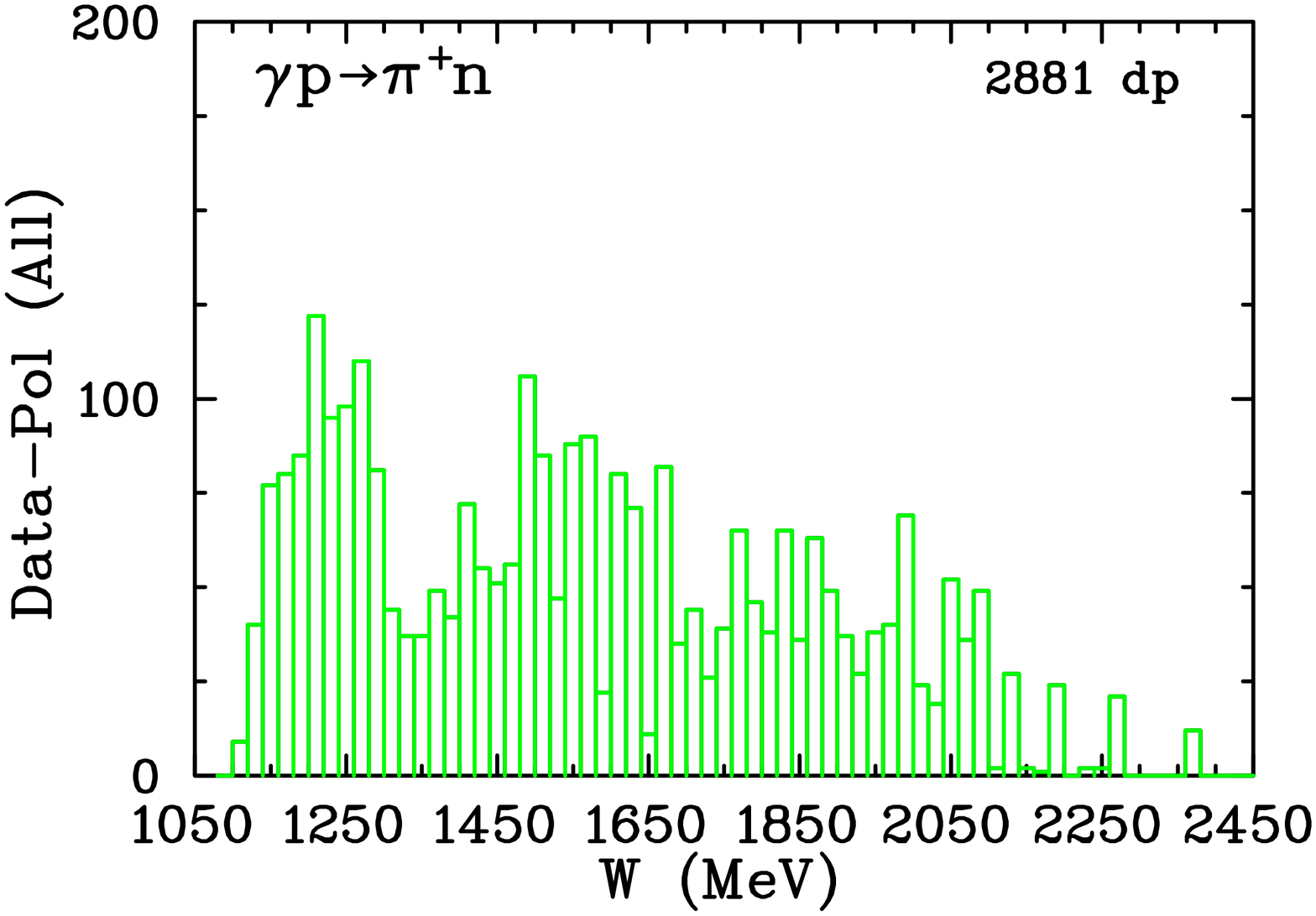}
\end{center}
\centerline{\parbox{0.80\textwidth}{
\caption[] {\protect\small (Color on-line) Data available for single pion 
	photoproduction off the proton as a function of 
	center-of-mass energy $W$ \cite{SAID}.  The number of data points (dp) 
	is given in the upper righthand side of each subplot.  Row 
	1: The first subplot (blue) shows 
	the total amount of $\gamma p\to\pi^0p$ data available for all observables, 
	the second plot (red) shows the amount of differential 
	cross-section ($d\sigma/d\Omega$) data available, the third 
	plot (green) shows the amount of $P$ observables data 
	available. Row 2: The first 
	subplot (blue) shows the total amount of $\gamma p\to\pi^+n$ data available for 
	all observables, the second plot (red) shows the amount of 
	differential cross-section ($d\sigma/d\Omega$) data available, 
	the third plot (green) shows the amount of $P$ observables 
	data available.} \label{fig:gamma_p.eps} } }
\end{figure}

\begin{figure}[htpb]
\begin{center} 
\includegraphics[angle=90, width=0.30\textwidth ]{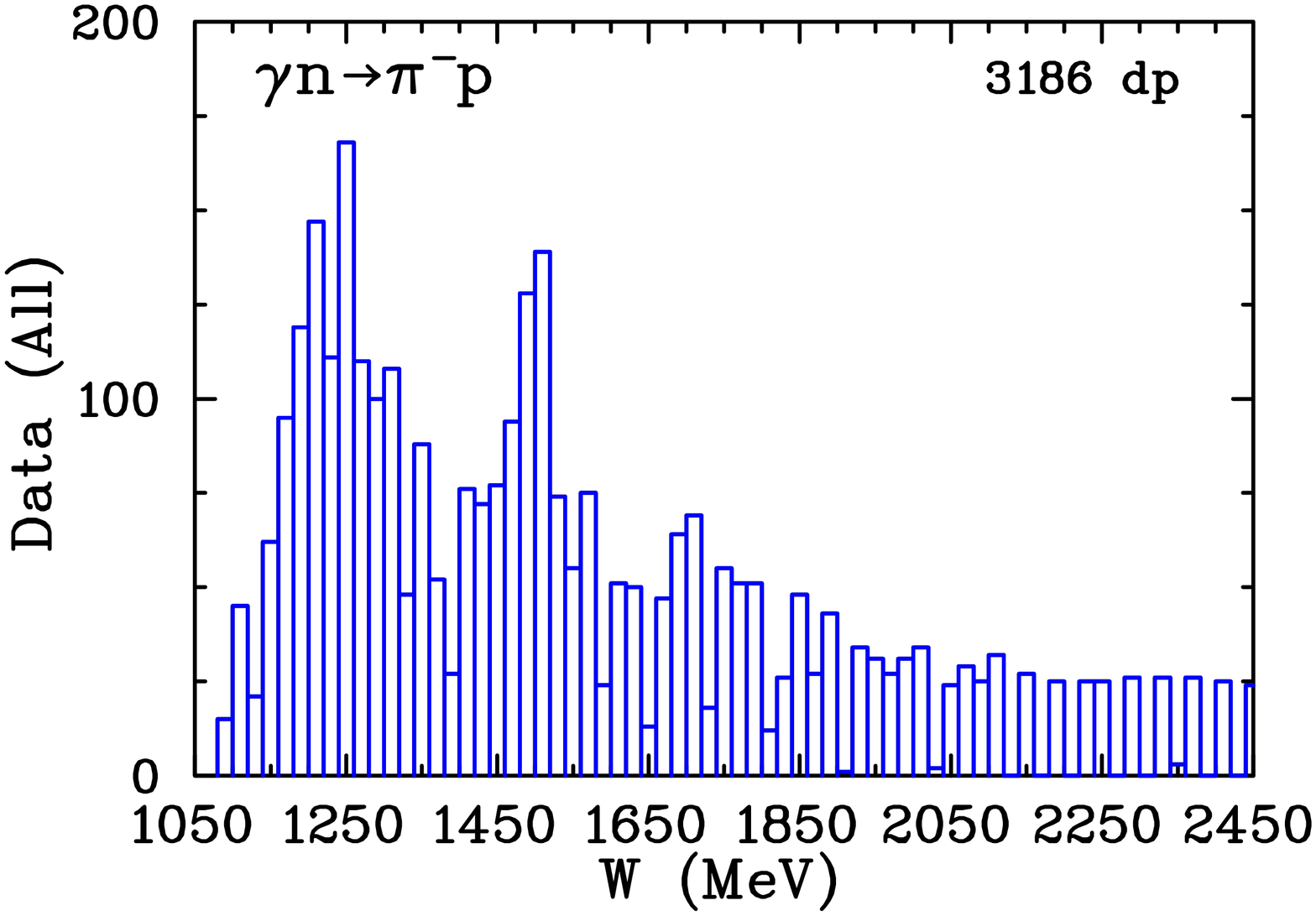}
\includegraphics[angle=90, width=0.30\textwidth ]{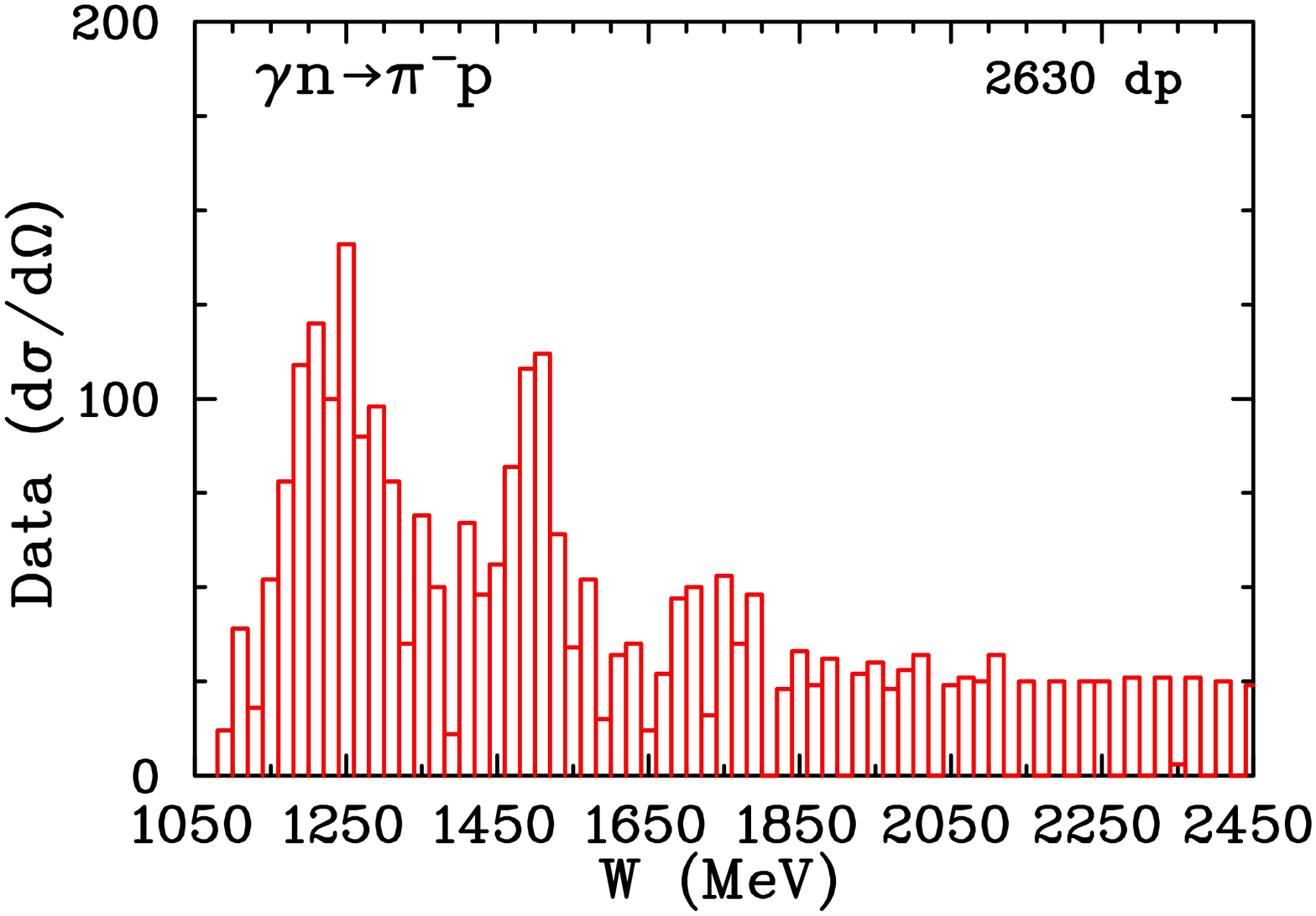}
\includegraphics[angle=90, width=0.30\textwidth ]{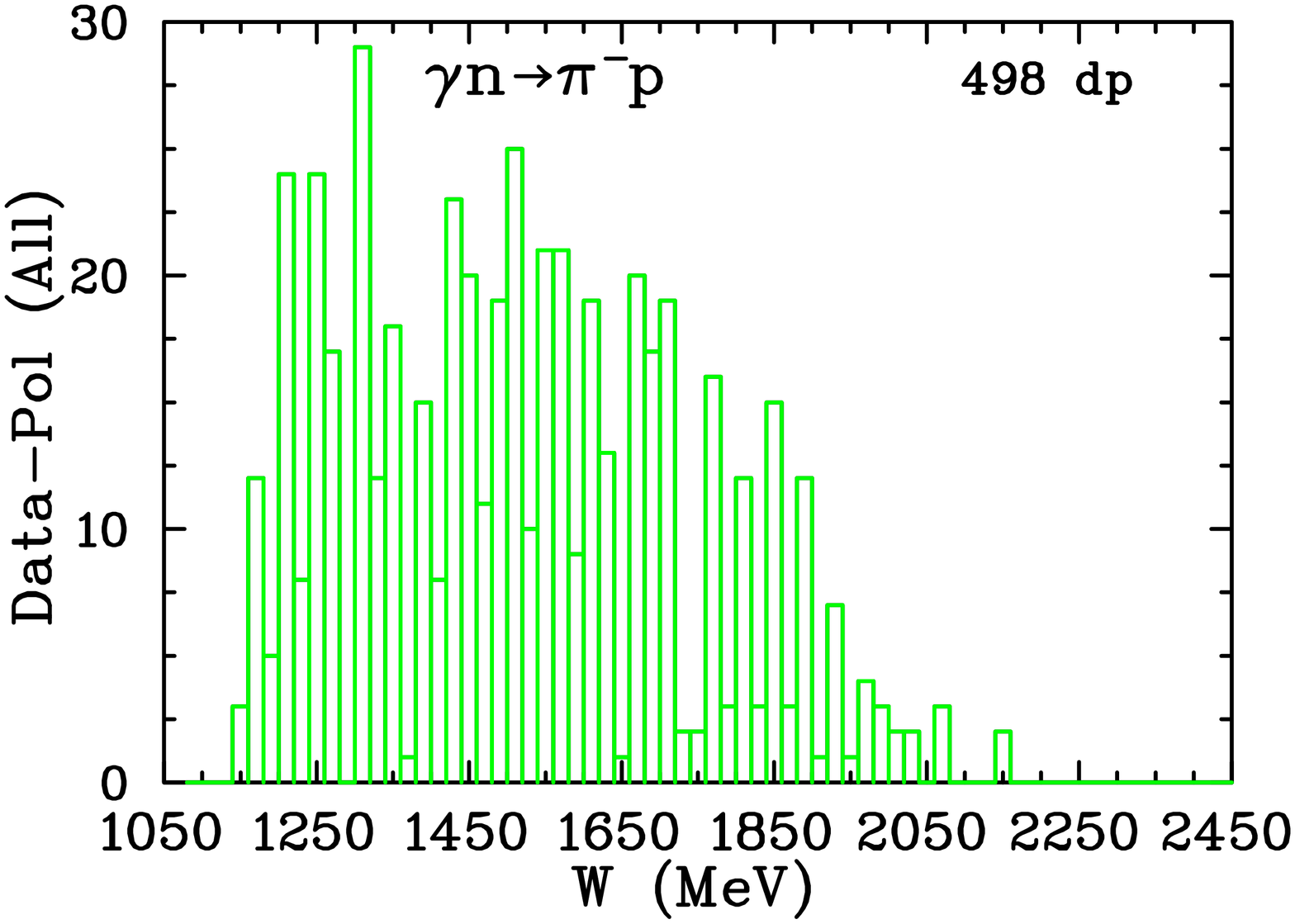}
\includegraphics[angle=90, width=0.30\textwidth ]{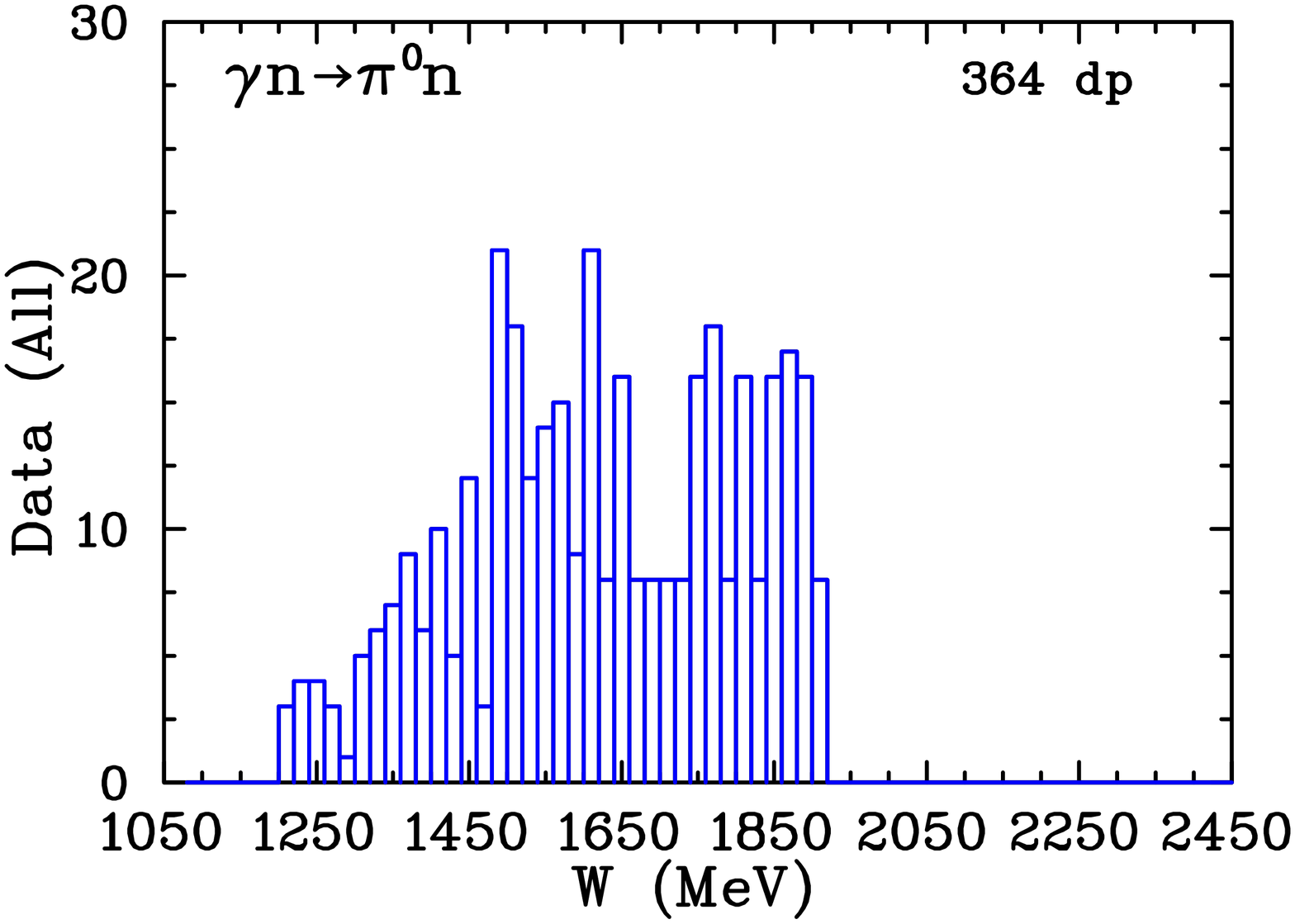}
\includegraphics[angle=90, width=0.30\textwidth ]{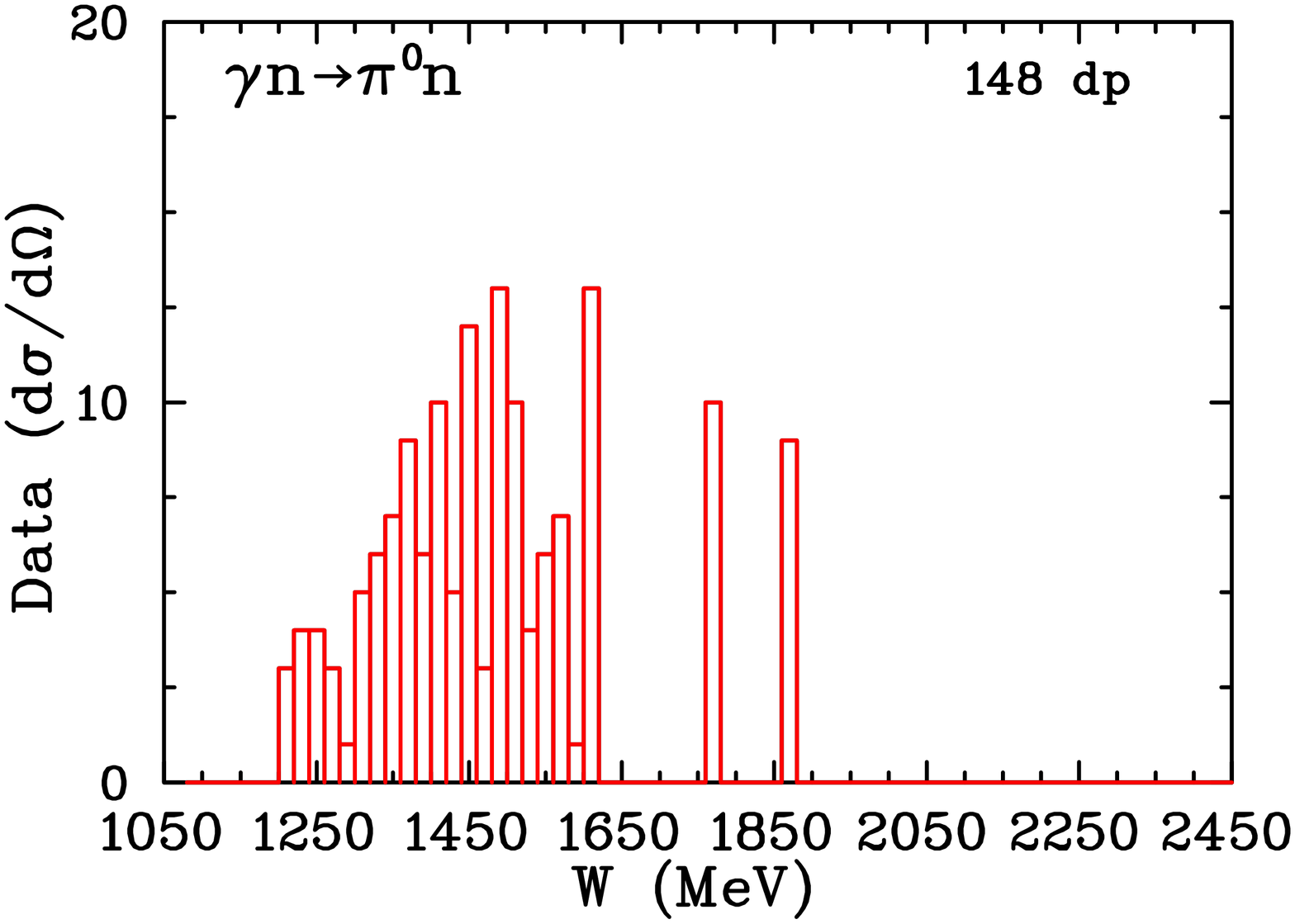}
\includegraphics[angle=90, width=0.30\textwidth ]{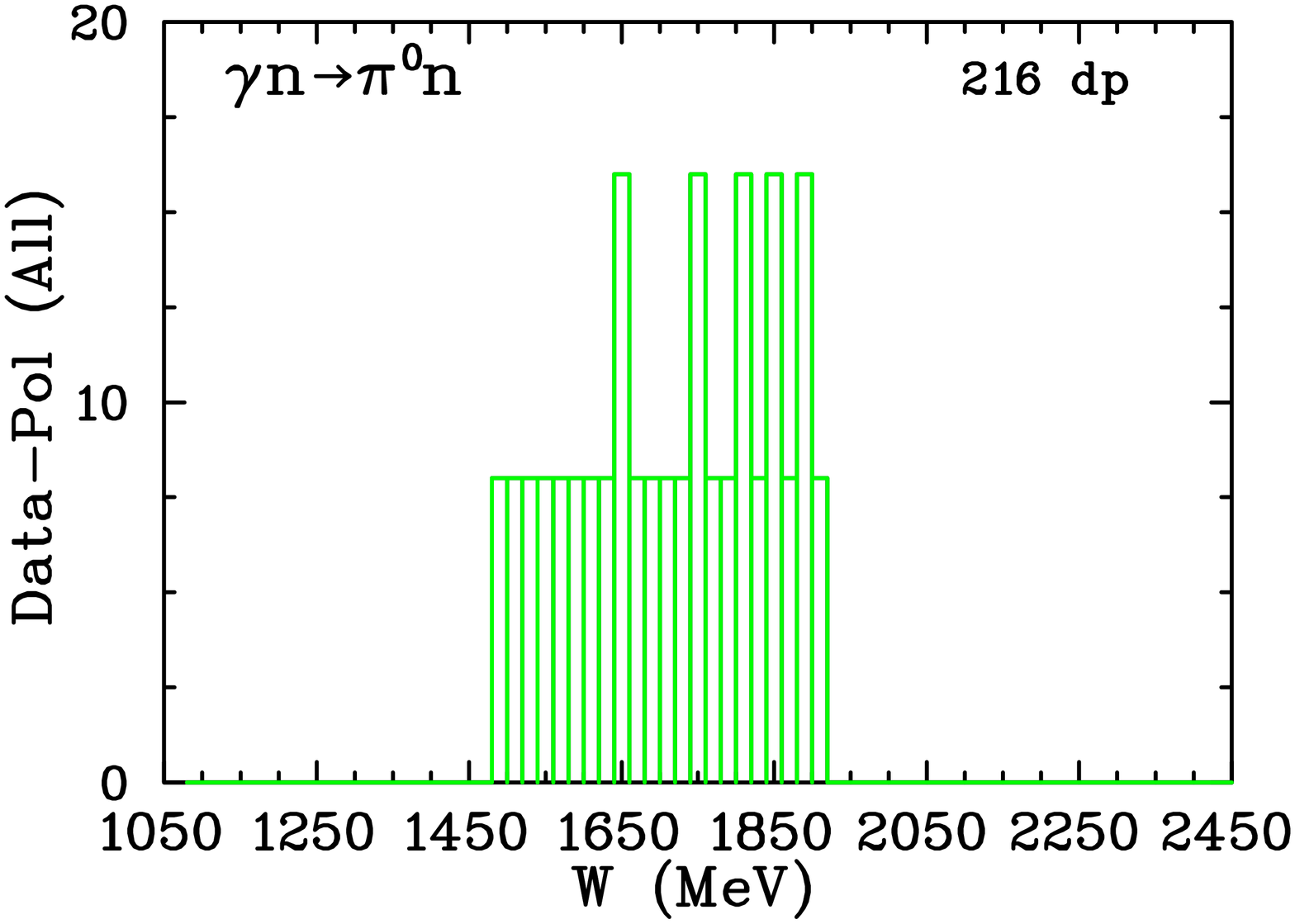}
 \end{center}
\centerline{\parbox{0.80\textwidth}{
 \caption[] {\protect\small (Color on-line) Data available for single pion 
	photoproduction off the neutron as a function of 
	center-of-mass energy $W$ \cite{SAID}.  The number of data points (dp) 
	is given in the upper righthand side of each subplot.  Row 
	1: The first subplot (blue) shows 
	the total amount of $\gamma n\to\pi^-p$ data available for all observables, the 
	second plot (red) shows the amount of differential 
	cross-section ($d\sigma/d\Omega$) data available, the 
	third plot (green) shows the amount of $P$ observables data 
	available. Row 2: The first subplot 
	(blue) shows the total amount of $\gamma n\to\pi^0n$ data available for all 
	observables, the second plot (red) shows the amount of 
	differential cross-section ($d\sigma/d\Omega$) data available, 
	the third plot (green) shows the amount of $P$ observables 
	data available.} \label{fig:gamma_n.eps} } }
\end{figure}

Figures~\ref{fig:gamma_p.eps} and \ref{fig:gamma_n.eps} summarize the 
available data below $W = 2.5$~GeV for single pion photoproduction on the 
proton and neutron, respectively.
Many high-precision data for these reactions have been measured recently. Their analysis allows a determination of accurate multipoles, which in turn provide information about electromagnetic baryon resonance properties. However, for the spectroscopy of missing states that couple only weakly to the $\pi N$ channel, these reactions are not ideally suited.
Of course, any reaction that can be studied using real photon beams can also be studied 
using virtual photons via electron scattering experiments; however, the 
analysis and interpretation of data from electron scattering is more 
complicated than that from photoproduction experiments because meson electroproduction involves six helicity amplitudes versus only four for photoproduction.

Reactions that involve the $\eta N$ and $K \Lambda$ channels
are notable because they have pure isospin-1/2 contributions:

\begin{tabular}{ll}
$\gamma p \to \eta p$,  ~~~~~~~~~~     & $\pi^- p \to \eta n$, \\
$\gamma n \to \eta n$,       &  \\
$\gamma p \to K^+ \Lambda$,  & $\pi^- p \to K^0 \Lambda$, \\
$\gamma n \to K^0 \Lambda$. &  \\
\end{tabular}

\begin{figure}[htpb]
\begin{center} 
	\includegraphics[angle=90, width=0.30\textwidth ]{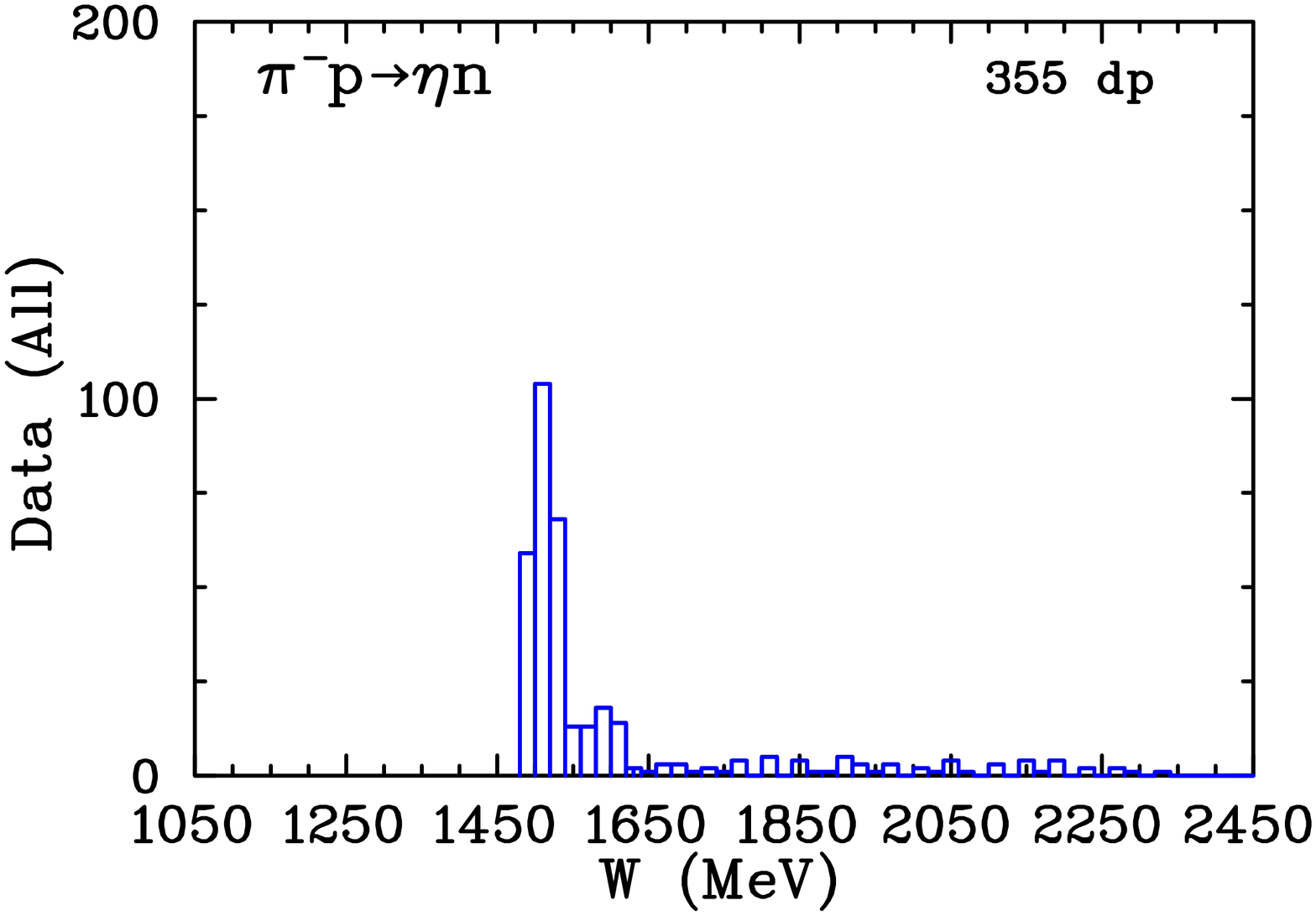}
	\includegraphics[angle=90, width=0.30\textwidth ]{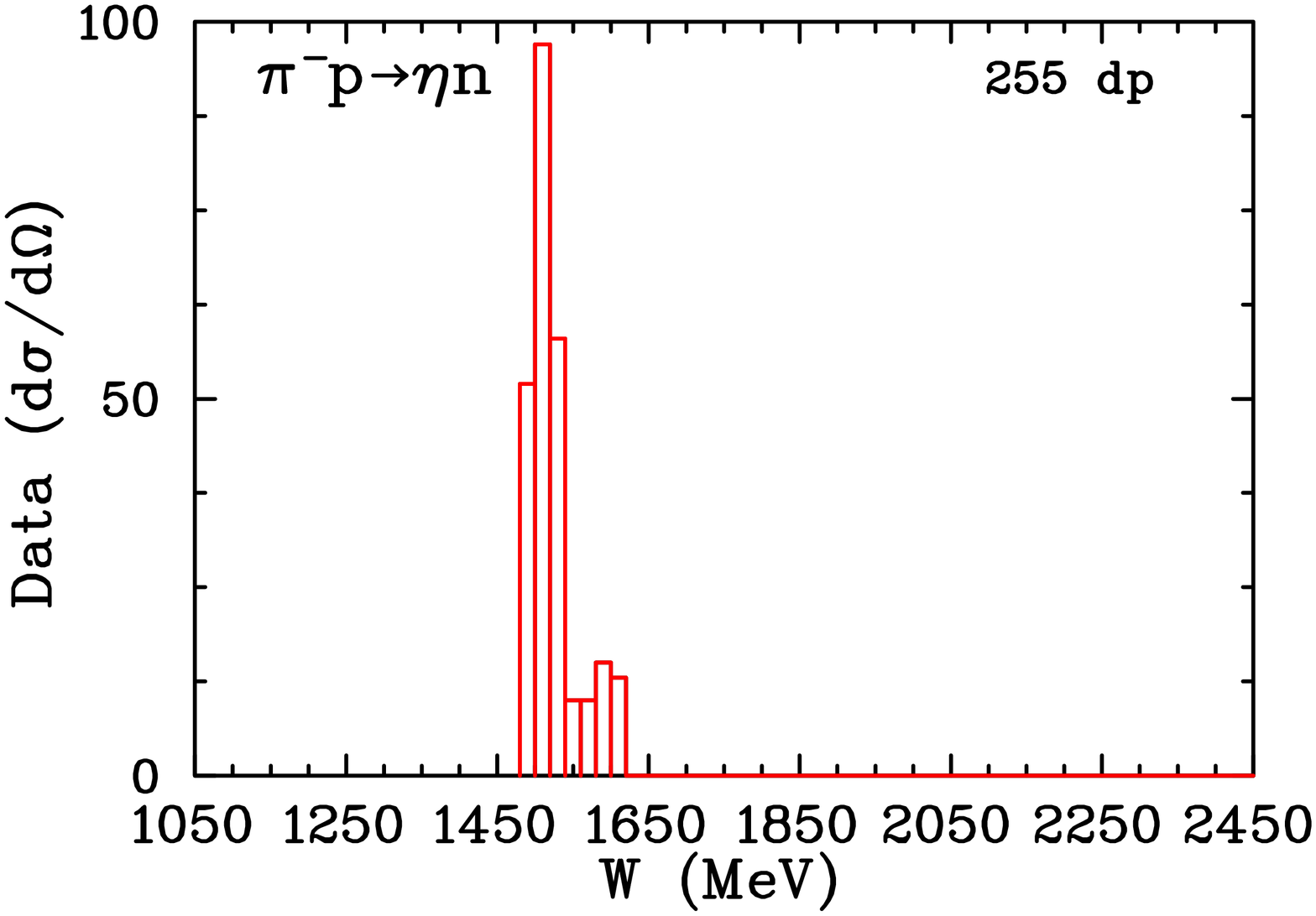}
	\includegraphics[angle=90, width=0.30\textwidth ]{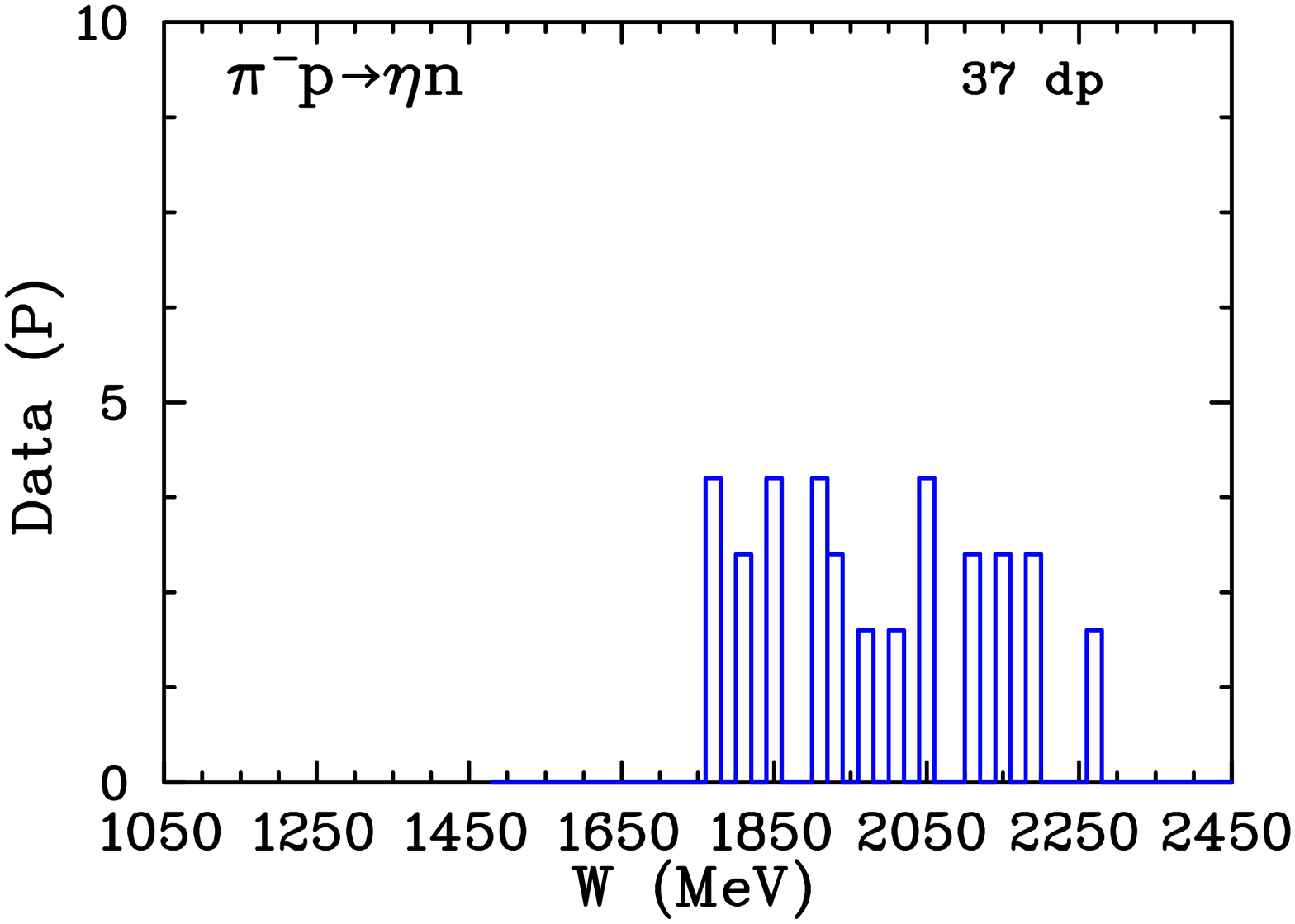}
\end{center}
\centerline{\parbox{0.80\textwidth}{
\caption[] {\protect\small (Color on-line) Data available for $\pi^-p\to\eta n$ as 
	a function of center-of-mass energy $W$ \cite{SAID}.  The number of data 
	points (dp) is given in the upper righthand side of each subplot.  
	The first subplot (blue) shows the total amount of data available 
	for all observables, the second plot (red) shows the amount of 
	differential cross-section ($d\sigma/d\Omega$) data available, 
	the third plot (blue) shows the amount of polarization data 
	available.} \label{fig:eta.eps} } }
\end{figure}

Figure \ref{fig:eta.eps} summarizes the available data below $W = 
2.5$~GeV for $\pi^-p\to\eta n$. The photoproduction reactions are 
especially significant because they provide an opportunity to search 
for ``missing resonances'' in reactions that do not involve the $\pi N$ 
channel. More generally, analyses of photoproduction combined 
with pion-induced reactions permit separating the EM and hadronic 
vertices.  It is only by combining information from analyses of both 
$\pi N$ elastic scattering and $\gamma N \to \pi N$ that make it possible 
to determine the $A_{1/2}$ and $A_{3/2}$ helicity couplings for $N^\ast$ 
resonances.

The reaction $\gamma p\to\eta p$ is one of the key reactions for which 
colleagues in the EM community hope to do a ``complete measurement''.  
Any coupled-channel analysis of those measurements will need precise 
data for $\pi^-p\to\eta n$. Most of the available data for that 
reaction come from measurements published in the 1970s, which have 
been evaluated by several groups as being unreliable above $W = 
1620$~MeV. 
Precise new data for $\pi^-p\to\eta n$ were measured more recently by 
the Crystal Ball Collaboration~\cite{prakhov05}, but these extend only 
up to the peak of the first S$_{11}$ resonance at 1535~MeV.  As Fig.~\ref{fig:eta.eps} 
shows, very few polarization data exist for these reactions.
In Ref.~\cite{Krusche:2014ava}
the imbalance of photon- vs.\ pion-induced $\eta$ production data is pointed out and
a call for improved data on $\pi^-p\to\eta n$, possibly from a HADES upgrade, is made.


Recently, the Giessen group concluded that ``further progress in 
understanding of the $\eta$-meson production would be hardly possible 
without new measurements of the $\pi N \to \eta N$ reaction''~\cite{shklyar13}.  
This task relates to the one-star $N$(1685) state that was recently added to 
the RPP Baryon Listings~\cite{pdg}. In 2007, a narrow structure at $W \approx 
1680$~MeV was reported in GRAAL measurements for quasi-free $\eta$ 
photoproduction on neutrons bound in a deuterium target \cite{kuznetsov07}. 
A narrow structure at this energy was also observed in inclusive measurements, 
$d(\gamma,\eta)pn$, performed at the LNS (now ELPH) at Tohoku University \cite{miyahara07}, 
and in quasi-free measurements of $\eta$ photoproduction 
on the neutron at Bonn~\cite{jaegle08,jaegle11} and at 
MAMI~\cite{werthmueller13,werthmueller14}. A narrow peak at $W \approx 1685$~MeV 
has also been observed in GRAAL measurements of quasi-free Compton scattering on 
the neutron~\cite{kuznetsov11}. This peak is not observed in $\gamma p \to \eta 
p$, and a good deal of speculation and controversy has arisen concerning its 
interpretation.   This state is unusually narrow for the non-strange sector, 
$\Gamma\leq 30$~MeV.  If it does exist, the understanding of its nature
is an attractive task for future measurements involving $\eta$ production. New 
experiments and analyses are in progress. 
The elementary reaction $\pi^-p \to \eta n$ could serve as
a new detection channel for the structure at $W \approx 1685$~MeV using the same $\eta n$ final state in which the structure appears in photoproduction experiments on the neutron. 
Measurements of $\pi^-p \to \eta n$, however,  would not be plagued by the problems associated with measurements on deuterium or $^3$He targets.  
The HADES collaboration will measure pion-induced reactions (Sec.~\ref{sec:current_projects}) but
not yet the $\eta n$ final state. A new J-PARC measurement will determine
only the $\pi^-p\to\eta n$ differential cross section. Clearly, to provide
conclusive answers to the puzzle tied to the structure at $W\approx 1685$~MeV, one needs a dedicated experiment with hadron beams and polarization measurements such as the proposed EIC facility at JLab could provide.



There are extensive data for $\gamma p\to K^+\Lambda$ but almost no data for the
reaction $\gamma n \to K^0\Lambda$ measured on the deuteron. Consequently, resonances with strong coupling to $K\Lambda$, weak coupling to $\gamma p$ but strong coupling to $\gamma n$
have been inaccessible in photoproduction to date.  
 A possible candidate is the narrow structure at $W \sim 1.65$~GeV discovered in $\eta$ photoproduction
on the neutron (see previous discussion).
That structure -- of unknown nature, and not necessarily a resonance -- is situated above the $K\Lambda$ threshold;
it has a known strong coupling to $\gamma n$, and it could be visible in $\pi^-p\to K^0\Lambda$. However, just around the energy of $W \sim 1.65$~GeV
the data in that reaction are plagued by systematic uncertainties and conflicting measurements (see, e.g., Fig.~19 in Ref.~\cite{Ronchen:2012eg}).  

Another reason to study the reaction $\pi^-p\to K^0\Lambda$ is given by the resonances from $K\Lambda$ photoproduction claimed by the Bonn-Gatchina
group~\cite{anisovich12a,anisovich12b} and others~\cite{Mart:2012fa}.
Because these states have a large branching ratio into $K\Lambda$, pion-induced  $K^0\Lambda$ production provides an entirely independent reaction
to confirm these states. So far, the data are not of sufficient quality to do so. These new states have only weak branching fractions into
the $\pi N$ channel, as they are less visible in elastic $\pi N$ scattering. One can, thus, expect a signal of moderate strength for the
reaction $\pi^-p\to K^0\Lambda$ and more precise data are called for.

A striking example of why improved data for pion-induced kaon production are necessary is given by the $N(1710)1/2^+$. Its properties and even its existence
have been debated intensively over time and seem to be intimately intertwined with the reaction $\pi^-p \to K^0\Lambda$, as argued in the overview given in Ref.~\cite{Burkert:2014wea}.
Similar arguments apply to the $N(1900)3/2^+$ as outlined in that reference. 
The latter resonance was also found in kaon photoproduction~\cite{Mart:2012fa},
which makes its study in the reaction $\pi^-p\to K^0\Lambda$ especially promising.

More precise data for the reaction $\pi^-p\to K^0\Lambda$ (in combination with $K^-p\to K^0\Xi^0$) would also enable the study of SU(3)
flavor symmetry and its breaking. In Ref.~\cite{Starostin:2001zz}
SU(3) flavor symmetry relations could be confirmed to surprising accuracy by comparing the reactions $\pi^-p\to\eta n$ and $K^-p\to\eta\Lambda$. Along
similar lines, the simultaneous study of both reactions $\pi^-p\to K^0\Lambda$ and $K^-p\to K^0\Xi^0$ (for which the data situation is
much less known close to threshold) would allow for another test of SU(3) flavor symmetry.

Another group of related reactions involve the $K \Sigma$ channel:

\begin{tabular}{ll}
$\gamma p\to K^+\Sigma^0$, ~~~~~~~~~~ & $\pi^-p\to K^0\Sigma^0$, \\
$\gamma p\to K^0\Sigma^+$,  & $\pi^-p\to K^+\Sigma^-$, \\
$\gamma n\to K^+\Sigma^-$,  & $\pi^+p\to K^+\Sigma^+$, \\
$\gamma n\to K^0\Sigma^0$. &  \\
\end{tabular}

Except for $\pi^+p\to K^+\Sigma^+$, these reactions involve a mixture 
of isospin 1/2 and 3/2.  
Although there have been a number of recent high-quality measurements 
involving $K \Sigma$ photoproduction, the status of complementary 
reactions measured with pion beams is rather dismal. There are 
generally fewer available data for $\pi^-p$ reactions with $K\Sigma$, 
$\eta ' N$, $\omega N$, and $\phi N$ final states than for 
$\pi^-p\to\eta n$.

\begin{figure}[htpb]
\begin{center} 
        \includegraphics[angle=0, width=0.49\textwidth ]{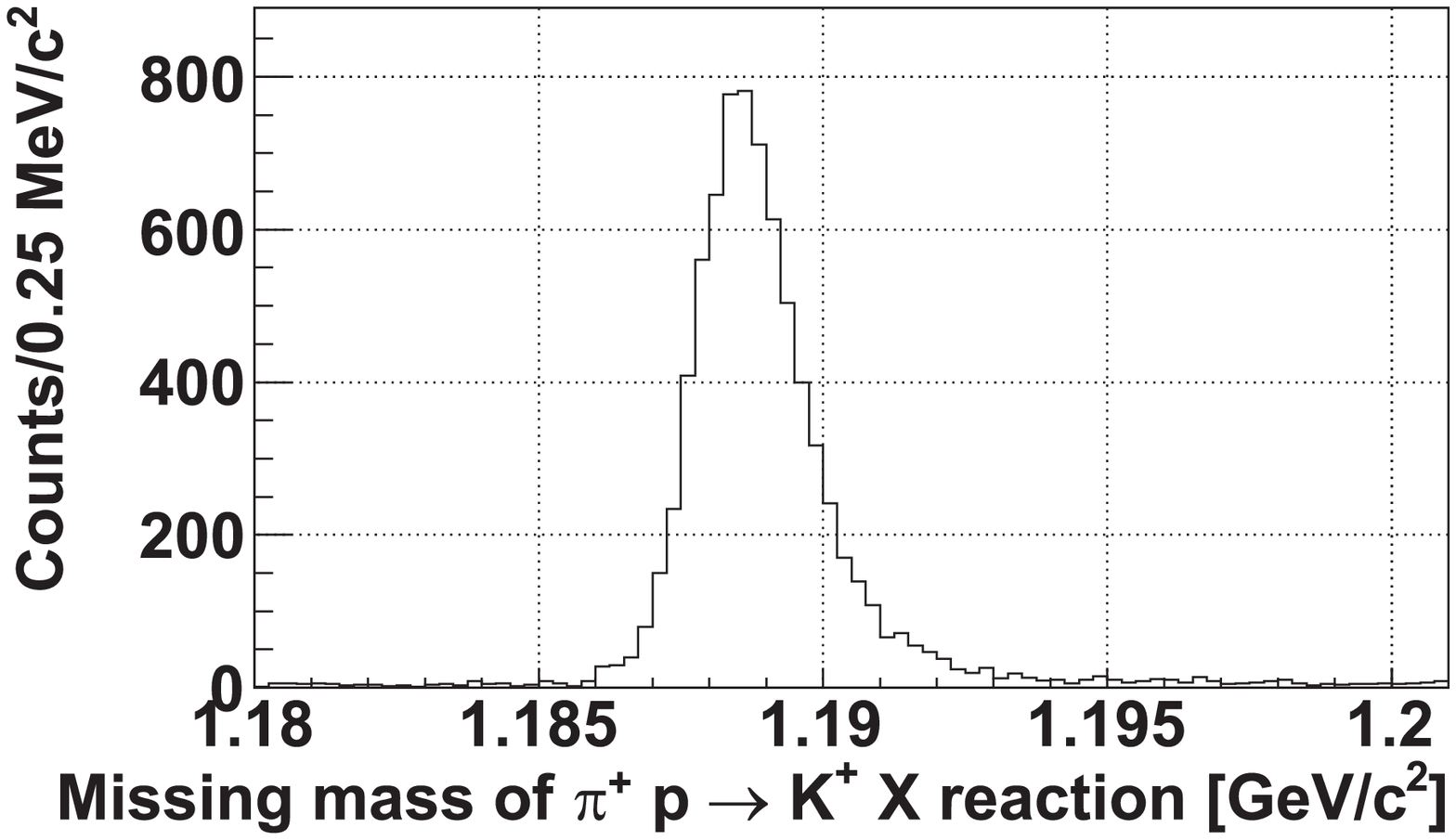}
        \includegraphics[angle=0, width=0.49\textwidth ]{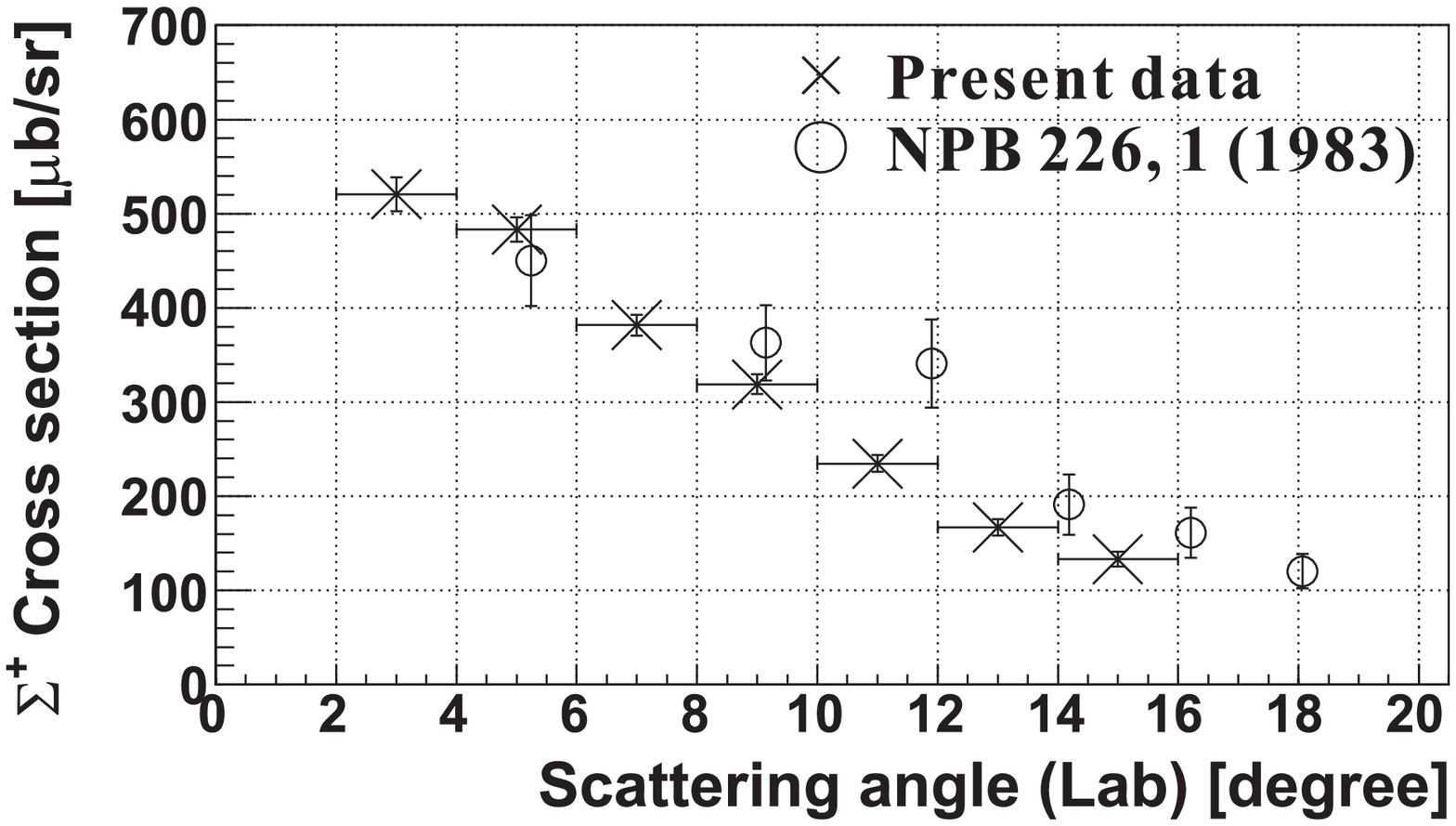}
\end{center}\centerline{\parbox{0.80\textwidth}{
\caption{Differential cross section of the reaction $\pi^+p\to K^+\Sigma^+$ 
	at $p_{\rm lab}=1.37$~GeV/$c$ from a recent J-PARC 
	experiment~\protect\cite{Shirotori:2012ka}, compared to the only 
	available older data~\protect\cite{Candlin:1982yv}.}
       \label{fig:kplus} } }
\end{figure}

To demonstrate recent experimental progress, Fig.~\ref{fig:kplus} shows new 
measurements of the pure isospin $I=3/2$ $K^+\Sigma^+$ final state by the E19 
Collaboration at J-PARC, compared to the best available older data. Note the 
error bars that are up to one order of magnitude more precise than in the 
previously available data. Also, angular coverage is significantly improved.

Measurements like this, over a more comprehensive energy range, will greatly 
improve PWAs of the $K\Sigma$ final state and, in return, help to extract 
the $S$-wave contribution needed, e.g.,  in approaches based on unitarized chiral 
perturbation theory (UChPT) (see Sec.~\ref{sec:UChPT}).

Regarding the $K\Sigma$ final state, the anomaly in $K\Sigma$ photoproduction 
recently discovered at ELSA deserves attention~\cite{Ewald:2011gw}. A surprisingly 
sudden drop in the $K^0\Sigma^+$ photoproduction cross section, in combination with 
a sudden change in the differential cross section was observed, with a UChPT 
explanation formulated in Ref.~\cite{Ramos:2013wua} that is tied to a resonance at $W=2.03$~GeV, which should also be visible in $\pi N\to K\Sigma$.  Better measurements of 
pion-induced reactions are needed to shed light on the issue. 
In particular, the $\pi^-p\to K^+\Sigma^-$ reaction is of utmost interest in this 
respect. No meson $t$-channel exchanges are possible for this reaction, which is 
particularly sensitive to $u$-channel exchanges. This reaction, in conjunction 
with the data situation, is discussed in depth in Ref.~\cite{Ronchen:2012eg}. 
Specifically, the model of Ref.~\cite{Ronchen:2012eg} cannot describe the total cross section around $W=2$~GeV, which might be a sign of new resonances.

From the self-analyzing property of hyperons, it is expected that the recoil 
polarization $P$ can also be measured with a greatly improved accuracy in future 
J-PARC experiments. The measurement of differential cross sections and recoil 
polarizations can be obtained at J-PARC with relatively modest investments; yet, the 
return is large and we make an explicit appeal to pursue this line of experimental 
activity.

In summary, the family of reactions $\pi N\to K^0\Lambda$, $K^0\Sigma^0$, $K^+\Sigma^-$, and $K^+\Sigma^+$ provides complementary and, in some cases even
exclusive, information compared to photoproduction. Data with larger angular coverage, smaller systematic uncertainties, and finer energy binning
are needed to confirm recently discovered resonances, to discover new resonances inaccessible in photoproduction, and to test theoretical
multichannel concepts.

Other important reactions that can be studied are those with $\pi\pi N$ 
final states.  The main reactions amenable to measurements include:

\begin{tabular}{ll}
$\gamma p\to \pi^0 \pi^0 p$, ~~~~~~~~~~ & $\pi^-p\to \pi^0 \pi^0 n$, \\
$\gamma p\to \pi^0 \pi^+ n$,  & $\pi^-p\to \pi^0 \pi^- p$, \\
$\gamma p\to \pi^+ \pi^- p$,  & $\pi^-p\to \pi^+ \pi^ - n$, \\
$\gamma n\to \pi^0 \pi^0 n$, &  $\pi^+p\to \pi^0 \pi^+ p$,\\
$\gamma n\to \pi^0 \pi^- p$, &  $\pi^+p\to \pi^+ \pi^+ n$,\\
$\gamma n\to \pi^+ \pi^- n$. &  \\
\end{tabular}

The analysis and interpretation of data from these 
reactions is more complicated because they involve three-body final 
states.  However, $\pi N\to\pi\pi N$ reactions have the lowest energy 
threshold 
of any inelastic hadronic channel and some of the largest cross sections.  For most established $N^\ast$ and 
$\Delta^\ast$ resonances, the dominant inelastic decays are to $\pi\pi N$ 
final states. Our main source of knowledge about $\pi N\to\pi\pi N$ comes 
from a 30-year-old study of 241,214 bubble-chamber events that were 
analyzed in an isobar-model PWA at center-of-mass energies from $W$= 1320 
to 1930~MeV~\cite{manley84}. For these reasons, one needs high-quality, 
high-statistics data for $\pi N\to\pi\pi N$ that can be analyzed together 
with complementary data for $\gamma N\to\pi\pi N$ channels.

Another reaction of interest is the photoproduction of vector mesons, such 
as the $\omega$ meson.  The few data now available for $\pi^-p\to\omega n$ 
lead to ambiguous solutions in PWAs of $\pi N \to\omega N$~\cite{karami79} and
are almost worthless in constraining PWAs of $\omega$ photoproduction.
The $\omega N$ threshold region is also especially attractive in searching
for new resonances because the reaction threshold is located at the
higher-energy edge of the third resonance region, in which the RPP \cite{pdg}
shows seven $N^\ast$ states with masses between 1650~MeV and 1720~MeV.  The next $N^\ast$ state
in the RPP is the two-star $N(1860)5/2^+$. It cannot be 
excluded that this energy range may contain unknown $N^\ast$ resonances 
that couple more strongly to $\omega N$ than to other meson-baryon 
channels. Such an advantage of the threshold region does not exist, for 
example, for the two most dominant channels, $\pi N$ and $\eta N$, which 
are strongly coupled to their near-threshold resonances, $\Delta(1232)$3/2$^+$ 
and $N(1535)$1/2$^-$~\cite{strakovsky}, respectively, in the first and second resonance regions where new resonances are not expected.

Signs for resonances with low star rating have recently been found in $\eta' N$ production~\cite{Huang2013,Huang:2014ynz}.
 More precise data for the pion-induced reaction $\pi^-p\to\eta' n$ are called for to confirm the existence of these states.

In $\phi$ photoproduction, 
a broad but pronounced structure for $E_\gamma=1.8$--2.3~GeV was found~\cite{Adhikari:2013ija}. This structure could come from a resonant state. In that case,
it should also be visible in pion-induced $\phi$ production, for which only very few data exist (see Ref.~\cite{Doring:2008sv} for a data compilation).
More precise data from pion beams would help clarify the situation.

Information on excited baryons is also contained in high-energy experiments such as COMPASS. Invariant-mass spectra for the baryon resonance regions
are available with unprecedented precision from the corresponding kinematical regions of the final states.  The principal problem is that these reactions are not elementary; i.e., there are one or more additional high-energy particles in the final state.
The Deck and related effects induce additional contributions that have to be disentangled and additional modeling must be done to control
these effects (see, e.g., Ref.~\cite{Lebiedowicz:2013vya}).
Given the high statistics, the analysis of excited baryons in invariant-mass spectra of high-energy experiments is thus a very worthwhile
but challenging project.
Elementary pion-induced reactions as proposed here, for which the center-of-mass energy is in the resonance region, do not suffer from the aforementioned problems.

In summary, having high-quality data (including polarized measurements) with both pion and photon beams has the potential to advance greatly our knowledge of baryon resonances and such data could potentially establish a number of new states to fill some of the gaps of  ``missing resonances''.

\subsection{Form-Factor Measurements}

Inverse pion electroproduction (IPE), $\pi^-p\to e^+e^-n$, plays a 
special role in resonance physics. IPE is the only process that allows 
the determination of EM nucleon and pion form-factors in the intervals 
$0<k^2<4M^2$ and $0<k^2<4m_\pi^2$, 
which are kinematically unattainable from the $e^+e^-$ initial state.  Here $M$ and $m_\pi$ denote the
nucleon and pion mass, respectively. IPE 
measurements will significantly complement electroproduction $\gamma^\ast 
N\to\pi N$ studies for the evolution of baryon properties with increasing 
momentum transfer by investigation of the case for the time-like virtual 
photon.   

Difficulties in the experimental study of IPE arise from the need for a reliable 
rejection of competitive processes:

(i) The cross section of $\pi^-p\to\pi^-p$ is $d\sigma/d\Omega\sim$~10$^{-27}$~cm$^2$/sr 
   and is concentrated in the forward direction. Therefore, the $e^-$ and $e^+$ of 
   IPE can be conveniently detected at ${\sim}90^\circ$ with respect to the $\pi^-$ 
   beam, where the elastically scattered hadrons are strongly reduced.

(ii) The cross section for $\pi^+$ production, i.e., $\pi^-p\to\pi^-\pi^+n$ is 
     about 1000 times greater than that of IPE. The corresponding pions at 
     $90^\circ$ are very soft and can be suppressed strongly by threshold 
     Cherenkov counters.

(iii) Reactions with a gamma ray converted into a Dalitz pair contribute 
    a rather unpleasant background.  The most important processes are 
    $\pi^-p\to\pi^0n$ and $\pi^-p\to\gamma n$, which contribute $\sim$60\%
    and 40\% of the counting rate due to capture in hydrogen of $\pi^-$ at 
    rest against 0.7\% from IPE.

The presence of an electron beam at the proposed EIC facility permits a unique opportunity to measure the 
pion EM form factor directly using an electron-pion collider. This is useful 
because the pion form factor serves as a paradigm for nonperturbative hadronic 
structure and is associated with chiral dynamics, gauge invariance, and 
perturbative QCD in a nontrivial fashion. Current methods to extract the form 
factor rely on extrapolation to the pion $t$-channel pole. 
Unfortunately, this procedure is not without ambiguity and remains 
controversial to this day.

The form factor is especially relevant in light of a more recent controversy 
concerning its approach to the expected perturbative QCD behavior. This issue 
is of fundamental importance since it questions the existence of perturbative 
QCD for exclusive processes~\cite{as}. Unfortunately, the experimental 
situation is confused, with the BaBar and Belle Collaborations obtaining 
results for $Q^2 |F_\pi(Q^2)|$ that appear to be in conflict~\cite{fpi_BaBar,fpi_Belle}.

\section{Spectroscopy of Hyperon Resonances}

Our current experimental knowledge of $\Lambda^\ast$ and $\Sigma^\ast$ 
resonances is far worse than our knowledge of $N^\ast$ and $\Delta^\ast$ 
resonances; however, within the quark model, they are no less fundamental.  
Clearly, there is a need to learn about baryon resonances in the ``strange 
sector'' to have a complete understanding of three-quark 
bound states.

The properties of multi-strange baryons ($\Xi^\ast$ and $\Omega^\ast$ 
states) are also poorly known. Only the ground states belonging to the 
SU(3) octet and decuplet are well identified as four-star states in the 
RPP~\cite{pdg}, whereas a few dozens of excited states are predicted based on 
quark-model calculations~\cite{isgur78,isgur79a,maltman80,chao81,capstick86}.

Flavor adds a lever arm to study strongly interacting QCD. For example, one 
can study the quark mass-dependent portion of the effective quark interaction. 
Specifically, the current understanding of the spin-orbit interaction is 
unclear, and even contradictory (between mesons and baryons).

Unlike in the cases described above, kaon beams are crucial to provide the 
data needed to identify and characterize the properties of hyperon 
resonances. The masses and widths of the lowest $\Lambda$ and $\Sigma$ baryons 
were determined mainly with kaon-beam experiments in the 1970s~\cite{pdg}.
First determinations of pole positions, for instance 
for $\Lambda(1520)$, were obtained only recently \cite{qiang2010}.
An intense kaon beam would open a window of opportunity not only to locate 
missing resonances but also to establish properties including decay channels 
systematically for higher excited states.

Hyperons can be produced directly and exclusively in both the $\overline{K}N$ 
formation process and in inelastic $KN$ reactions. Some states that couple 
strongly to the $\overline{K}N$ channel have been studied in formation 
experiments. (See Ref.~\cite{zhang13} for a recent overview.) Together with missing- and invariant-mass reconstruction 
techniques, production cross sections give precise information of properties, 
for example decay widths, in an elementary process of production. In addition, 
missing and/or unknown resonances can be searched for using the missing-mass 
technique in $KN$ reactions. By introducing strangeness with kaon beams, 
hyperon production in kaon-induced reactions is not OZI suppressed and has a 
significant cross section even for excited states. High-statistics 
measurements are essential to disentangle observables because the spectrum is dense and contains
broad states.

Low-momentum kaon beams provide an opportunity to search for strange exotic 
states. The lineshape of $\Lambda$(1405)$1/2^-$ can be studied in $K^-p$ 
and $K^-d$ reactions. 
The $H$-dibaryon, which has a quark configuration of $uuddss$, will be searched 
for in the ($K^-$,~$K^+$) reaction \cite{Sako2014}.  
The measured $\pi\Sigma/\pi\pi\Sigma$ branching ratio for the $\Sigma(1670)$ 
produced in the reaction $K^- p \to \pi^- \Sigma(1670)^+$ depends strongly on 
momentum transfer, and it has been suggested that there exist two $\Sigma(1670)$ 
resonances with the same mass and quantum numbers, one with a large $\pi\pi\Sigma$ 
branching fraction and the other with a large $\pi\Sigma$ branching 
fraction~\cite{pdg}.  This $\Sigma$(1670) puzzle could be solved 
using future production experiments with kaon beams. The spectroscopy of $\Xi$ 
and $\Omega$ baryons can be investigated with high-momentum kaon beams. The 
lowest excited states of cascade baryons are thought to be analogous to the 
$\Lambda$(1405)$1/2^-$.

\subsection{Status of Data and Analyses for Specific Reactions}

Neutral hyperons $\Lambda^\ast$ and $\Sigma^\ast$ have been systematically 
studied in the following formation processes by several groups~\cite{gopal77,zhang13,zhang13a,zhong13,Kamano:2014zba,gao2012,liu2012,shi2015}:

\begin{tabular}{ll}
$K^-p\to K^- p$,        ~~~~~~~~~~   & $K^-p\to\pi^+\Sigma^-$, \\
$K^-p\to \overline{K^0}n$, & $K^-p\to\pi^0\Sigma^0$, \\
$K^-p\to \pi^0 \Lambda$,   & $K^-p\to\pi^-\Sigma^+$. \\
\end{tabular}

In recent analyses of the reaction
$K^- p \to \pi^0 \Lambda$, fits by different groups tend to agree for the
differential cross section and polarization, but they largely differ
for the spin rotation parameter $\beta$~\cite{anisovich12b,zhang13}.  Data 
for $\beta$ are limited to measurements at only seven energies~\cite{bell83}.  
This observable is more sensitive
to contributions from high partial waves than the differential cross section 
or polarization.
These facts show the need for new measurements with a polarized target.

In addition, $\Sigma^{*-}$ can be produced in $K^-n$ reactions with a 
deuteron target:

\begin{tabular}{l}
$K^-n\to\pi^-\Lambda$, \\
$K^-n\to\pi^0\Sigma^-$, \\
$K^-n\to\pi^-\Sigma^0$. \\
\end{tabular}

The PWA method is powerful for disentangling overlapping states with 
large widths, especially above 1.6 (1.7)~GeV/$c^2$ for $\Lambda^\ast$ 
($\Sigma^\ast$) resonances.

Note that $\Lambda$(1405)$1/2^-$ and $\Sigma$(1385)$3/2^+$ lie below 
the $\overline{K}N$ threshold; therefore, properties of these states 
can be obtained only through production processes such as

\begin{tabular}{l}
$K^-p\to\pi^-\Sigma^{*+}\to\pi^-\pi^+\Lambda^\ast$. \\
\end{tabular}

A $t$-channel process of the reaction provides a ``virtual'' $\overline{K^0}$ 
beam, which enables us to produce $\Sigma^{*+}$. The states  $\Lambda$(1405)$1/2^-$ 
and $\Sigma$(1385)$3/2^+$ were identified in decay channels of $\Sigma\pi$ and 
$\Lambda\pi$/$\Sigma \pi$, respectively.


Recently, the spin and parity of $\Lambda(1405)$ were measured
in the $\gamma  p \to K^+ \Lambda(1405)$ reaction and
confirmed to be $1/2^-$ as expected theoretically \cite{Moriya2014}.
However, the nature of $\Lambda(1405)1/2^{-}$ is still an issue.
It has been pointed out that the lineshape of $\Lambda(1405)$ depends on reaction channels
\cite{Borasoy:2005ie,Kaiser:1995eg,Jido:2003cb}.
Therefore, a comparison between pion- and kaon-induced reactions together
with photoproduction is important. In addition, the lineshape of $\Lambda(1405)$ differs
in $\pi^{+}\Sigma^{-}$ and $\pi^{-}\Sigma^{+}$ decay channels as a result of the
isospin interference between different $\pi\Sigma$ channels. The $\pi^{\pm}\Sigma^{\mp}$
spectra were measured in $\pi^{-} p \to K^{0}\Lambda(1405)$ \cite{thomas73} and
$K^{-} p \rightarrow \pi^{+}\pi^{-}\Lambda(1405)$ \cite{hemingway85} reactions. The neutral
$\pi^{0}\Sigma^{0}$ channel was measured in the $K^{-} p \to \pi^{0}\pi^{0}\Sigma^{0}$
reaction \cite{prakhov04}. The observed peak position is located near 1405~MeV in charged $\pi\Sigma$
spectra, whereas the peak in the neutral channel is closer to 1420~MeV.
All three $\pi\Sigma$ channels were studied recently in a single experiment
using the $\gamma p \rightarrow K^{+}\Sigma\pi$ reaction \cite{Moriya2013}. Efforts in the
lineshape study are ongoing at facilities such as JLab and J-PARC.


One of the reactions of interest is formation of $\Lambda$(1670)1/2$^-$ in the 
$K^-p\to\eta\Lambda$ reaction, which is closely related by SU(3) symmetry 
to $N$(1535)1/2$^-$ formation in the $\pi^-p\to\eta n$ reaction. The branching 
fraction of the near-threshold S-wave $\Lambda$(1670)1/2$^-$ into 
$\eta\Lambda$ is much larger than that of other $\Lambda^\ast$ states, 
just as the branching fraction of $N$(1535)1/2$^-$ into $\eta N$ is much larger 
than that of other $N^\ast$ states. On the other hand, the branching 
ratio of $\Sigma(1620)$1/2$^-$ into $\eta\Sigma$ is not known, 
whereas that of $\Sigma(1750)$1/2$^-$ is significant. One possible 
explanation is the $ud$ diquark correlation, which has isospin zero in 
the negative-parity $\Lambda$ while it is one for the $\Sigma$. The $K^-p 
\to\eta\Lambda$ and $K^-p\to\eta\Sigma^0$ reactions are analog reactions 
to $\pi^-p\to\eta n$ and in all cases the $\eta$ meson serves as an 
``isospin filter'' that requires any intermediate resonances to have the 
same isospin as the final-state baryon.

Cascade baryons could be intensively studied with high-momentum kaon 
beams and modern multiparticle spectrometers.  Most of our knowledge 
about multi-strange baryons was obtained from old data measured with 
bubble chambers.  The lack of appropriate beams and detectors in the 
past greatly limited our knowledge. Currently only the cascade ground 
states of spin-1/2 and spin-3/2 are well identified. For excited states, 
possible production reactions with kaon beams are the following:

\begin{tabular}{l}
$K^-p\to K^+\Xi^{*-}$, \\
$K^-p\to K^{*+}\Xi^{*-}$, \\
$K^-p\to K^{*0}\Xi^{*0}$. \\
\end{tabular}

Model-independent guidance for analyzing these processes and some theoretical results can by found in Refs.~\cite{jackson2014,Jackson:2015dva}.

There are other production processes with single or multi pions. For 
example:

\begin{tabular}{l}
$K^-p\to K^+\pi^+\pi^-\Xi^{*-}$, \\
$K^-p\to K^+\pi^-\Xi^{*0}$. \\
\end{tabular}

By tagging pions and/or using specific reactions accompanying $K^\ast$, 
mass measurements of the cascade baryons can be carried out with the 
missing-mass technique. Of course, it is desirable to use a detector 
to identify decay products. An analysis of the decay vertex is thought 
to be very efficient for suppressing background processes. Masses, widths, and decay modes 
will be studied at J-PARC.  These measurements will be 
complementary to planned measurements at CLAS12 to study $\Xi^\ast$ 
states via several possible reactions such as $\gamma p \to K^+ K^+ 
(\Xi^{*-})$ and $\gamma p \to K^+ K^+ \pi^- (\Xi^{*0})$ \cite{clas12}.

High-momentum kaon beams are also crucial for producing $\Omega$ 
baryons. For example, they could be studied in the inclusive reaction

\begin{tabular}{l}
$K^-p\to\Omega^{*-}X$
\end{tabular}

by measuring decay particles. The $\Omega^-$ 
production mechanism should be quite specific since it is the first 
baryon with  constituents of which none could come from the target 
proton. These measurements will be complementary to planned measurements 
at CLAS12 to measure $\Omega$ photoproduction on the proton 
target~\cite{clas12}. Specific plans are to make the first precise 
measurement of the $\Omega^-$ differential cross section in $\gamma 
p\to\Omega^-K^+K^+K^0$ and to search for $\Omega^\ast$ resonances.

\section{Meson Spectroscopy}

Although it was light hadron spectroscopy that led the way to the 
discovery of color degrees of freedom and Quantum Chromodynamics, much 
of the field remains poorly understood, both theoretically and 
experimentally \cite{pennington13}.  The availability of pion and kaon 
beams provide an 
important opportunity to improve this situation.  Experimentally, meson spectroscopy can be investigated by using PWAs to determine quantum numbers from the angular distributions of final-state particle distributions.  Such methods will be used to analyze data from future measurements at CLAS12.  Pion beams with c.m.\ energies up to 5~GeV should be adequate for a complementary program.  Such energies correspond to beam momenta of about 13~GeV/$c$.
We note that meson beam experiments may not be ideal for the study of meson resonances; however, this approach has been taken at BNL and COMPASS and a closely related one is being pursued by the GlueX collaboration. We therefore briefly review some of the open issues in light meson spectroscopy in this section.

The chief areas of interest in spectroscopy are light scalar mesons and 
multiquark states, glueballs, and hybrids. The last three represent new 
forms of matter over the familiar quark-antiquark and three-quark states 
that comprise the majority of mesons and baryons. It has long been 
guessed that such states can exist, but whether the dynamics of QCD 
permits them, or whether they are observable, remains an open and 
contentious issue today. Because mesons produced by $t$-channel exchange 
from nuclear targets access all mesonic quantum numbers these exotic 
states are open to experimental investigation.  Experimental effort with meson beams will complement the GlueX experiment at JLab, which seeks to explore the properties of  hybrids with a photon beam.

\subsection{Multiquarks}

Speculation about multiquark states started more that 40 years ago with a 
claim that a dynamical scalar isoscalar resonance in $\pi\pi$ scattering 
is predicted by current algebra, unitarity, and crossing symmetry~\cite{brown}. 
A related idea was proposed by Jaffe, who noted that $qq\bar q\bar q$ 
states could make up a scalar nonet [$\sigma$, $\kappa$, $f_0(980)$, 
$a_0(980)$]~\cite{jaffe}. This hypothesis has been a rich source of ideas 
and controversy ever since. Only recently, with the work of Ref.~\cite{sigma}, 
has it been uniformly accepted that a $\sigma$ resonance even exists. The 
interpretation of these states, and the existence of the strange analog 
state, $\kappa$, remain open issues \cite{pennington07}. In the intervening 
decades, the idea 
of multiquark states has been applied to a host of additional states -- $a_0$ 
and $f_0$ ($K\overline{K}$)~\cite{KK1,KK2}, $f_1(1420)$ 
($K^\ast\overline{K}$)~\cite{f1}, $f_2(2010)$ ($\phi\phi$)~\cite{f2}, 
$f_0(1770)$ ($K^\ast\overline{K^\ast}$)~\cite{f0}, 
$\psi(4040)$ ($D^\ast\overline{D^\ast}$)~\cite{4040}, $X(3872)$ 
($D\overline{D^\ast}$)~\cite{ess} -- with many similar guesses for the current 
crop of ``X, Y, Z'' states~\cite{ess2}.

\subsection{Glueballs}

Conjectures on the possible existence and properties of glueballs (states 
comprised primarily of nonperturbative gluons) date from the beginnings 
of QCD~\cite{gb-old1,gb-old2,gb-old3}. Early speculation has given way to specific 
calculations in lattice gauge theory, which indicate a rich spectrum of 
states~\cite{chen}. Unfortunately, these calculations are beset by 
statistical noise and the spectrum (see Table~\ref{tab:gb}) is only known 
in the ``quenched'' approximation where the effects of mixing with quarks 
is neglected \cite{boglione97}. The phenomenology associated with this 
mixing remains murky, 
and progress must rely on a dramatically improved experimental situation.

\begin{table}[ht]\centering
\begin{tabular}{ll}
\hline\hline
$J^{PC}$ & mass (MeV) \\
\hline
$0^{++}$ & 1710 (50)(80)\\
$2^{++}$ & 2390 (30)(120)\\
$0^{-+}$ & 2560 (35)(120)\\
$1^{+-}$ & 2989 (30)(140)\\
\hline\hline
\end{tabular}
\centerline{\parbox{0.80\textwidth}{
 \caption[] {\protect\small (Color on-line) Lattice glueball spectrum below 
	3~GeV~\protect\cite{chen}. Errors are the continuum 
	extrapolation plus anisotropy errors and the scale error.}
	\label{tab:gb} }}
\end{table}

It is thus unfortunate that no glueballs have been definitively identified. 
A promising earlier candidate called the $\xi(2200)$ has not withstood 
careful analysis. At  present, the best candidate is the $f_0(1500)$ [or 
possibly the $f_0(1710)$], which appears as a supernumerary state in the 
enigmatic scalar meson sector~\cite{ac}. Further information on the 
experimental and theoretical status of glueballs can be found in 
Refs.~\cite{gb-reviews1,gb-reviews2,gb-reviews3}.

\subsection{Hybrids}

Hybrid mesons are postulated to be bound states that contain quarks and gluons. 
As with multiquarks and glueballs, speculation on the existence and properties 
of these states date to the start of QCD~\cite{hybrid-old1,hybrid-old2,hybrid-old3}. In a fashion 
reminiscent of other exotic states, it is not even known what form the gluonic 
degrees of freedom take. They can, for example, manifest as effective 
constituent gluons, ``flux tube'' degrees of freedom, or something else.

An important feature of hybrid mesons is that the extra degree of freedom 
provided by the valence glue permits quantum numbers that are not accessible 
to $q\bar q$ states. In particular, the parity and charge conjugation quantum 
numbers for fermion-antifermion systems of spin $S$ and angular momentum $L$ 
are $(-)^{L+1}$ and $(-)^{L+S}$, respectively. This implies that the quantum 
numbers
\begin{equation}
	0^{--},\qquad ({\rm odd})^{-+}, \qquad {\rm and}\  \qquad ({\rm even})^{+-}\nonumber
\end{equation}
are exotic. In particular, the discovery of such a ``quantum number-exotic'' 
state implies that it is a multiquark, glueball, or hybrid.

\begin{figure}[htpb]
\begin{center}
	\includegraphics[angle=0, width=0.8\textwidth]{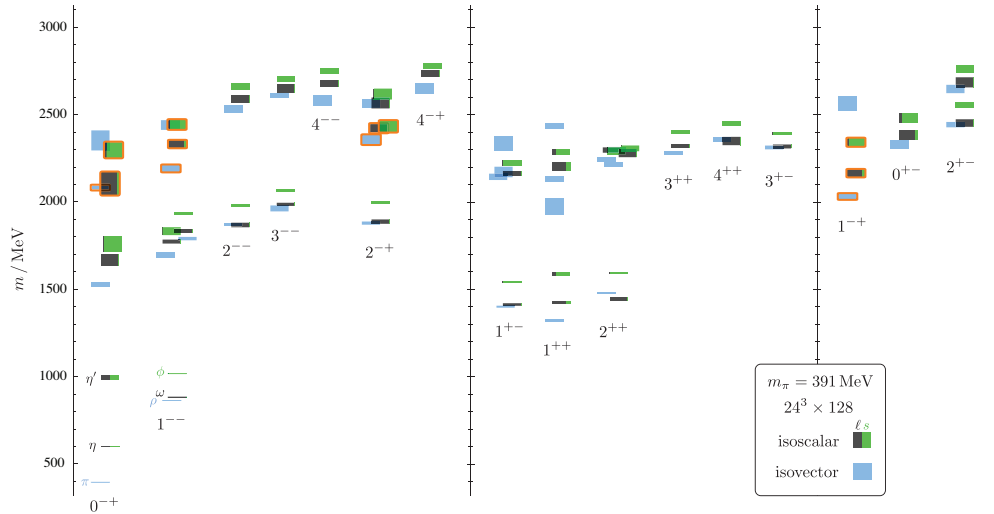}
\end{center}
\centerline{\parbox{0.80\textwidth}{
\caption[] {\protect\small (Color on-line) Lattice computation of the light meson 
	spectrum~\protect\cite{dudek1}. Boxes outlined in red indicate 
	states with significant gluonic content. The three columns to 
	the right are quantum number-exotic and gluon-rich. This plot is 
	reproduced with permission of the authors.} \label{fig:spectrum} } }
\end{figure}

Lattice gauge theory has contributed substantially to the understanding of 
hybrids in the past 15 years. Early efforts obtained the spectrum of the 
gluonic degrees of freedom in the presence of static quarks~\cite{jkm}, 
which assists in modelling heavy hybrids (such as $b\bar b g$ states). 
More recently, improved techniques permit realistic unquenched computations 
of the light meson spectrum. Since lattice gauge theory automatically 
incorporates all gluodynamics, some of these states are hybrid in nature. 
A summary of the results from Ref.~\cite{dudek1} is contained in 
Fig.~\ref{fig:spectrum}. Important points to note are that the pion mass 
is 391~MeV, which indicates that quark masses are too high, that mixing in 
the isoscalar states can be measured and is in agreement with phenomenology, 
and that a large spectrum of hybrid states (indicated in red) exists. The 
extrapolation to physical quark masses remains difficult, however recent 
quenched and unquenched lattice data point to a exotic $1^{-+}$ hybrid mass 
of approximately 1650~MeV~\cite{meyer}.

Unlike glueballs, there is extensive experimental evidence for light hybrid 
mesons. Unfortunately, as we shall see, this evidence is  controversial and 
the status of hybrids as a viable manifestation of QCD dynamics remains 
open. The most studied resonance is called the $\pi_1(1400)$ -- the notation 
implies that the state is an isovector $1^{-+}$ resonance. As can be seen in 
Table~\ref{tab:pi1400}, it has been sighted by several experiments, albeit 
with widely varying mass and width~\cite{meyer}. A generic problem in all 
the experimental analyses is that poorly understood angular acceptance can 
lead to feed down from well-known resonances that populates exotic channels. 
In addition, meson-meson interactions and coupled channel effects can mimic 
resonance behavior in exotic channels such as $\pi^0\eta$ in P-wave. This 
possibility was examined in Ref.~\cite{sz}, where it was claimed that the 
$\pi_1(1400)$ signal vanishes under reasonable modelling assumptions.

\begin{table}[ht]\centering
\begin{tabular}{crrc} \hline\hline
Mode & Mass (GeV) & Width (GeV) & Experiment \\ 
\hline
$\eta\pi^{-}$ & $1.405\pm 0.020 $ & $0.18\pm 0.02$     & GAMS  \\
$\eta\pi^{-}$ & $1.343\pm 0.0046$ & $0.1432\pm 0.0125$ & KEK  \\
$\eta\pi^{-}$ & $1.37\pm 0.016$ & $0.385\pm 0.040$     & E852  \\
$\eta\pi^{0}$ & $1.257\pm0.020$ &$0.354\pm 0.064$      & E852  \\
$\eta\pi$     & $1.40\pm 0.020$ & $0.310\pm 0.050$     & CBAR  \\
$\eta\pi^{0}$ & $1.36\pm 0.025$ & $0.220\pm 0.090$     & CBAR  \\
$\rho\pi$     & $1.384\pm 0.028$& $0.378\pm 0.058$     & Obelix \\
$\rho\pi$     & $\sim 1.4$      & $\sim 0.4$           & CBAR \\
\hline
$\eta\pi$     & $1.351\pm 0.030$& $0.313\pm 0.040$     & RPP \\
\hline\hline
\end{tabular}
\centerline{\parbox{0.80\textwidth}{
\caption[]{\protect\small Reported masses and widths of the $\pi_{1}(1400)$. 
	Table data from Ref.~\protect\cite{meyer}.} \label{tab:pi1400} }}
\end{table}

A similar story has played out with the $1^{-+}$ $\pi_1(1600)$ state. 
Although, this resonance lies closer to model and lattice expectations, and 
has been seen by the VES, E852, and COMPASS Collaborations~\cite{meyer}, an 
analysis with a large dataset has not found evidence for the state~\cite{dz}. 
Countering this is a recent result from COMPASS that claims a $1^{-+}$ 
exotic state decaying into $\pi\rho$ at 1660~MeV with a width of 
269~MeV~\cite{compass-exotic}.

Finally, a heavier exotic candidate, the $\pi_1(2015)$, decaying to 
$f_1\pi$ and $b_1\pi$, has been reported by E852~\cite{meyer}.

Despite this murky experimental situation, confidence in the existence of 
hybrid mesons remains high. The coupling of these states to their production 
and decay channels is clearly important to their eventual discovery. Although 
nascent lattice computations on hybrid decay have been made, this important 
phenomenology currently relies on models. The predictions of two models are 
presented in Table~\ref{tab:decays}. One sees typical hadronic widths for the 
majority of the states. An important feature of many of the decay modes is 
that they proceed via a combination of S-wave and P-wave states. This implies 
that multiple pions typically appear in the final state, which in turn means 
that hermetic detectors are important.

\begin{table}[ht]\centering
\begin{tabular}{ccccc}\hline\hline
  &  &  &  & \\ 
Name & $\mathbf{J^{PC}}$ & \multicolumn{2}{c}{Total Width (MeV)} & Large Decay Modes\\ 
         &                   & PSS & IKP & \\ \hline
$\pi_{1} $   & $1^{-+}$ &  $81-168$ & $117$ & 
$b_{1}\pi$, $\rho\pi$, $f_{1}\pi$, $a_{1}\eta$, \\
         & & & & $\eta(1295)\pi$, $K_{1}^{A}K$, $K_{1}^{B}K$ \\
$\eta_{1}$  & $1^{-+}$ &  $59-158$ & $107$ &
$a_{1}\pi$, $f_{1}\eta$, $\pi(1300)\pi$, \\
         & & & &  $K_{1}^{A}K$, $K_{1}^{B}K$ \\
$\eta^{\prime}_{1}$ & $1^{-+}$ &  $95-216$ & $172$ &
$K_{1}^{B}K$, $K_{1}^{A}K$, $K^\ast K$ \\ \hline
$b_{0}$  & $0^{+-}$ & $247-429$ & $665$ &
$\pi(1300)\pi$, $h_{1}\pi$ \\
$h_{0}$     & $0^{+-}$ & $59-262$  & $94$  &
$b_{1}\pi$, $h_{1}\eta$, $K(1460)K$ \\
$h^{\prime}_{0}$    & $0^{+-}$ & $259-490$ & $426$ &
$K(1460)K$, $K_{1}^{A}K$, $h_{1}\eta$ \\ \hline
$b_{2}$     & $2^{+-}$ &    $5-11$ & $248$ &
$a_{2}\pi$, $a_{1}\pi$, $h_{1}\pi$ \\
$h_{2}$     & $2^{+-}$ &    $4-12$ & $166$ &
$b_{1}\pi$, $\rho\pi$ \\
$h^{\prime}_{2}$    & $2^{+-}$ &    $5-18$ &  $79$ &
$K_{1}^{B}K$, $K_{1}^{A}K$, $K^\ast_{2}K$, $h_{1}\eta$ \\ 
\hline\hline 
\end{tabular}
\centerline{\parbox{0.80\textwidth}{
\caption[]{\protect\small Exotic quantum number hybrid widths. The 
	column labelled PSS lists predictions from the vector flux 
	tube decay model of Ref.~\protect\cite{pss}, while IKP 
	denotes predictions from the Isgur-Kokoski-Paton model as 
	computed in~\protect\cite{pss}.  $K_{1}^{A}$ represents the 
	$K_{1}(1270)$ while $K_{1}^{B}$ represents the $K_{1}(1400)$.} 
	\label{tab:decays} }}
\end{table}

\subsection{Physics Opportunities}

In spite of previous work at hadron facilities such as Brookhaven National 
Laboratory and IHEP at Protvino, and current work at COMPASS at CERN and 
J-PARC at KEK, much remains to be done in light-meson spectroscopy. Indeed, 
the previous discussion illustrates that substantial effort is required 
before even a partial understanding of multiquarks, hybrids, or glueballs 
can be claimed.

\begin{table}[ht]\centering
\begin{tabular}{l|ll}
\hline\hline
$E_\pi$ (GeV) & meson  & comment \\
\hline
3  &  $\sigma(500) $ & chiral multiquark state? \\  
   &  $\kappa(800)$  & chiral multiquark strange partner \\
   &  $a_0/f_0(980)$ & tetraquark/ $K\overline{K}$ molecule \\
   &  $\pi(1300)$    & three-body resonance?  \\
   &  $f_0(1370)$    & superfluous state? \\
   &  $\eta(1405)$   & superfluous state? \\
   &  $\pi_1(1400)$  & possible hybrid state \\
   &  $f_0(1500)$    & glueball candidate \\
   &  $\pi_1(1600)$  & hybrid candidate \\
\hline
4  & $\omega_2(1650)$& missing state \\             
   & $\rho_2(1650)$  & missing state \\
   & $\pi_1(1650)$   & predicted $1^{-+}$ hybrid \\
   & $f_0(1720)$     & predicted $0^{++}$ glueball? \\ 
   & $\pi(1800)$     & radial pion/ superfluous? \\
   & $\phi(1800)$    & missing state \\
   & $\pi_1(1850)$   & (alt) predicted $1^{-+}$ hybrid \\
   & $K_2^\ast (1850)$& missing state \\
   &  $\pi_1(1900)$  & possible hybrid candidate \\
   & $\phi_2(1900)$  & missing state \\
\hline
6  & $\rho_3(2200)$  & missing state \\
   & $f_2(2300)/ f_2(2340)$ & superfluous states? \\
   & $\rho_5(2350)$  & missing state \\
   & $K_5^\ast (2380)$& missing state \\
   & $f_0(2400)$     & predicted radial $0^{++}$ glueball \\
   & $f_2(2400)$     & predicted $2^{++}$ glueball \\
   & $a_6(2450)$     & missing state \\
\hline\hline
\end{tabular}
\centerline{\parbox{0.80\textwidth}{
\caption[]{\protect\small Pion beamline energy required to create various 
	meson states of interest.} \label{tab:states} }}
\end{table}

The first step in establishing such an understanding must be filling out
the regular $q\bar q$ meson spectrum. It is highly deplorable that
dozens of low-lying states still remain undiscovered~\cite{godfrey2}.  Some of
these states, and a collection of other states of interest are listed in
Table~\ref{tab:states}. 
This table lists states according to the nominal energy required to produce them. Higher beam energies may be required to permit effective kinematic separation of mesonic and baryonic resonances.
Of course, this is only a representative sample.
For example, if a $\pi_1$ hybrid meson is discovered, one expects light
and $s\bar s$ isoscalar mesons 100 and 300~MeV higher in mass. There
should also be nearby $0^{-+}$, $1^{--}$, $0^{+-}$, and $2^{+-}$
multiplets. Thus an entire spectrum of novel states awaits discovery.
Similarly, glueball states must be embedded in the spectrum somewhere,
and if quenched lattice gauge theory computations are a reliable guide
to the physical spectrum, then several glueballs should be within reach
of a 6~GeV beamline.

Of course, hunting for resonances is only the first (very important) step.
Measuring the production and decay properties of a resonance, along with
its EM couplings, provides vital information on the strongly interacting 
regime of QCD. It is only through efforts like this that progress toward 
understanding a crucial part of the Standard Model can be achieved.

%
%
%
%
%
%
%

\section{Chiral Dynamics}
\subsection{Chiral Perturbation Theory and Low-Energy Pion-Nucleon Dynamics}
\label{sec:ChPT}
Chiral perturbation theory (ChPT) allows for the model-independent extraction 
of the amplitude close to the $\pi N$ threshold. Through the separation of 
energy scales the long-range dynamics can be taken explicitly into account 
while short-range dynamics is absorbed in counter-terms parametrized with 
unknown low-energy constants  (LECs)~\cite{Bernard:1995dp,Bernard:1996gq}, 
which allow for an order-by-order renormalization of the chiral expansion. 
At a given chiral order, the LECs parametrize the different structures of 
the interaction Lagrangian that are compatible with chiral 
symmetry~\cite{Fettes:2000gb}.  As these constants are well defined within 
a given renormalization scheme, they can be used in reactions other than 
elastic $\pi N$ scattering, e.g., $\pi N\to\pi\pi 
N$~\cite{Fettes:1999wp,Siemens:2014pma}.  This allows for model-independent 
predictions that have contributed to the success of ChPT.

The precise knowledge of the LECs lies at the heart of ChPT. Their values 
are determined through fits to elastic $\pi N$ partial-wave amplitudes, as 
pioneered by Fettes and Mei{\ss}ner~\cite{Fettes:1998ud} up to fourth
order~\cite{Fettes:2000xg}, including the $\Delta$-isobar explicit degree of
freedom~\cite{Fettes:2000bb} and isospin breaking up to third
order~\cite{Fettes:2001cr}. See also the
fundamental analysis of Becher and Leutwyler \cite{Becher:2001hv} for
the analysis of low-energy pion scattering.
It should be noted that the considered isospin
breaking effects lead to genuinely new interactions beyond mass differences 
and Coulomb corrections that are incorporated in many phenomenological 
analyses. In any case, the extraction of the LECs is only as good as the 
determination of partial-wave amplitudes, which in turn depend on the 
quality of the data. Although the GWU/SAID analysis includes the world 
database and applies Coulomb corrections, the experimental situation is 
still not satisfactory. 
New precision measurements 
close to the $\pi N$ threshold that cover a wide angular range and allow for 
an energy scan are needed. 
%


At the precision frontier of low-energy chiral dynamics, it is necessary to have
consistent data.  Single-energy solutions usually represent the 1-$\sigma$
confidence interval for a given partial wave~\cite{arndt06}.  
Uncertainties in LECs are often determined from fits to these partial waves.  However,
for a statistically meaningful error propagation from data to LECs, the
correlation  between different partial waves must be known to allow for
correlated $\chi^2$ fits. The determination of the corresponding covariance
matrices requires high-precision data. One of the problems is the 
difficult-to-control systematic uncertainty that may even vary within the 
same experiment as
seen previously.  Additionally, the database consists of data sets from many
different experiments, several of them with individual normalization issues.
Such unknowns induce systematic problems in the statistical analysis. A new
measurement with maximal angular coverage and energy resolution will remove the
systematic bias and allow for a statistically sound determination of
uncertainties and correlations of LECs, thus advancing the low-energy chiral
dynamics and better quantifying the uncertainty in ChPT predictions. Work in this direction is already being carried out: LECs up to fourth order have been determined directly from data recently~\cite{Wendt:2014lja}.

Low-energy pion-nucleon scattering data are also critical input for precision 
studies of dispersive
 pion-nucleon dynamics. A fundamental quantity is the
pion-nucleon $\sigma$ term that measures the  explicit chiral symmetry breaking
in the nucleon mass and that requires the evaluation  of the amplitude in the
unphysical region. Extrapolations of this kind are particularly sensitive to
the available scattering data, usually in combination with measurements of 
pionic deuterium and its analysis~\cite{Gotta,Baru:2011bw,Baru:2010xn}.
Imposing crossing symmetry, unitarity, and analyticity via Roy-Steiner
equations, better determinations of $\pi N$ amplitudes and the $\pi N$
$\sigma$-term are of high theoretical interest and currently being 
developed~\cite{Ditsche:2012fv,Elvira:2014wma}.


As for the reaction $\pi N\to\pi\pi N$, the database is discussed in 
Sec.~\ref{sec:reactions}. This reaction has been studied in ChPT for many
years, see, e.g., Refs.~\cite{Fettes:1999wp,Siemens:2014pma}. Compared to
elastic pion-nucleon scattering that depends on one kinematic variable at a
given energy (scattering angle), the three-body final state depends on four 
(we neglect here initial polarizations). They can be parametrized 
in terms of one invariant mass, one scattering angle with the spectator, and two
decay angles of the particles forming the invariant mass. This implies that  
 experimental information with high 
statistics and angular coverage is required for determination of the amplitude. Indeed, as shown in 
Ref.~\cite{Siemens:2014pma}, there is substantial information on total 
cross sections to which the Heavy Baryon ChPT  and manifestly covariant 
ChPT can be compared, but little information on double and higher 
differential cross sections. The amplitude is, of course, particularly 
sensitive to the latter. Improved data, preferably in the form of events, 
is necessary to determine the LECs, in particular, those that are tied to 
this reaction such as $g_{\Delta\Delta\pi}$. Indeed,  the LECs from elastic 
$\pi N$ scattering are used in Ref.~\cite{Siemens:2014pma} to predict the 
$\pi N\to\pi\pi N$ data. More precise experimental information is crucial to 
constrain parameters further and to allow for combined analyses of elastic 
$\pi N$ and $\pi\pi N$ production in the future. 


\subsection{Unitarized Chiral Perturbation Theory}
\label{sec:UChPT}

Beyond the perturbative regime, the convergence of ChPT becomes slow until 
the occurrence of  resonance poles in the complex 
plane prohibits the perturbative expansion at all.  The $\Delta$ resonance 
has been explicitly included in the perturbative expansion~\cite{Fettes:1999wp,
Siemens:2014pma}, including  the nucleon-$\Delta$ mass splitting as an
additional small scale. For heavier resonances this is no longer possible.
However, resummed schemes can be constructed that can be matched to ChPT
order-by-order. Exact unitarity can be implemented but not crossing symmetry. Resonance poles appear in Unitarized Chiral Perturbation
Theory (UChPT) through the summation of interaction kernels to all orders in 
a Bethe-Salpeter equation.  The full four-dimensional Bethe-Salpeter
equation with interactions at the one-loop level is not yet tractable.
However,  solutions using the NLO contact terms and the gauge-invariant photon
interaction were derived some time ago~\cite{Borasoy:2005ie}. 

One of the main interests in studying the meson-baryon system in UChPT is the
predicted SU(3) flavor structure of resonances. At lowest order, the hadronic
and EM properties of dynamically generated resonances are full predictions, 
while higher orders in the chiral  expansion generate LECs that have to be fitted 
to data. Pioneering UChPT predictions for the structure of the $N(1535)1/2^-$ and 
the $\Lambda(1405)$1/2$^-$~\cite{Kaiser:1995eg,Kaiser:1995cy,Oset:1997it,Oller:2000ma,
Oller:2000fj,Lutz:2001yb} have generated an entire field of theoretical 
activity predicting and analyzing the baryon spectrum~\cite{Nieves:2001wt,
Inoue:2001ip,Meissner:1999vr,Lutz:2001mi,GarciaRecio:2003ks,Baru:2003qq,
Doring:2005bx,Doring:2006pt,Doring:2007rz,Doring:2009uc,Oset:2009vf,
Khemchandani:2011et,Khemchandani:2011mf,Doring:2010fw,Borasoy:2006sr,
Borasoy:2007ku,Bruns:2010sv,Ruic:2011wf,Mai:2012wy}. It is common to most 
UChPT approaches to formulate the amplitudes in coupled channels to study 
the SU(3) flavor structure of excited baryons.  In many cases resonances are 
generated in the vicinity of thresholds such that the state is bound with 
respect to heavier channels but resonant to lighter channels. Such a quasi-bound 
state is, for example, the $N(1535)1/2^-$ that is above the $\pi N$ and 
$\eta N$ threshold but below the $KY$ thresholds. A strong attraction in these 
heavier channels, predicted from the chiral Lagrangian, leads to the formation 
of the state. The chiral amplitude has also direct consequences for the OZI 
prohibited $\phi$ production~\cite{Doring:2008sv} and EM 
properties~\cite{Jido:2007sm}. The second S11 resonance, $N(1650)1/2^-$, has also been explained in terms of chiral dynamics~\cite{Nieves:1999bx,Bruns:2010sv,Garzon:2014ida}.

As the UChPT amplitude extends to energies above the heavier thresholds, it 
can be directly tested with data. It is thus of importance to  
measure the $\pi N\to K\Lambda$ and $\pi N\to K\Sigma$ transitions accurately.  More 
specifically, the UChPT prediction concerns predominantly the $S$-wave for 
these reactions. It is, in principle, possible to describe P-waves with 
unitarized chiral interactions~\cite{Mai:2012wy}. However, for P-waves, new 
chiral operators enter which increase the available degrees of freedom 
without further constraining the chiral dynamics of the S-wave. 

Thus, to test and refine UChPT calculations, the S-wave of the amplitude needs 
to be isolated, which requires PWA. The reactions $\pi N\to\eta N$, $\pi N\to 
K\Lambda$ and $\pi N\to K\Sigma$ have been analyzed by many groups
\cite{Ronchen:2012eg,Doring:2010ap,shrestha12,shrestha12a,Anisovich:2013vpa,
anisovich12a,anisovich12b, Cao:2013psa,shklyar13,shklyar05}  
but in several cases no consensus of 
the S-wave strength has been reached, because the data  are far from being 
sufficiently good for this purpose. 

To demonstrate the importance of data input for chiral unitary calculations, 
one has to mention the connection to QCD simulations on a lattice (cf.\ Sec.~\ref{sec:lattice}). These 
simulations employ unphysical quark masses and, thus, a chiral extrapolation 
to the physical world is required. Unitarized ChPT provides a reliable 
means to provide such an extrapolation in the second and third resonance region.

\subsection{Strangeness in UChPT}
\label{sec:strange_UChPT}
A very similar call for improved hadronic measurements can be made for the 
strangeness $S=-1$ sector. Here, the $\Lambda(1405)$1/2$^-$ appears as a state 
generated from the $\overline{K} N$ and $\pi\Sigma$ channels. The channel 
dynamics is more complex and has led to the prediction of a two-pole nature 
of the $\Lambda(1405)$1/2$^-$~\cite{Oller:2000fj, Jido:2003cb}. This hypothesis is currently 
debated and tested in various experiments. 

To demonstrate the dependence of the actual pole positions of the two 
$\Lambda(1405)$1/2$^-$ on data, we just quote the recent result of 
Ref.~\cite{Mai:2012dt} in which non-canonical pole positions have been obtained 
and yet the available total cross-section data have been described. 
In Ref.~\cite{Mai:2014xna} it has been shown that the quality of the hadronic data does not allow for a precise determination of the pole positions.
In Refs.~\cite{Roca:2013av,Roca:2013cca}
the impact of the new photoproduction data on the $\Lambda(1405)$ lineshape~\cite{Moriya2013} has been quantified. (See also 
Refs.~\cite{Ikeda:2012au,Doring:2011xc,Mai:2014uma} for an update on the kaon-deuteron 
scattering length.)

The UChPT approach is not only able to describe the bulk features of the 
kaon-induced $\overline{K}N$ and $\pi\Sigma$ data. Also, within the same 
approach the $\Lambda$(1670)1/2$^-$ is dynamically generated. That resonance 
appears as a quasi-bound $K\Xi$ state~\cite{Oset:2001cn,Doring:2010rd}. A 
measurement of the $K\Xi$ final state is called for, in particular, when 
considering the poor data situation~\cite{Sharov:2011xq}. 

In summary, in UChPT approaches, the $\overline{K}$-induced final states 
$\pi\Sigma$, $\overline{K}N$, $\eta \Lambda$, and $K\Xi$ are important in the sense that their S-wave contribution needs to be 
isolated to allow direct tests of the UChPT hypotheses. It is imperative to have improved measurements of these final states. 

\begin{figure}[t]
\begin{center} 
	\includegraphics[angle=0, width=0.8\textwidth ]{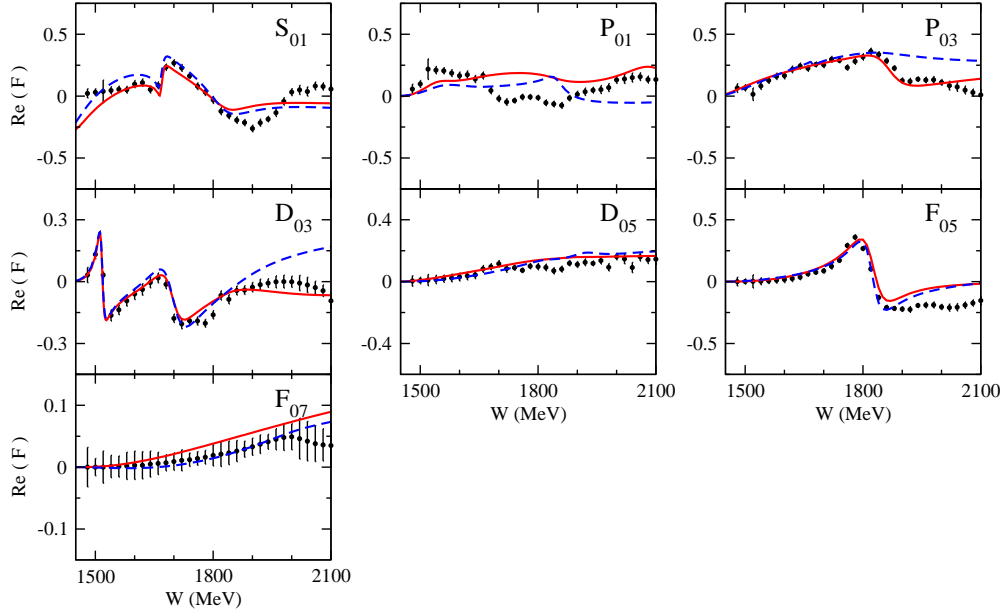}
\end{center}
\centerline{\parbox{0.80\textwidth}{
\caption{(Color on-line) Real part of the $\overline{K}N\to\overline{K}N$ amplitude in the 
	isospin channel of the $\Lambda(1405)$1/2$^-$ ($I=0$). The graph is 
	taken from Ref.~\protect\cite{Kamano:2014zba} and shows two solutions 
	of the ANL/Osaka group compared to the single-energy solution of the 
	Kent State group~\protect\cite{zhang13}. The near-threshold 
	$S$-wave amplitude $S_{01}$ is crucial for UChPT calculations.}
	\label{fig:anl} } }
\end{figure}

As an example of how current PWAs disagree on the near-threshold S-wave 
components, we show in Fig.~\ref{fig:anl} the recent result of S-wave 
extraction from the ANL/Osaka group~\cite{Kamano:2014zba} compared to the 
single-energy solution of the Kent State group~\cite{zhang13}.

As stated in the analysis of Ref.~\cite{Kamano:2014zba}, the data situation 
clearly does not allow to further pin down the partial-wave content. Like for 
the $\pi N$ case, measured data are old; a clear case can be made to remeasure 
the $\overline{K}$-induced reactions to provide more precise input to literally 
hundreds of theoretical studies on the $\overline{K}N$ interaction. 
Polarized measurements are of special relevance to this end. 

So far, we have discussed the hadronic properties of the 
$N$(1535)$1/2^-$ and the $\Lambda$(1405)$1/2^-$ in the context of chiral 
dynamics. While these resonances may be the most prominent examples of dynamical 
generation in the meson-baryon sector, much more theoretical effort has been 
dedicated to excited baryons. For example, the interaction of pseudoscalar mesons 
with the baryon decuplet leads to the prediction of several baryonic states, some 
of which were identified with known resonances~\cite{Sarkar:2004jh,Kolomeitsev:2003kt}. In 
particular, the EM properties were evaluated and found to be in good agreement 
with experimental measurements of the $\pi^0\eta p$ final 
state~\cite{Doring:2005bx,Ajaka:2008zz,Doring:2010fw}. For further investigation, 
the experimental study of pion-induced two-meson production is 
crucial~\cite{Doring:2006pt} 
and also of relevance for three-body
models of resonance generation~\cite{MartinezTorres:2007sr,Khemchandani:2008rk}.
Similarly, the interaction of vector mesons with 
baryons has lead to the prediction of excited states~\cite{Oset:2009vf}, with a 
dynamically generated $I=1/2$ $J^P=3/2^-$ resonance (first predicted in 
Ref.~\cite{Doring:2009yv}). Such interactions lead necessarily to strong 
contributions in the $\pi\pi N$ channel.

In summary, unitary ChPT leads to the prediction of several baryonic states. 
Some of them have not yet been found, and others await experimental 
confirmation through improved measurement of pion-induced reactions. While 
EM properties have been evaluated with great success, it is ultimately the 
precise knowledge of the pion- and kaon-induced family of reactions that can 
clearly rule out or confirm UChPT predictions.

\subsection{Lattice QCD}
\label{sec:lattice}
Lattice QCD simulations for excited baryons are considerably more complicated 
than for excited mesons due to signal-to-noise and combinatorial 
problems of contractions of three quarks instead of two. The first true 
extraction of pion-nucleon phase shifts in the $J^P=1/2^-$ sector was
achieved by the Graz group~\cite{Lang:2012db}. For the baryon problem in the 
bound-state case, see results from the Hadron Spectrum and BGR 
Collaborations~\cite{Edwards:2011jj,Engel:2013ig}. 

Chiral extrapolation for the $J^P=1/2^-$ sector was made in 
Ref.~\cite{Doring:2013glu} for two typical lattice setups. In particular, it 
was shown that unphysical quark masses lead to a re-ordering of thresholds.
Thus, the excited baryon spectrum is much more difficult to disentangle; hidden 
poles of the amplitude appear in the unphysical regime, qualitatively changing 
the spectrum of excited states. 

The chiral extrapolation to physical quark masses is also problematic in other cases. For example, the unquenched lattice QCD spectrum shows eigenvalues
close to the physical point that are 300 or 400 MeV too high compared to the $N(1440)1/2^+$ Roper resonance position~\cite{Edwards:2011jj, Roberts:2013ipa}. This is the so-called Roper puzzle. 
More precise experimental input is required to improve identification of lattice eigenvalues with physical states, especially from pion-induced reactions.

\section{Current Hadronic Projects}
\label{sec:current_projects}

Past measurements involving pion and kaon scattering measurements were made at a variety of laboratories, mainly in the 1970s and 1980s when experimental techniques were far inferior to the standards of today.  In the United States, pion beams in the momentum range 190~MeV/$c$ to 730~MeV/$c$ were available at the ``meson factory'' LAMPF in Los Alamos.  This means that the maximum c.m.\ energy for baryon spectroscopy measurements at LAMPF was only $W \approx 1500$~MeV. LAMPF was a linear accelerator for 1000~$\mu$A of protons at 800~MeV. The meson factory SIN (now PSI) near Zurich was a sector-focused cyclotron capable of 100~$\mu$A of protons at 600~MeV, and the meson factory TRIUMF in Vancouver was a sector-focused cyclotron for negative hydrogen ions up to 100~$\mu$A at 500~MeV \cite{ericson}. Most of the data available today for $\pi^- p \to \pi\pi N$ were collected with a bubble chamber at the Saturne accelerator in Saclay during the early 1970s, while most of the world's previous data for $\pi^- p \to K^0 \Lambda$ and several other reactions were collected using optical spark chambers during the late 1970s at the 7-GeV proton accelerator NIMROD in the UK. 
Most of the previous measurements for $\pi^- p \to e^+ e^- n$ were performed in Dubna, but the statistical uncertainties are so large that a high-quality analysis cannot be performed.  The flux of kaon beams is typically a factor of 500 or more less than that of pion beams.
 By today's standards, these accelerators had low beam intensities and, of course, detector technologies have advanced greatly since that time. 

It is important to recognize that current and forthcoming hadronic projects are largely complementary to the proposed hadron beam facility. We summarize the status of the J-PARC, HADES, COMPASS, and PANDA efforts here.

J-PARC provides separated secondary beam lines up to 2~GeV/$c$.
The primary beam intensity is currently 25\ kW, and will be increased to 
100\ kW in three years. This corresponds to 
$\sim10^{9}$ ppp (protons per pulse) for pion beam intensity and to $\sim10^{6}$ ppp for kaon beam intensity.
The $K/\pi$ ratio is expected to be close to 10, which is realized with
double-stage electrostatic separators.
There are plans to collect data for $\pi^\pm p\to\pi^\pm p$,
$\pi N\to\pi\pi N$, and $\pi N\to KY$ in 2016 or later at the 30-GeV
proton synchrotron \cite{hicks}. The delay will depend on the resolution of the
accident that occurred in the summer 
of 2013 \cite{JPARC_accident}.  
Recovery work at the Hadron Facility is continuing with an aim to start experimental programs soon, especially using intense kaon beams. In early 2015, the spectral information of $\Lambda(1405)$ in three different charge decay modes will be minutely inspected in the $K d \to \Lambda(1405) n$ reaction. Also multi-hadron states such as the $H$ dibaryon and kaonic nuclei will be studied in $(K^-,K^+)$ and $^{3}{\rm He}(K^-,n)$ reactions, respectively.
In the spring of 2016, a high-momentum beam line will be newly
constructed.
It is designed as a primary beam line and also unseparated secondary beam
line up to 20~GeV/$c$. 
 There are plans to study meson production in $\pi^{-}p$ reactions leading to $\eta n$, $\eta' n$, $\rho^0 n$, $\omega n$, and $\phi n$ final states using the higher momentum pion beams. In addition, charmed baryons will be studied systematically in the $\pi^{-}p \to D^{*} \Lambda^{*}_{c}$ reaction and cascade baryons will be studied using tagged kaons in the $K^{-}p \to K^+ \Xi^{*-}$ reaction.
 As a future project, an extension of 
the Hadron Facility is now being intensively discussed. A key feature 
will be a separated kaon beam line up to 10~GeV/$c$, which will be 
realized with RF-type separators. One of the main programs will be a systematic study of $\Omega$ baryons.

HADES at GSI collected unpolarized data for $\pi^-p\to\pi^-p$, $\pi^+\pi^-n$, $\pi^0\pi^-p$,
$\pi^0\pi^0n$, $e^+e^-n$ in August and September of 2014.
Data taking is planned to continue from 2017 on when
SIS~18 will be available again. The addition of a calorimeter is being considered to access better the $\eta n$ final state.
EPECUR at ITEP collected unpolarized differential cross-section data for $\pi^\pm p\to\pi^\pm p$ back to 2009--2011.
There is no chance to continue this program due to the accident with the
ITEP 10-GeV proton synchrotron \cite{ITEP_accident}.

The COMPASS experiment at the CERN SPS is focused on the study of hadronic structure and spectroscopy. The primary tools are a high intensity muon beam and a 190 GeV pion beam. Currently, hadron structure is being probed by Drell-Yan measurements with transversely polarized protons.  Measurements of generalized parton distributions and semi-inclusive deep inelastic scattering will start in 2015 and run through 2017 \cite{Sandacz}.

The PANDA experiment will be one of the key projects at the Facility for Antiproton and Ion Research (FAIR) currently under construction at GSI.  PANDA is focused on studies of hadron structure, strange baryon spectroscopy, and hadron interactions.  Antiprotons produced by a primary proton beam will be filled into the High Energy Storage Ring (HESR), where they will undergo collisions with the fixed target inside the PANDA detector.  There is special interest in investigating the time-like form factor of the proton,
searches for glueballs, hybrids, molecules, and tetraquarks, and investigations of in-medium
effects. The HESR with PANDA and Electron Cooler will allow the storage of
$10^{10}$ - $10^{11}$ antiprotons with momentum resolution $dp/p < 4 \times 10^{-5}$.  The momentum range for
antiprotons will cover 1.5 to 15 GeV/$c$ and the electron range will be up to
9~GeV/$c$.

\section{What is Needed for Hadron-Induced Reactions}
\label{sec:what_is_needed}

The current run plans at these facilities will greatly improve the data
base; however, there are no plans for polarized measurements.  As mentioned previously, what is truly needed is a dedicated state-of-the-art facility with secondary pion and kaon beams.  Such a facility would need a large-acceptance detector and the availability of a polarized target. The creation of a hadron physics complex at the proposed JLab EIC would be a very efficient use of the expertise and infrastructure such a facility could provide.
 In particular, such a dedicated facility should be able to provide the
features listed in the following, together with a short summary of key
arguments made in this White Paper:

\begin{itemize}
\item
A polarized target. As argued, the simultaneous measurement of the spin-rotation parameters $A$ and $R$ on a polarized target, together with the cross section and the recoil polarization $P$ obtained from a conventional, unpolarized target, provides complete information about the amplitudes (which also removes any sign ambiguity from the constraining relation $A^2+R^2+P^2=1$). 
\item
Near-to-complete angular coverage for analyses to deliver reliable
partial waves.  Possibility of energy scan through the
resonance regions from $\pi N$ threshold up to $W \sim 2.5$~GeV.
\item
Detection of the final-state $\pi N$: A complete experiment is within
surprisingly close reach if a polarized target is available. Elastic
$\pi N$ scattering already has the most extensive database of all
pion-induced reactions. More precise low-energy data on $\pi N \to \pi N$ are required for chiral perturbation theory.
\item
Detection of the final-state $\eta$: Crucial for the coupled-channel
approach. The data are of very inferior quality beyond the
$N(1535)1/2^-$, precisely in the region of the much debated structure in
$\eta$ photoproduction on the neutron. New data can provide ``smoking
gun'' evidence for its nature and quantum numbers. Also, the $\eta N$
channel needs to be understood for the control of inelasticities in
partial-wave analyses. In photoproduction, the $\eta p$ channel is one
of the prime candidates for a complete experiment, and so could be the isospin-selective reaction  $\pi^- p \to \eta n$.
\item
Detection of the final-state $K\Lambda$: Through the self-analyzing
nature of the $\Lambda$, recoil polarization measurements are easy to
achieve. In combination with a polarized target, one would have a data
set that puts very tight constraints on partial-wave analyses, which could confirm resonances seen in $\gamma p \to K^+ \Lambda$ through an independent reaction.
\item
Detection of the final-state $K\Sigma$: The data situation is rather
poor, which does not allow for a reliable partial-wave analysis. Such
partial waves are urgently needed to test the nature of some resonances
claimed to be hadronic molecules.
\item 
Detection of final-state $\pi\pi N$ channels: Crucial for the coupled-channel approach, in particular in the context
of hybrid baryons~\cite{hicks}.  The available data are extremely sparse and precise new data for $\pi^+\pi^- n$, $\pi^\pm \pi^0 p$, and $\pi^+ \pi^+ n$ measured with polarized and unpolarized targets are critically needed to determine $\pi\Delta$, $\rho N$, and other couplings in combination with photoproduction data.
\item Detection of final-state $e^+ e^- n$ channel: Inverse pion electroproduction measurements will significantly complement electroproduction $\gamma^\ast N \to \pi N$ studies for the evolution of baryon properties with increasing momentum transfer by investigation of the case for the time-like virtual photon.

\end{itemize}

These specifications are rather qualitative at this point. It will
require both experimental and theoretical simulations to proceed further
and formulate quantitative requirements. This task is beyond the scope of
this White Paper.


%
%
%

\section{Summary}


The goals of current EM facilities would benefit greatly from having hadron-beam data of a quality similar to that of electromagnetic data. To this end, it is commonly recognized that a vigorous U.S.\ program in hadronic physics requires a modern facility with pion and kaon beams.  A pion beam and a facility in which $\pi N$ elastic scattering and the reactions $\pi^- p \to K^0 \Lambda$, $\pi^- p \to K^0 \Sigma^0$, $\pi^- p \to K^+ \Sigma^-$, and $\pi^+ p \to K^+ \Sigma^+$ can be measured in a complete experiment with high precision would be very useful.  Full solid angle coverage is required to study inelastic reactions such as $\pi^- p \to \eta n$, $\pi^+ p \to \pi^0 \pi^+ p$, or strangeness production (among many other reactions).  Such a facility ideally should be able to allow baryon spectroscopy measurements up to center-of-mass energies $W$ of about 2.5~GeV, which would require pion beams with momenta up to about 2.85~GeV/$c$.  The 2~GeV/$c$ pion beam at J-PARC will allow baryon spectroscopy measurements up to $W \approx 2150$~MeV.

In this White Paper, we have outlined some of the physics programs that could be advanced with a hadron-beam facility. These include studies of baryon spectroscopy, particularly the search for ``missing resonances'' with hadronic beam data that would be analyzed together with photo- and electroproduction data using modern coupled-channel analysis methods. A hadron beam facility would also advance hyperon spectroscopy and the study of strangeness  in nuclear and hadronic physics.

Furthermore, searches for highly anticipated, but never unambiguously observed, exotic states such as multiquarks, glueballs, and hybrids, would be greatly enhanced by the availability of a hadron beam facility. 
Simply observing many of the missing low-lying meson states would also assist in constructing new models of the emergent properties of QCD, thereby improving our understanding of this strongly coupled quantum field theory. Improved
low-energy $\pi N$ and $\overline{K}N$  scattering data are also critically needed to provide input for model-independent chiral perturbation theory analyses. In particular, precise data will allow for a statistically sound determination of low energy constants and their uncertainties.

An electron-pion collider would open exciting new opportunities to measure the pion's EM form factor directly, while a pion beam alone would allow detailed studies of inverse pion electroproduction, which is the only process that allows the determination of EM nucleon and pion form factors in the case of time-like virtual photons.

Finally, a state-of-the-art  hadron beam facility could be used to investigate a much wider range of physics than baryon and meson spectroscopy alone.  For example, it could be used for studies of pion diffractive dissociation to two jets ($\pi + A \to$~2~jets~$+ X$), pion double-charge exchange ($A(\pi^+,\pi^-$)) at high energies, hypernuclear spectroscopy, inelastic scattering of mesons on nuclei to study in-medium effects, neutrino physics using neutrinos from the decays of pions and kaons, physics with muons produced from the decays of pions and kaons (e.g., for studies of lepton number violation using $\mu^+ \to e^+ \gamma$ or $\mu^+ + e^- \to \mu^- + e^+$), physics with $K^+$ and $K_L^0$ beams, meson-$A$ interactions of mesons with nuclei outside of the valley of stability, and dibaryons.  

We include at the end of this White Paper a list of endorsers who have expressed support for the initiative described herein.



\section*{Acknowledgements}
 
The authors are grateful to all of our colleagues who made suggestions about improving this paper, especially Drs. Yakov Azimov, Reinhard Beck, David Bugg, Daniel Carman, Frank Close, Evgeny Epelbaum, Alessandra Filippi, Avraham Gal, Gary Goldstein, Christoph Hanhart, Robert Jaffe, Hiroyuki Kamano, Nikolai Kivel, Franz Klein, Friedrich Klein, Eberhard Klempt, Boris Kopeliovich, Vladimir Kopeliovich, Anna Krutenkova, Bastian Kubis, Matthias Lutz, Maxim Mai, Terry Mart, Ulf-G.\ Mei{\ss}ner, Volker Metag, Gerald Miller, Viktor Mokeev, Ulrich Mosel, Takashi Nakano, Kanzo Nakayama, Yongseok Oh, Eulogio Oset, Jose Pelaez, Michael Pennington, Raquel Molina Peralta, John Price, Beatrice Ramstein, James Ritman, Deborah R\"onchen, Mikhail Ryskin, Piotr Salabura, Carlos Salgado, Andy Sandorfi, Andrei Sarantsev, Hartmut Schmieden, Vitaly Shklyar, Cole Smith, Greg Smith, Eugene Strokovsky, Joachim Stroth,  Antoni Szczurek, Kazuhiro Tanaka, Ulrike Thoma, Willem van Oers, Gerhard Wagner, Colin Wilkin, Ron Workman, Stan Yen, Yuhong Zhang, Vladimir Zelevinsky, and Bing-Song Zou.  This material is based upon work partially supported by the U.S.\ Department 
of Energy, Office of Science, Office of Nuclear Physics, under Award Numbers 
DE-FG02-99-ER41110, DE-FG02-00ER41135, and DE-FG02-01-ER41194, by the 
National Science Foundation under award number 1415459, and by the National Research Foundation of Korea, Grant No.\ NRF-2011-220-C00011.


\appendix
\section*{Endorsements}

\begin{center}
Igor Alekseev$^1$, Moskov Amaryan$^2$, Annalisa D'Angelo$^3$, Makoto Asai$^{4}$, Harut Avakian$^5$, Yakov Azimov$^6$, Marco Battaglieri$^7$, David Bugg$^8$, Daniel Carman$^5$, Sasa Ceci$^9$
% ...
\end{center}

\hspace{0.15in}

$^1$ Institute for Theoretical and Experimental Physics, Moscow, Russia\\
$^2$ Old Dominion University, Norfolk, Virginia, USA \\
$^3$ University Roma Tor Vergata, Rome, Italy \\
$^4$ SLAC National Accelerator Laboratory, Menlo Park, California, USA \\
$^5$ Jefferson Lab, Newport News, Virginia, USA \\
$^6$ Petersburg Nuclear Physics Institute, Gatchina, Russia \\
$^7$ Istituto Nazionale di Fisica Nucleare, Genova, Italy \\
$^8$ Queen Mary University of London, London, England, UK \\
$^9$ Rudjer Bo\v{s}kovi\'c Institute, Zagreb, Croatia \\
% ...


\end{document}